\documentstyle[12pt,epsf]{article}

\newcommand{\be}{\begin{equation}}
\newcommand{\ee}{\end{equation}}
\newcommand{\ba}{\begin{eqnarray}}
\newcommand{\ea}{\end{eqnarray}}

\newcommand{\leps}{ \epsilon }
\newcommand{\reps}{ \bar{\epsilon} }

\begin{document}
                                             \begin{flushright}
                                             SPbU-IP-97-13 \\
                                             hep-ph/9707xxx
                                             \end{flushright}

\thispagestyle{empty}
\vskip 2.5cm

{\large
\centerline{
                Phenomenological
        $\pi N \to \pi \pi N$
                  Amplitude
                and Analysis of Low Energy Data
           }
\centerline{
                on Total Cross Sections and
                1--D Distributions
           }
}
\vskip0.3cm
\centerline{ \bf
                A.A.~Bolokhov
           }
\centerline{\sl
                Sankt-Petersburg State University,
                Sankt-Petersburg, 198904, Russia }

\centerline{ \bf
                 M.V.~Polyakov and S.G.~Sherman }
\centerline{ \sl
St. Petersburg Institute for Nuclear Physics,
Sankt--Petersburg, 188350, Russia }
\vskip 0.5cm

\centerline{\bf Abstract }
\vskip0.2cm

\noindent
                We develop the phenomenological amplitude
                of the
        $\pi N \to \pi \pi N$
                reaction
                describing the exchanges of
        $\Delta$
                and
        $N_{*}$
                along with the OPE mechanism.
                The contribution of the latter contains
                4 independent low energy parameters
                (up to
        $O(k^{4})$
                order).
                The terms of the polynomial background
                are added to stand for far resonances and for
                contact terms originating from the
                off--mass--shell interactions.
                These terms are introduced with the account of
                isotopic, crossing,
        $ C $,
        $ P $
                and
        $ T $
                symmetries of strong interactions.

                The data consisting of total cross sections
                in the energy region
        $ 0.300 \le P_{\rm Lab} \le 500 $~MeV/c
                and 1D distributions from the bubble--chamber
                experiments
                for three reaction channels
                were undergoing fittings to determine
                free parameters of the amplitude.
                The best solutions are characterized by
        $ \chi^{2}_{\rm DF} =  $ 1.16.
                At the
                considered energies
                the isobar exchanges are found to be more
                important than OPE.
                The obtained solutions reveal the need in more
                precise data and/or in polarization
                measurements because of large correlations
                of isobar parameters with the OPE ones.\par
                The theoretical solutions were used for modeling
                the Chew--Low extrapolation
                and the Olsson--Turner threshold approach.
                It is shown that the noncritical application
                of the former results in
                100\%
                theoretical errors,
                the extracted values being in fact the random
                numbers.
                The results of the Olsson--Turner method
                are characterized by significant systematic
                errors coming from unknown details
                of isobar physics.

\vskip 3.2cm
\centerline{
                     Sankt--Petersburg }
\centerline{
                           1997 }

\newpage
\section{ Introduction }


                The pion--production reactions play
                the important role in the low energy physics
                of elementary particles and nuclei.
                The elementary processes
        $\pi N \to \pi \pi N$,
        $\gamma N \to \pi \pi N$
                of pion production on nucleons are
                currently in the focus of investigations
                due to the progress
                of Chiral perturbation theory
        (ChPT)
                the foundation of which
                was created by the Weinberg's work
\cite{Weinberg79}
                and principal steps were done in the series of
                papers by Gasser and Leutwyler
\cite{GasserL8485}.
                (For more references and reviews
                of this approach as well as more recent trends
                one can use the book by Donoghue, Golowich
                an Holstein
\cite{DonoghueGH92}
                and the review papers by Meissner
\cite{Meissner93}
                and Pich
\cite{Pich95};
                a deep insight into the contemporary
                interpretation of
                ChPT
                and the role of spontaneous symmetry breaking
                is provided by a series of preprints by
                Leutwyler
\cite{Leutwyler94}.)

                The predictions of
                ChPT
\cite{GasserL8485}
                (summary of the most interesting
                predictions as well as
                forthcoming experimental tests
                might be found in the talk
\cite{Pocanic94})
                explain the very sharp interest in the
                values of
        $ \pi \pi $--scattering lengths
                and other parameters of the
        $ \pi \pi $
                scattering.
                Since it is
                not possible
                experimentally
                to create the pionic target
                or the colliding pion beams
                there are only indirect ways
                for obtaining experimental data
                on the
        $ \pi \pi $
                scattering.
                The reactions
        $  \pi N \to \pi \pi N $
                and
        $  K \to \pi \pi e {\nu} $
                are considered as the most important sources
                of the (indirect) information
                on low energy characteristics of the
        $ \pi \pi $
                interaction.
                (The former reaction
                in what follows will be simply referred as
        $\pi 2 \pi$
                when possible.)

                The review of methods of extracting the
                latter characteristics from
        $\pi 2 \pi$
                data
                being in use previously
                and details of their applications
                might be found in the paper
\cite{Leksin70}
                by Leksin.
                The methods considered for application to
                the modern experiments
                are the following
\cite{BolokhovSChD94}.

          {\bf 1.}
                The Chew--Low extrapolation procedure
                by Goebel, Chew and Low
\cite{ChewL}
                is an apparently model--independent approach.
                It can provide the complete information on the
        $ \pi \pi $
                cross section
                provided the OPE dominates
                and the interval
                of the nucleon momentum transfer
     $ \tau $
                (which equals the mass of the virtual pion)
                allows an unique extrapolation.
                The last condition and the need for
                sufficient statistics shifts the region
                of application of the Chew--Low procedure
                to rather large values of energy
     ($ \approx $ 1 -- 4  GeV).
                When comparing the phase space
                of momentum transfer
     $ -20\mu^2 < \tau < -0.2 \mu^2 $
                at
     $ P_{\rm Lab}=500 $~MeV/c
                with the distance of extrapolation
     $ \approx \mu^2 $
                it becomes obvious
                that provided there are enough statistics,
                the kinematics of the
        $ \pi 2 \pi $
                reaction itself does not prevent
                the use of the Chew--Low
                procedure at much smaller energies.
                It is the presence of contributions like that of
     $ \Delta $ and $ N_{*} $
                isobars
                which makes a straightforward extrapolation
                difficult at moderate energies
                due to the perturbation of the simple
     $ \tau $--dependence
                of the OPE graph.
                The absolute values of all other contributions
                are killed at the extrapolation point
     $ \tau = \mu^2 $,
                but the result of extrapolation is known to be
                sensitive to the shape of the extrapolating
                curve
\cite{MartinMS76,Leksin70}.

          {\bf 2.}
                In view of importance of concurrent mechanisms
                at intermediate energies
                the approach might be changed to determining
                the OPE parameters directly
                in the physical region of
                the reaction --- this was implemented
                by the model of Oset
                and Vicente--Vacas
\cite{OsetV85}.
                It is clear that the neglect of a specific
                resonance contribution and/or the account
                of another one
                are capable to provide a lot of derivatives of
                the Oset--Vicente model.

                There is the energy region below
     $ P_{\rm Lab}=500 $~MeV/c,
                where the
                variation of
        $ \tau $
                is sufficient to detect the OPE contribution
                since the contributions of the concurrent
                processes (being nonnegligent)
                are smooth enough.
                The model
\cite{BolokhovVS91}
                takes these features into account
                and
                naturally completes the Oset--Vicente approach
                in this specific energy domain.

          {\bf 3.}
                The investigations by Olsson and Turner
\cite{OlssonT686972}
                are confined
                to the threshold  of
     $ \pi 2 \pi $
                reactions.
                Since the
                phase space of
        $ \tau $
                variable shrinks to the point
        $ \tau_0 = - 2.31 \mu^2 $
                the application of the Chew--Low procedure
                is  impossible there.
                The idea is to take advantage of Chiral
                Dynamics at the
        $ \pi 2 \pi $
                threshold.

                The important results of the approach are
                the formulae expressing the
        $ \pi \pi $--scattering
                lengths
                in terms of the
                threshold characteristics of the
                pion--production reactions.
                These formulae have gained a broad scale of
                application, especially in the recent years when
                new data on the
        $ \pi N \to \pi \pi N $
                reactions in the close--to--threshold
                energy region became available
\cite{Kernl-0p89,Kernl+-n89,Kernl++n90,KernelEtAl91},
\cite{SeviorEtAl91,Lowe_00n91,PocanicEtAl94}.

                The evidence of the importance
                of next--to--leading order terms
                of Chiral Lagrangian
                for the
        $\pi N$
                interaction and, in particular, for the
        $\pi 2 \pi$
          amplitude
\cite{Meissner93},
                makes it necessary to modify
                the Olsson--Turner method.
                Recently
                the approach of heavy baryon approximation
                was used to derive corrections
                to the Olsson--Turner formulae
                and to make direct predictions of
                ChPT
                for the threshold
        $\pi 2\pi $
                amplitude itself
\cite{BernardKM94}.

\vskip0.5cm

                The recent results
\cite{KnechtMSF95}
                of the so called
                Generalized ChPT approach
\cite{JSternSF93}
                and the progress
                in the two--loop ChPT calculations
\cite{BijenesCEGS96,KnechtMSF95}
                are claiming for more precise
                experimental information on the
        $\pi\pi$
                interaction at low energies at
        $ O ( k^{6} ) $
                order.
                Some experiments listed in
                ref.
\cite{Pocanic94}
                had already been finished:
                BNL and LAMPF results on total cross
                sections have been published
\cite{Lowe_00n91},
\cite{PocanicEtAl94},
                1D-distributions have appeared recently in
                the WWW
                (home pages
\newline
\centerline{
     {\it http://helena.phys.virginia.edu/\~{}pipin/E1179/E1179.html},
           }
\centerline{
     {\it http://helena.phys.virginia.edu/\~{}pipin/E857/E857.html}),
           }
                higher distributions are in progress,
                the off--line treatment of experimental tapes
                of the TRIUMF experiment
\cite{SmithEtAl95}
                will be completed soon.
                Therefore,
                it is timely to make the solution
                which theoretical method
                from the above list
                is capable to provide more reliable
                treatment of the modern
        $ \pi 2 \pi $
                experiments.


                The main goal of the present work
                is to provide grounds for comparison
                of the listed approaches.
                For this purpose we develop
                the most extensive phenomenological
                amplitude of the considered reaction
                suitable for near--threshold and
                intermediate energy regions.

                We are basing on the approach
                {\bf 2.}
                since both the Chew--Low extrapolation
                and the Olsson--Turner threshold formulae
                can not provide any hint
                for cross--checking of the rest methods.
                We try to fix the phenomenological
                amplitude by fitting the data on
                total cross sections and distributions
                of the reaction in question
                in the energy region from threshold up to
     $ P_{\rm Lab} \le 500 $~MeV/c.

                To avoid any doubt in the results
                in respect to correctness of acceptances,
                systematic errors, etc.
                the distribution data are chosen to be
                the bubble--chamber ones.
                This leaves us with
                rather
                old experiments
                (which are discussed in sect. 3).
                However,
                the significant part of the data has never
                been published and most its part eluded
                strong theoretical analysis
                (apart authors' checks of some
                isobar--like models).
                Therefore,
                it seems important
                to develop the tools of theoretical treatment
                of such data
                for
                determination of characteristics of
                pion--pion and pion--nucleon interactions
                along with other parameters
                of the phenomenological amplitude.

                It is worth noting that the considered
                reaction via unitarity relations is directly
                connected to the fundamental process of
                elastic pion--nucleon scattering
                at intermediate energies
                and other processes like
        $ \gamma N \to \pi \pi N $
                which gain the raising interest
                in the
                ChPT
                approach.
                It also enters
                the description of pion--nuclei
                scattering
                as an elementary process.
                Therefore,
                the structure of
        $\pi 2\pi $
                amplitude
                is of great importance for nuclear and
                particle physics.

                The paper is organized as follows.
                The content of sect. 2 reminds
                the basics of the low energy phenomenology
                of the
        $ \pi N \to \pi \pi N $
                reaction
                and describes the structure of our amplitude.
                Sect. 3 provides the summary of experimental
                data on distributions and total cross sections
                which we are analyzing.
                Sect. 4 is devoted to specifics of the
                fitting procedure and main results of
                analysis.
                The results of modeling
                the Olsson--Turner and Chew--Low approaches
                are exposed in sects. 5, 6
                along with the discussion of some properties
                of our amplitude and theoretical solutions.
                The summary, the concluding remarks
                and the discussion of the perspectives of
                the further development
                are given in Conclusions.
\newpage
\vspace{1.cm}
\section{
                Model of
               $\pi N \to \pi \pi N$
               Amplitude
}
\vspace{0.5cm}

                The principal features
                of the near--threshold phenomenology of the
        $ \pi N \to \pi \pi N $
                reaction had been already discussed
                in the paper
\cite{BolokhovVS91};
                in the quoted work
                the smooth background amplitude
                (+ the OPE one)
                had been derived
                for the energy domain bounded
                by the reaction threshold and
                the threshold
        $ P_{\rm Lab} \approx 500 $~MeV/c
                of the
        $ \Delta $--isobar
                creation in the final state.
                However,
                the statistically significant data on
        $ \pi 2 \pi $
                distributions
                (described in the next section)
                exist just for the boundary
                (and slightly above)
                of the pointed energy region.
                This makes necessary to modify
                the amplitude elaborated in ref.
\cite{BolokhovVS91}
                since the smoothness assumption
                is hardly to be valid there.

                In the current section
                we recall the phenomenology of
        $ \pi 2 \pi $
                processes to be taken into account
                (subsect. 2.2)
                and principal parts
                of the modified amplitude
                (subsect. 2.3)
                the parameters of which must be determined
                from the data fittings.
                We start with the brief description
                of the spin--isospin structure
                of the discussed amplitude.

\vspace{0.5cm}
\subsection{
            $\pi N \to \pi \pi N$
               Amplitude
}

\subsubsection{
                General Structure
}

                        In the isotopic space the amplitude
                $M^{abc}_{\beta \alpha} ( \lambda_f ; \lambda_i )$
                of the reaction
\be
\label{React}
                \pi^{a}(k_1) + N_{\alpha}(p;\lambda_i) \
                        \rightarrow   \
                        \pi^{b}(k_2) + \pi^{c}(k_3)
                        + N_{\beta}(q;\lambda_f)  \ \
\ee
                has 4 degrees of freedom and might be expressed
                either in terms of the definite isospin
                amplitudes or in terms of the isoscalar ones.
                Separating the nucleon spinor wave functions
\be
\label{SAmp}
                M^{abc}_{\beta \alpha} ( \lambda_f ; \lambda_i )
                \ = \ \bar{u}(q;\lambda_f)
                \hat{M}^{abc}_{\beta \alpha} (i \gamma_5)
                {u}(p;\lambda_i) \ ,
\ee
                where the
        $(i \gamma_5)$
                multiplier ensures the
                correct
        $P$--parity
                properties of the considered
                amplitude
                one can define the isoscalar amplitudes
        $ \hat{A} $,
        $ \hat{B} $,
        $ \hat{C} $,
        $ \hat{D} $
                by
\be
\label{IDec}
                \hat{M}^{abc}_{\beta \alpha} \ = \
                \hat{A} \tau^{a}_{\beta \alpha} \delta^{b c} +
                \hat{B} \tau^{b}_{\beta \alpha} \delta^{a c} +
                \hat{C} \tau^{c}_{\beta \alpha} \delta^{a b} +
                \hat{D} i \epsilon^{abc} \delta_{\beta \alpha} \ .
\ee

                        The analysis
\cite{BolokhovVS87}
                of the spinor properties of the amplitude
        (\ref{SAmp})
                allows to express each of the isoscalar functions
                $A,B,C,D$ in terms of 4 independent form factors
                in the crossing--covariant way
\ba
\nonumber
                \hat{A} \ = \ S_A +
                        \bar{V}_A \hat{k} +
                        {V}_A \hat{\bar{k}} +
                        i/2 \ T_A [ \hat{k} , \hat{\bar{k}} ] \ ;
\\
\nonumber
                \hat{B} \ = \ S_B +
                        \bar{V}_B \hat{k} +
                        {V}_B \hat{\bar{k}} +
                        i/2 \ T_B [ \hat{k} , \hat{\bar{k}} ] \ ;
\\
\nonumber
                \hat{C} \ = \ S_C +
                        \bar{V}_C \hat{k} +
                        {V}_C \hat{\bar{k}} +
                        i/2 \ T_C [ \hat{k} , \hat{\bar{k}} ] \ ;
\\
\label{IAmp}
                \hat{D} \ = \ S_D +
                        \bar{V}_D \hat{k} +
                        {V}_D \hat{\bar{k}} +
                        i/2 \ T_D [ \hat{k} , \hat{\bar{k}} ] \ .
\ea
                Here,
        ${k} , \bar{k}$
                are the crossing--covariant combinations
                of pion momenta
\ba
\label{vark}
                k =  - k_1 + \leps k_2 + \reps k_3  \ ;
        \; \; \;
                \bar{k} = - k_1 + \reps k_2 + \leps k_3  \ ,
\ea
                where
     $ \leps = \exp (2 \pi i / 3 )
                \ = \ - 1 / 2 + i \sqrt{3} / 2 $,
     $ \reps = \leps^{*} \ = \ - 1 / 2 - i \sqrt{3} / 2 $.
                These combinations together with
                the independent crossing--invariant ones
\be
\label{varQP}
                Q \equiv  - k_1 + k_2 + k_3 \
                = \ p - q \
                \; ; \; \;
                P \equiv  p + q \
\ee
                are used to define 5
                independent crossing--covariant
                scalar variables
\ba
\nonumber
                \tau = Q^2 \ ; \ \
                \theta = Q \cdot k \ ; \ \
                \bar{\theta} = Q \cdot \bar{k} \ ;
\\
\label{vars}
                \ \
                         \nu = P \cdot k \ ; \ \
                \bar{\nu} = P \cdot \bar{k} \ ,
\ea
                which completely
                determine the
                point in the phase space of the considered
                reaction.
                The expressions of all scalar products of
                particles' momenta are given in the paper
\cite{BolokhovVS91}
                for the case of the unbroken
                isotopic symmetry.
                The definitions
(\ref{vark}),
(\ref{varQP}),
(\ref{vars})
                are assumed for the physical particles
        ($ k_{2} $
                is the
        $ \pi^{-} $
                momentum in the reactions
        $\{ - + n \}$
                and
        $\{ - \, 0 \, p \}$;
                in the
        $\{ + \, 0 \, p \}$
                case
        $ k_{2} $
                is the
        $ \pi^{+} $
                momentum).
                All actual kinematical calculations
                in the computer programs are being processed
                with the isotopic symmetry breaking
                due to the particles' masses ---
                this complicates the expressions
                given in the
                quoted paper.
                For simplicity we shall hold on the
                unbroken isotopic symmetry case in the
                illustrations and in the discussions
                which follow.

                The
        $\tau$
                variable
                coincides with the mass of the virtual pion
                of the OPE graph.
                The
        $ 4 \pi $
                vertex of this graph
                is characterized also
                by the Mandelstam variables.
                The discussion of the off--shell
                dependence of the
        $ 4{\pi}  $ vertex
                on these variables
                is given in the paper
\cite{BolokhovVS88pi}.
                To avoid ambiguity
                we use only the dipion invariant mass
\be
         s_{\pi\pi}
                \equiv (k_{2} + k_{3})^{2}
\ee
                in the discussion below.

                        The amplitudes of the observable
                channels of the reactions
        $ \pi N \to \pi \pi N $
                with the
                convention for the normalization
                of the particle states
                adopted in
\cite{BolokhovVS91}
                are provided by relations
                (the nontrivial statistical factors
                are taken into account in the cases of the
        $\{0 \, 0 \, n\}$
                and
        $\{++n\}$
                channels):
\ba
\nonumber
                \hat{M}_{\{-+n\}} \ = \
                        \sqrt{2} / 2 \ (\hat{A} + \hat{C}) \ ;
        \; \;
                \hat{M}_{\{00n\}} \ = \
                       \ \ 1 / 2 \ (\hat{A}) \ ;
        \; \;
                \hat{M}_{\{++n\}} \ = \
                        1 / 2 \ (\hat{B} + \hat{C}) \ ;
\\
\label{ChAmp}
                \hat{M}_{\{-0p\}} \ = \
                        1 / 2 \ (\hat{C} - 2 \hat{D}) \ ;
        \; \;
                \hat{M}_{\{+0p\}} \ = \
                        1 / 2 \ (\hat{C} + 2 \hat{D}) \ .
\ea
                They have the same form as in eqs.
        (\ref{IAmp})
\be
\label{ChAmpform}
                \hat{M}_{X} \ = \ S_X +
                        \bar{V}_X \hat{k} +
                        {V}_X \hat{\bar{k}} +
                        i/2 \ T_X [ \hat{k} , \hat{\bar{k}} ] \ ;
                \;\;
                        X = (
                                \{-+n\},
                                \{-0p\},
                                \{00n\},
                                \{++n\},
                                \{+0p\}
                            )
                \; ,
\ee
                where
                the spinor structures are defined according
                to the expansions
        (\ref{ChAmp}), e.g.
        $
                S_{\{-+n\}} \ = \ \sqrt{2} / 2
                                  (
                                        S_A + S_C
                                  )
                \; ,
        $
                etc.
                In practice,
                the following combinations of the vector
                structures
\be
\label{VRVI}
                V^{R}_{X} \equiv ( V_{X} + \bar{V}_X ) / 2
                \; ; \; \;
                V^{I}_{X} \equiv ( V_{X} - \bar{V}_X ) / (2i)
                \;
\ee
                are being used in the course of calculations.

\subsubsection{
               Cross Section
}
                The experimental data of the channel
        $ X $
                are compared with the theoretical
                cross sections
        $ \sigma_{(\alpha)}^{\rm Th}  $
                for a given experimental point
        $ (\alpha) $
\begin{equation}
\label{cs}
  \sigma_{(\alpha)}^{\rm Th} =  \frac{\sigma_c}
                                     {4J}
           \int
                 \frac {d^3 k_{2}}  {(2\pi)^3 2 k_{2 0}}
                 \frac {d^3 k_{3}}  {(2\pi)^3 2 k_{3 0}}
                 \frac {d^3 q}  {(2\pi)^3 2 q_{0}}
           (2\pi)^4 \delta^4 (p+k_1-q-k_2 -k_3)
                        ||M||^2
                        \Theta_{\alpha}
        \; .
\end{equation}
                Here,
        $ \sigma_c \equiv (\hbar c)^2 =
          0.38937966(23) [{\rm GeV}^2 \, m{\rm barn}] $
                is the conversion constant,
        $$ 4J = 4\sqrt{(p \cdot k_{1})^2  - m_p^2 \mu_1^2} $$
                stands for normalization of the initial state
                and the characteristic function
        $  \Theta_{(\alpha)} = $
        $  \Theta_{(\alpha)} (p,k_1, $
        $  q,k_2 ,k_3) $
                of the bin
        $ (\alpha) $
                describes the appropriate cuts in the
                phase space
                (if any;
                in the case of the total cross section
                this function is equal to 1).
                The notation
        $ ||M||^{2} $
                stands for the squared modulus of the amplitude
                summed over polarizations of the final nucleon
                and averaged over ones of the initial proton
                (we shall call it simply
                {\it matrix element}).
                The statistical factor
                (equal to the product of
        $ 1 / n_{\nu}! $
                over subsets of identical particles)
                for calculation of the cross section
                was included into definitions
(\ref{ChAmp})
                of the physical reaction amplitudes.


                The matrix element
        $ ||M||^{2} \equiv ||M_{X}||^{2} $
                is the quadratic form of the
                vector of spinor structures
        $ ( S_{X}, V_{X}, \bar{V}_X, T_{X} ) $
                (or, the same, of the vector
        $ ( S_{X}, V^{R}_{X}, {V}^{I}_X, T_{X} ) $):
\ba
\nonumber
                {\| M_{X} \|}^2
         \equiv
                1/2 \ \sum_{\lambda_f, \lambda_i}
                  \left[
                           \bar{u}(q;\lambda_f) \hat{M}_{X}
                           (i \gamma_5) {u}(p;\lambda_i)
                  \right]
                                \;
                  \left[
                           \bar{u}(q;\lambda_f) \hat{M}_{X}
                           (i \gamma_5) {u}(p;\lambda_i)
                  \right]^{*}
\\
\label{SqAmp}
         =
        \left(
\begin{array}{c}
                S_{X}
\\
                V_{X}
\\
                \bar{V}_{X}
\\
                T_{X}
\end{array}
        \right)^{\dagger}
                          G
        \left(
\begin{array}{c}
                S_{X}
\\
                V_{X}
\\
                \bar{V}_{X}
\\
                T_{X}
\end{array}
        \right)
                           \ , \;
             (X=\{-+n\},\{-0 \ p\},\{0 \ 0 \ n\},
                                \{++n\},\{+0 \ p\}) \ ;
\ea
\ba
\label{Gmat}
                G
        & \equiv &
                \frac{1}{2}
                        {\rm Sp}
                        \left[
                          (\hat{q} + m)
        \left\{
\begin{array}{c}
                \hat{1}
\\
                \hat{k}
\\
                \hat{\bar{k}}

\\
                \frac{i}{2}
                [
                \hat{k},
                \hat{\bar{k}}
                ]
\end{array}
        \right\}
                          (\hat{p} - m)
                        \gamma_{0}
        \left\{
\begin{array}{c}
                \hat{1}
\\
                \hat{k}
\\
                \hat{\bar{k}}

\\
                \frac{i}{2}
                [
                \hat{k},
                \hat{\bar{k}}
                ]
\end{array}
        \right\}^{\dagger}
                        \gamma_{0}
                        \right]
                           \ .
\ea

                For the simplified case of equal pion masses
                the matrix
        $  G $
                of the above form
                is given in the paper
\cite{BolokhovVS91}.
                In practice,
                the matrix for every channel had been
                calculated separately with the physical
                masses of all particles in the considered
                channel.

                To plot the data and the theoretical results
                we define the
                {\it quasi--amplitude}
        $  \langle M_{(\alpha)} \rangle $
                which is the square root of the cross section
(\ref{cs})
                divided by the phase space
                in the case of the total cross section data;
                in the case of distributions both
                the cross section and the phase space are
                independently normalized to 1
                --- we call this quantity
                {\it normalized}
                quasi--amplitude
        $  \langle M_{(\alpha)} \rangle_{\rm norm} $:
\be
\label{qamp}
        \langle M_{(\alpha)} \rangle
               \equiv
                       \sqrt{
                            \frac{  \sigma_{(\alpha)}
                                    ( \| M \|^2 )
                                 }
                                 {  \sigma_{(\alpha)} (1) }
                             }
        \; , \; \;
        \langle M_{(\alpha)} \rangle_{\rm norm}
               \equiv
                       \sqrt{
                            \frac{  \sigma_{(\alpha)}^{n}
                                    ( \| M \|^2 )
                                 }
                                 {  \sigma_{(\alpha)}^{n} (1) }
                             }
        \; . \;
\ee
                Here, the phase space
        $  \sigma (1) $
                is the theoretical cross section
(\ref{cs})
                obtained with the unit matrix element.

\subsubsection{
               Threshold Properties
}

                At the threshold of the reaction there
                are considerable simplifications in the
                representations
        (\ref{IAmp})
                or
        (\ref{ChAmpform})
                since:
%
                a)
                the
                momenta of the outgoing pions are equal to each
                other
        $
                      (  \hat{k}_2 \ = \ \hat{k}_3 )
        $,
%
                b)
                the contribution of the
        $\hat{D}$
                amplitude to the amplitudes
        (\ref{ChAmp})
                becomes zero
                and
%
                c)
        $
                        \hat{B} \ = \ \hat{C}
        $.

                Another simplification takes place
                in the sum over final polarizations
                and the average over
                initial ones of the amplitude
        (\ref{SAmp})
                squared
                modulus.
                At the threshold
                it degenerates to
\ba
\label{SqAmp0}
                {\| M_{X} \|}^2 \ = \
                        (-\tau_0) [
                                   S^{0}_{X} -
                                   ( 2 m + 3 \mu )
                                   ( \bar{V}^{0}_X + V^{0}_X )
                                  ]^2
                           \ ,
\\
\nonumber
             (X=\{-+n\},\{-0 \ p\},\{0 \ 0 \ n\},
                                \{++n\},\{+0 \ p\}) \ ,
\ea
                where
        $
                \tau_0 \ = \
                - 3 \mu^2 m / ( m + 2\mu )
        $
                is the threshold value of the
                variable
        $\tau$.
                This provides the grounds to introduce the
                {\it threshold amplitudes}.

                        The isotopic
                threshold amplitudes are
\ba
\nonumber
                A^0
                & = & \sqrt{-\tau_0}
                        [S^{0}_{A} -
                           ( 2 m + 3 \mu )
                           ( \bar{V}^{0}_A + V^{0}_A ) ]
                           \ ;
\\
\label{ABC0}
                B^0
                & = & \sqrt{-\tau_0}
                        [S^{0}_{B} -
                           ( 2 m + 3 \mu )
                           ( \bar{V}^{0}_B + V^{0}_B ) ]
                           \ ;
\\
\nonumber
                C^0
                & = & \sqrt{-\tau_0}
                        [S^{0}_{C} -
                           ( 2 m + 3 \mu )
                           ( \bar{V}^{0}_C + V^{0}_C ) ]
                           \ ,
\ea
                where all form factors are to be calculated at
                the threshold values of kinematical variables
                --- this provides
\be
\label{B0C0}
                        B^0 \ = \ C^0 \ .
\ee

               To link the isotopic threshold amplitudes
        (\ref{ABC0})
               with the experimental information
               let us construct the quantities $M^0$
\ba
\nonumber
              M^0_{\{-+n\}} \ = \ (A^0 + B^0) \sqrt{2}/2 \ , \ \
              M^0_{\{-0p\}} \ = \ B^0 / 2 \ , \ \
              M^0_{\{00n\}} \ = \ A^0 / 2 \ ,
\\
\label{M0}
              M^0_{\{++n\}} \ = \ (B^0 + C^0) / 2
              \ = \ B^0 \ , \ \
              M^0_{\{+0p\}} \ = \ B^0 / 2 \ ,
\ea
                which include both the isotopic and
                the statistical
                factors.

                        Then the absolute values of the
                threshold amplitudes
        $M^0_X$
                might be expressed in terms of threshold limits
                of the experimental quasi--amplitudes
(\ref{qamp}):
\ba
\nonumber
        \langle M_{\{-+n\}} \rangle_{|_{s \to s_{0}}} =
                           | A^{0} + B^{0} | / \sqrt{2} \; , \;\;
        \langle M_{\{-0p\}} \rangle_{|_{s \to s_{0}}} =
                           | B^{0} | / 2 \; , \;\;
        \langle M_{\{00n\}} \rangle_{|_{s \to s_{0}}} =
                           | A^{0} | / 2 \; ,
\\
\label{Rmod}
        \langle M_{\{++n\}} \rangle_{|_{s \to s_{0}}} =
                           |  B^{0} |  \; ,  \;\;
        \langle M_{\{+0p\}} \rangle_{|_{s \to s_{0}}} =
                           | B^{0} | / 2 \; .
\ea

                    The  threshold amplitudes
        (\ref{M0})
                are dimensional and
                in the following their numerical values
                will be given in
        $[({\rm GeV})^{-1}]$.

                        The threshold limits
        (\ref{M0}),
                being
                determined by only two isotopic threshold
                amplitudes
        $A^0$
                and
        $B^0$,
                must satisfy three relations.
                The first two are straightforward:
\be
\label{R425}
              | M^{0}_{\{++n\}} | \ = \
              2 | M^{0}_{\{-0p\}} | \ = \
              2 | M^{0}_{\{+0p\}} | \ .
\ee

                From the definition
        (\ref{M0})
                it follows that
\be
\label{R413}
        M^0_{\{++n\}} = \sqrt{2}  M^0_{\{-+n\}}
                        - 2  M^0_{\{00n\}}  \ .
\ee
                This implies that,
                depending on the relative sign of
        $ A^0 $
                and
        $ B^0 $,
                the relation between
                positive quantities
        (\ref{Rmod})
                might be either
\be
\label{R41m3}
       | M^0_{\{++n\}} | = \sqrt{2} | M^0_{\{-+n\}} |
                        - 2 | M^0_{\{00n\}} |
\ee
                or
\be
\label{R41p3}
       | M^0_{\{++n\}} | = \sqrt{2} | M^0_{\{-+n\}} |
                        + 2 | M^0_{\{00n\}} |
\ee
                or
\be
\label{R4m1p3}
       | M^0_{\{++n\}} | = - \sqrt{2} | M^0_{\{-+n\}} |
                        + 2 | M^0_{\{00n\}} | \ .
\ee

                Now it is necessary to recall the known
                properties of
        $ \pi 2 \pi $
                cross sections.
                Since in the near threshold region the known
                cross sections
        $ \sigma_{\{00n\}}$
                of the
        $\{ 0 \; 0 \; n \}$
                channel are approximately equal
                to the cross sections
        $ \sigma_{\{-+n\}}$
                this excludes the possibility
        (\ref{R41m3})
                because it results in the negative RHS.
                In the whole region
        $ P_{\rm Lab} \leq 0.5 {\rm GeV/c} $
                the cross sections
        $ \sigma_{\{++n\}}$
                are considerably smaller than
        $ \sigma_{\{-+n\}}$
                and/or
        $ \sigma_{\{00n\}}$;
                so, the variant
        (\ref{R41p3})
                is impossible.
                Thus,
                combining with the previous two
                relations,
                we can state:
\be
\label{R431}
              2 | M^{0}_{\{-0p\}} | \ = \
              2 | M^{0}_{\{+0p\}} | \ = \
       | M^0_{\{++n\}} | \ = \
                         2 | M^0_{\{00n\}} |
                - \sqrt{2} | M^0_{\{-+n\}} | \ .
\ee

\subsubsection{
                Remarks on
                Sign Ambiguity
}

                To resume the discussion of
                the general structure of the
        $ \pi 2 \pi $
                amplitude
                let us consider briefly the problem of the
                sign ambiguity in the theoretical
                amplitude
                which is used to fit the
                experimental
                cross section data.
                The examination of relations
(\ref{Rmod})
                shows that already in this simplified
                threshold case
                apart the overall sign ambiguity of
                isotopic threshold amplitudes
        $ A^0 $,
        $ B^0 $
                their relative sign might also become
                indefinite
                depending on the accuracy
                of the experimental information.

                In what follows
                we call the solution
                {\it physical}
                ({\it unphysical})
                when
        $ A^0 $
                and
        $ B^0 $
                are of different (equal) signs.
                In terms of the
        $ \pi \pi $--scattering
                lengths
                the two cases of signs correspond
                to the different (equal) signs of
        $ a^{I=0}_{0} $
                and
        $ a^{I=2}_{0} $.

                In general there are no grounds to wait for
                a simplification to take place at a distance
                from the threshold.
                Indeed, the expression
(\ref{SqAmp})
                for the matrix element
        $ {\| M_{X} \|}^2  $
                might be brought to the diagonal form
                (for example,
                in terms of the analogues of the diagonal
                derivative amplitudes of Rebbi
\cite{Rebbi68})
                in which it becomes the sum of four squared
                modules of the orthogonal amplitudes.
                The abundance of solutions found
                in the course of data fittings
                should be explained in part by a variety
                of choices of the signs in the above four
                terms in the cross section of every channel.

\vspace{1.cm}
\subsection{
               Near--Threshold
        $\pi N \to \pi \pi N$
                Phenomenology
                Guidelines
}
\vspace{0.5cm}

                1. The current knowledge of physics
                of the pion--nucleon interactions provides
                no evidence of possible mechanisms or
                processes resulting in the strong variation
                of the amplitude of the
        $ \pi N \to \pi \pi N $
                reaction
                in the phase space at energies up to
        $   \approx 2 m_{\pi}$
                above threshold.
                All resonances but
        $N_{*}$
                and
        $\Delta$
                are located outside the phase space.

                2. The contribution of the near--threshold
        $\Delta$
                pole
                is only due to the chain
        $\pi N \to \Delta \to \pi \pi N$.
                It is suppressed
                1) by the negligible width of the decay
        $\Delta \to \pi \pi N$;
                2) by Quantum mechanics:
        $\Delta$
                is located precisely at the
        $\pi 2 \pi$
                threshold where the process
        $\pi N \to \pi \pi N$
                must proceed through the waves
        $P_{11}$
                and
        $P_{31}$
                of the initial pion--nucleon system
                (in the
        $L_{2I 2J}$
                notation)
                while
        $\Delta$
                can be created only in the
        $P_{33}$
                wave of the
        $\pi$--$N$
                system.

                3. The closest to the phase space contribution
                of the meson--resonance type is that of OPE ---
                all the other are very distant at the
                discussed energies.

                4. The tails of resonances result
                in some constant background
                in the physical domain
                and, at most,
                in some slow variation of the amplitude.
                (One would need the extraordinary precision of
                experimental data to distinguish between
                a linear function in the corresponding variable
                and the far pole.)

                5. The number of free (or unknown) parameters in
        $\pi \pi$, $\pi N$, $\pi \Delta$, $\pi N_*$
                interactions exceeds the number of degrees of
                freedom of the smooth near--threshold amplitude.

                6. All resonances contribute with terms
                built of various scalar products of momenta.
                There are at most five such products
                which might be considered to be independent or,
                the same, they are built of
                five invariant variables
                (at the fixed initial energy,
                four variables left).
                Hence, the arbitrary linear function
                is parametrized with 6 parameters:
                a constant + 5 coefficients at linear variables.
                This applies to any of 4 independent isospin
                amplitudes which describe 5 observable channels
                of the reaction.

                7. There are four different spinor structures
                in the
         $\pi 2 \pi$
                amplitude.
                Even in the case of the unpolarized experiment
                they provide the specific dependence
                on variables
                which might well be nonnegligent
                at some distance from the threshold.

                8. The Bose statistics, the isospin symmetry,
                the
        $C$--invariance
                and the crossing, being the exact properties
                of the amplitude derived according to
                the rules of the quantum field theory,
                restrict the amplitude dependence on variables,
                making some parameters vanish  and
                relating the rest parameters.

                9. As a result there might be
                about 20 parameters for the near--threshold
                description
                of all channels
                and, in particular,  no more
                than 15 for the case of the
        $\pi^{-} p \to \pi^{-} \pi^{+} n$
                process.
                By the unitarity conditions
                which relate the nonvanishing
        $\pi 2 \pi$
                isospin amplitudes only with the
        $P_{3 1}$
                and
        $P_{1 1}$
                waves of elastic
        $\pi N$
                amplitudes,
                the above parameters must be approximately
                real.
                Only few of 20 imaginary parts of the
                polynomial background are expected
                to be of importance.

                10. The discussed  structures
                must be combined with the explicit term
                describing the OPE amplitude
                and isobar contributions.
                This will provide about
                20 additional parameters if
        $ D $
                waves of the
        $\pi \pi$
                scattering
                and all structures of the vertices
        $\pi \pi N \Delta$,
        $\pi \pi N_{*} \Delta$
        $\pi \pi N N_{*} $
                are important.

                11. The cuts due to thresholds of another
                inelastic processes like the
        $3\pi$
                production, the
        $\eta$
                production, etc.
                are conspired by nearby isobars.

                12. The analysis of the paper
\cite{Manley84}
                makes it evident that isobar resonances
                saturate the existing data
                on total cross sections below 1 GeV.
                Therefore, the imaginary part of the
        $\pi 2 \pi$
                amplitude might be described by the
                Breit--Wigner form of isobar contributions
                and the discussed above parameters
                of the imaginary background
                might appear
                to be negligible.

\subsection{
               Resonance Contributions to
            $\pi N \to \pi \pi N$
               Amplitude
}


                The
        $\pi 2\pi $
                amplitude
                to be developed
                must cover the energy interval
                from the threshold to the isobar region.
                Therefore, the approach
        {\bf 2}
                mentioned in Introduction
                is the most suitable one
                for the discussed purpose
                since it admits the
                due account of isobar physics
                and can provide
                the maximal model independence
                of the obtained results.

                Let us first enumerate
                the resonance--exchange graphs
                which are proper to our reaction
                (it is suitable to use the two--particle
                channel for labelling).
                Three representatives
        $ {\rm SE}_{\pi N} $,
        $ {\rm SE}_{\pi \pi} $,
        $ {\rm SE}_{NN} $
                of
                inequivalent crossing--related
                single--pole graphs
                are shown
                in Fig. 1.
                Every four--particle vertex of
                the above graphs can be expanded in
                at most two different ways
                providing no more than three
                types of inequivalent
                double-pole graphs
                which we mark in
        $ {\rm DE}_{\pi N, \pi N} $,
        $ {\rm DE}_{\pi N, \pi \pi} $,
        $ {\rm DE}_{NN, \pi \pi} $
                (see Fig. 1).
                In what follows
                we shall use the self--explanatory notations
                like
        $ {\rm SE}_{\pi N} \{ \Delta \} $,
        $ {\rm DE}_{NN, \pi \pi} \{ \omega, \rho \} $
                when discussing contributions of classes
                of crossing--related graphs.
                It is evident
                that all distinct resonances
                and particles responsible for pole
                contributions to the amplitude
                enter already the single--pole scheme;
                the particles which might be relevant to
                the intermediate energy amplitude
                are listed in the Table
\ref{SEres}.
\begin{table}
\begin{center}
\begin{tabular}{||c|c||c|c||c|c||}
\hline
     $ {\rm SE}_{\pi N} $
                        &
        $ I(J^{P} ) $
                        &
                    $ {\rm SE}_{\pi \pi} $
                        &
        $ I^{G}(J^{P C_{n}}) $
                                                &
                                     $ {\rm SE}_{NN} $
                        &
        $ I^{G}(J^{P C_{n}}) $
\\
\hline
\hline
     $ N = ( p, n ) $
                        &
     $ \frac{1}{2}({\frac{1}{2}}^{+}) $
                        &
                  $ \sigma = f_{0} (400 \!-\!\! 1200) $
                        &
     $ {0}^{+}({0}^{++}) $
                                                &
                                 $ \pi = ( \pi^{\pm}, \pi^{0} ) $
                        &
     $ {1}^{-}({0}^{-+}) $
\\
     $ \Delta = \Delta (1232) $
                        &
     $ \frac{3}{2}({\frac{3}{2}}^{+}) $
                        &
                  $ \rho = \rho (770) $
                        &
     $ {1}^{+}({1}^{--}) $
                                                &
                                  $ \omega = \omega (782) $
                        &
     $ {0}^{-}({1}^{--}) $
\\
     $ N_{*} = N (1440) $
                        &
     $ \frac{1}{2}({\frac{1}{2}}^{+}) $
                        &
                  $ f = f_{2} (1270) $
                        &
     $ {0}^{+}({2}^{++}) $
                                                &
                                  $ A = a_{1} (1260) $
                        &
     $ {1}^{-}({1}^{++}) $
\\
     $ {N}' = N (1520) $
                        &
     $ \frac{1}{2}({\frac{3}{2}}^{-}) $
                        &
                  $ \rho' = \rho (1450) $
                        &
     $ {1}^{+}({1}^{--}) $
                                                &
                                 $ \pi' = \pi (1300) $
                        &
     $ {1}^{-}({0}^{-+}) $
\\
     $ \not\!\!{N} = N (1535) $
                        &
     $ \frac{1}{2}({\frac{1}{2}}^{-}) $
                        &
        $   \cdots $
                        &
                        &
        $   \cdots $
                                                &
\\
        $   \cdots $
                                                &
                                                &
                                                &
                                                &
                                                &
\\
\hline
\end{tabular}
\end{center}
\caption[SE graphs]{ Resonances and particles from PDG--96
\cite{PDG96}
                        responsible for pole contributions
                        to the low energy
        $  \pi N \to \pi \pi N $
                        amplitude. }
\label{SEres}
\end{table}
                There are enough representatives
                of the lowest spin--isospin states
                in the lists
        $ {\rm SE}_{\pi \pi} $,
        $ {\rm SE}_{N N} $.
                In the case of the
        $ {\rm SE}_{\pi N} $
                channel
                the list
                is far from being complete.
                In the present paper we limit ourselves
                by contributions from particles
        $ N $,
        $ \Delta $,
        $ N_{*} $,
        $ \sigma $,
        $ \rho $,
        $ \pi $,
        $ \omega $
                and
        $ A $;
                in fact,
                the interactions of
        $ \sigma $,
        $ \omega $
                and
        $ A $
                given below in the lists of lagrangians
                for the purpose of the forthcoming discussion
                were omitted from the actual analysis.



\begin{figure}[h]
   \centerline{\epsffile{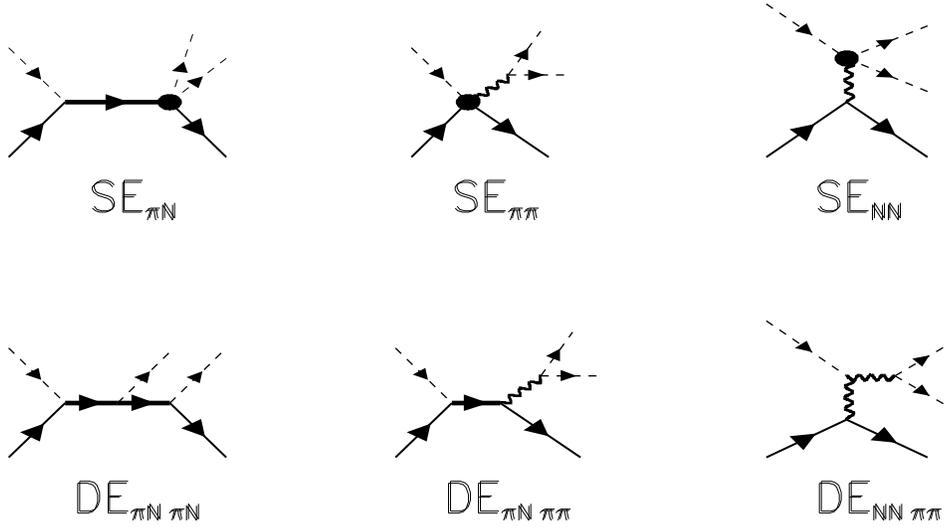}}
   \vspace{-2.cm}
   \centerline{
               \parbox{13cm}{
                              \caption{
                                        \label{grF}
                Representatives of graphs
                of resonance contributions.
                                      }
                             }
              }
\end{figure}

                Apart the OPE contribution our model is
                described by
                the effective interaction lagrangian
                which is used to construct the tree--level
                amplitude;
                all terms of the lagrangian are collected
                in Tables
\ref{SE4v},
\ref{DE3v}.
                In general,
                we are trying to use
                the minimal derivative coupling
                especially when particles
                of the nontrivial spin
                (for example, in the
        $ \rho N \Delta $
                vertex)
                are being involved.
                In the contrary case
                the terms of too high order in the momentum
                are inevitable to be present in the amplitude.
                However,
                in the case of the three--particle vertices
                containing the pion,
                like
        $ \pi N N $,
        $ \pi N \Delta $,
                etc.,
                the vertices are brought
                to the derivative--coupling form.
                Having in mind the complete equivalence of
                nonderivative and derivative couplings
                in such vertices
                (in the difference of the corresponding
                amplitudes the
                propagators
                always become contracted)
                we assume that only other types of
                vertices
                (for example,
        $ 4\pi $,
        $ \pi \pi N N $,
                etc.)
                are responsible for the explicit
                breaking of Chiral
                Symmetry.

                Even in the brief summary of properties
                of the considered interactions
                we must point
                that the form of the four--particle vertices
                is chosen to represent all spin--isospin
                structures of the given vertex.
                This makes
                impossible to
                fix the lagrangian parameters from the
                decay data
                (for example,
                the known width for
        $ N_{*} \to \pi \pi N $
                provides only bounds for 4 parameters
                of the
        $ \pi \pi N N_{*} $
                lagrangian given in the Table
\ref{SE4v}).
                However,
                we prefer to avoid any doubt in respect to
                a possible model dependence of the results
                for
        $ \pi \pi $--scattering
                lengths
                or, at least,
                to leave a chance to check the presence
                of such dependence.

                The central contribution to our amplitude
                is supposed to come from the OPE graph
        $ {\rm SE}_{N N} (\pi) $.
                It is parametrized basing on
                the cross--symmetric
                threshold expansion of the
        $ 4\pi $
                vertex
                elaborated in the paper
\cite{BolokhovEtAl96}.
                Apart the account of the imaginary part
                the expansion
                in the
        $ O (k^{4}) $
                order
                is equivalent to the form
                described in the papers
\cite{BolokhovVS87,BolokhovVS91}.
                The results of the present work
                which will be discussed below
                confirm that the precision of the
                currently available
                experimental data does not allow to
                consider higher terms of the
        $ \pi \pi $
                amplitude.


                We list in Tables
\ref{SE4v},
\ref{DE3v}
                below the interactions most part of
                which had been used for the construction
                of the phenomenological amplitude
                entering the data fittings.
                Since all 3--particle vertices
                of the SE series of graphs
                are present also in the DE graphs
                we do not describe them separately
                --- see Table
\ref{DE3v}.
                In Table
\ref{SE4v}
                the symbol
        $  V_{4 \pi} $
                stands for the
        $ {\pi \pi} $
                interaction
                --- this contribution is taken
                in the direct
                amplitude form.

\begin{table}
\begin{center}
\begin{tabular}{|c|c|}
\hline
\mbox{
\begin{tabular}{c}
    Graph\\
    \mbox{[Vertex]}
\end{tabular}
     }
                        &     Lagrangian
\\
\hline
\hline
        $ {\rm SE}_{\pi N} \{ N \} $
                        &
\\
        $
           [ {\pi \pi N N} ]
        $
             &
        $
                   g_{\pi \pi N N}^{1}
                \bar{N}
                {\delta}_{a b}
                N
                       \pi^{a}
                       \pi^{b}
                  +
                   g_{\pi \pi N N}^{2}
                \bar{N}
                {\delta}_{a b}
                N
                                \partial^{\mu} \pi^{a}
                                \partial_{\mu} \pi^{b}
        $
\\
 &
        $
              +
                   g_{\pi \pi N N}^{3}
                \bar{N}
                  i {\epsilon}_{a b c} \tau^{c}
                        \gamma_{\mu}
                N
                        [
                                \partial^{\mu} \pi^{a}
                                               \pi^{b}
                                -
                                               \pi^{a}
                                \partial^{\mu} \pi^{b}
                        ]
        $
\\
   &
        $
              +
                   g_{\pi \pi N N}^{4}
                \bar{N}
                  i {\epsilon}_{a b c} \tau^{c}
                       [ \gamma_{\mu} ,
                          \gamma_{\nu}
                       ]
                N
                                \partial^{\mu} \pi^{a}
                                \partial^{\nu} \pi^{b}
        $
\\
\hline
        $ {\rm SE}_{\pi N} \{ N_{*} \} $
                        &
\\
        $
          [ {\pi \pi N N_{*}} ]
        $
             &
        $
                   g_{\pi \pi N N_{*}}^{1}
                \bar{N}
                {\delta}_{a b}
                N_{*}
                       \pi^{a}
                       \pi^{b}
                  +
                   g_{\pi \pi N N_{*}}^{2}
                \bar{N}
                {\delta}_{a b}
                N_{*}
                                \partial^{\mu} \pi^{a}
                                \partial_{\mu} \pi^{b}
        $
\\
  &
        $
              +
                   g_{\pi \pi N N_{*}}^{3}
                \bar{N}
                  i {\epsilon}_{a b c} \tau^{c}
                        \gamma_{\mu}
                N_{*}
                        [
                                \partial^{\mu} \pi^{a}
                                               \pi^{b}
                                -
                                               \pi^{a}
                                \partial^{\mu} \pi^{b}
                        ]
        $
\\
  &
        $
              +
                   g_{\pi \pi N N_{*}}^{4}
                \bar{N}
                  i {\epsilon}_{a b c} \tau^{c}
                       [ \gamma_{\mu} ,
                          \gamma_{\nu}
                       ]
                N_{*}
                                \partial^{\mu} \pi^{a}
                                \partial^{\nu} \pi^{b}
                + \{ {\rm H.C.} \}
        $
\\
\hline
        $ {\rm SE}_{\pi N} \{ \Delta \} $
                        &
\\
        $
          [ {\pi \pi N \Delta} ]
        $
             &
        $
                \bar{N}
                (
                   g_{\pi \pi N \Delta}^{1} F^{0}_{d b c}
                + 3
                   g_{\pi \pi N \Delta}^{2} F^{1}_{d b c}
                )
                        i \gamma_{5}
                             {\Delta}^{d}_{\mu}
                                \partial^{\mu} \pi^{b}
                                               \pi^{c}
        $
\\
             &
        $
                +
                \bar{N}
                (
                   g_{\pi \pi N \Delta}^{3} F^{0}_{d b c}
                + 3
                   g_{\pi \pi N \Delta}^{4} F^{1}_{d b c}
                )
                        \gamma_{\nu} \gamma_{5}
                             {\Delta}^{d}_{\mu}
                                \partial^{\mu} \pi^{b}
                                \partial^{\nu} \pi^{c}
                + \{ {\rm H.C.} \}
          \; ;
        $
\\
             &
        $
             F^{0}_{d b c}
              =
                 i {\epsilon}_{d b c}
                 + {\delta}_{d b} \tau_{c}
                 - {\delta}_{d c} \tau_{b}
          \; ;  \;\;\;
             F^{1}_{d b c}
              =
                 {\delta}_{d b} \tau_{c}
                 + {\delta}_{d c} \tau_{b}
                 - 2 / 3 ~{\delta}_{b c} \tau_{d}
        $
\\
\hline
\hline
     $ {\rm SE}_{N N} \{ \pi \}$
                        &
\\
        $
          [ {\pi \pi \pi \pi} ]
        $
             &
        $
                V_{4 \pi}
        $
\\
\hline
     $ {\rm SE}_{N N} \{ \omega \}$
                        &
\\
        $
          [ {\pi \pi \pi \omega} ]
        $
             &
        $
                g_{\pi\pi\pi \omega}
                  \omega_{\mu}
                  \partial_{\nu} \pi^{a}
                  \partial_{\alpha} \pi^{b}
                  \partial_{\beta} \pi^{c}
                  i {\epsilon}_{a b c}
                  i {\epsilon}^{\mu \nu \alpha \beta}
        $
\\
\hline
     $ {\rm SE}_{N N} \{ A \}$
                        &
\\
        $
          [ {\pi \pi \pi A } ]
        $
             &
        $
                g_{\pi\pi\pi A}^{1}
                  A^{\mu}_{b}
                  \pi^{b}
                  \pi^{a}
                  \partial_{\mu} \pi^{a}
        $
\\
             &
        $
               +
                g_{\pi\pi\pi A}^{2}
                  A^{\mu}_{b}
                  \partial_{\mu} \pi^{b}
                  \partial_{\nu} \pi^{a}
                  \partial^{\nu} \pi^{a}
               +
                g_{\pi\pi\pi A}^{3}
                  A^{\mu}_{b}
                  \partial_{\nu} \pi^{b}
                  \partial_{\mu} \pi^{a}
                  \partial^{\nu} \pi^{a}
        $
\\
\hline
\hline
     $ {\rm SE}_{\pi \pi} \{ \sigma \}$
                        &
\\
        $
          [ {\pi \sigma N N } ]
        $
             &
        $
                g_{\pi \sigma N N}
                \bar{N}
                        \tau^{a}
                        \gamma_{\mu}
                        \gamma_{5}
                                N
                  \partial^{\mu} \pi_{a} \sigma
        $
\\
\hline
     $ {\rm SE}_{\pi \pi} \{ \rho \}$
                        &
\\
        $
          [ {\pi \rho N N } ]
        $
             &
        $
                g_{\pi \rho N N}^{1}
                \bar{N}
                        \gamma_{\mu}
                        \gamma_{\nu}
                        \gamma_{5}
                                N
                (
                  \partial^{\mu} \pi_{a} \rho^{\nu}_{a}
                 -
                  \partial^{\nu} \pi_{a} \rho^{\mu}_{a}
                )
                        / 2
        $
\\
  &
        $
              +
                g_{\pi \rho N N}^{2}
                \bar{N}
                        \tau^{a}
                        \gamma_{5}
                                N
                  \partial_{\mu} \pi_{b} \rho^{\mu}_{c}
                   i {\epsilon}_{a b c}
        $
\\
\hline
\hline
\end{tabular}
\end{center}
\caption[SEx graphs]{
                Lagrangians for 4--particle vertices
                of the
     $ {\rm SE} $
                graphs.
                         }
\label{SE4v}
\end{table}

\begin{table}
\begin{center}
\begin{tabular}{|c|c|}
\hline
\mbox{
\begin{tabular}{c}
    Graph\\
    \mbox{[Vertex]}
\end{tabular}
     }
                        &     Lagrangian
\\
\hline
\hline
        $ {\rm DE}_{\pi N, \pi N} \{ N, N \} $
                        &
\\
        $
           [ {\pi N N} ]
        $
             &
        $
                g_{\pi N N}
                \bar{N}
                       \tau_{a}
                        \gamma_{\mu} \gamma_{5}
                                N
                                \partial^{\mu} \pi^{a}
        $
\\
\hline
        $ {\rm DE}_{\pi N, \pi N} \{ N, N_{*} \} $
                        &
\\
        $
           [ {\pi N N_{*}} ]
        $
             &
        $
                g_{\pi N N_{*}}
                \bar{N}
                       \tau_{a}
                        \gamma_{\mu} \gamma_{5}
                                N_{*}
                                \partial^{\mu} \pi^{a}
                + \{ {\rm H.C.} \}
        $
\\
\hline
        $ {\rm DE}_{\pi N, \pi N} \{ N, \Delta \} $
                        &
\\
        $
           [ {\pi N \Delta} ]
        $
             &
        $
                g_{\pi N \Delta}
                (
                  \bar{\Delta}^{a}_{\mu}
                        P_{a b}
                             N
                                \partial^{\mu} \pi^{b}
                +
                  \bar{N}
                        P^{\dag}_{a b}
                             {\Delta}^{a}_{\mu}
                                \partial^{\mu} \pi^{b}
                )
        $
\\
\hline
        $ {\rm DE}_{\pi N, \pi N} \{ N_{*}, N_{*} \} $
                        &
\\
        $
           [ {\pi N_{*} N_{*}} ]
        $
             &
        $
                g_{\pi N_{*} N_{*}}
                \bar{N_{*}}
                       \tau_{a}
                        \gamma_{\mu} \gamma_{5}
                                N_{*}
                                \partial^{\mu} \pi^{a}
        $
\\
\hline
        $ {\rm DE}_{\pi N, \pi N} \{ N_{*}, \Delta \} $
                        &
\\
        $
           [ {\pi N_{*} \Delta} ]
        $
             &
        $
                g_{\pi N_{*} \Delta}
                (
                  \bar{\Delta}^{a}_{\mu}
                        P_{a b}
                             N_{*}
                                \partial^{\mu} \pi^{b}
                +
                  \bar{N_{*}}
                        P^{\dag}_{a b}
                             {\Delta}^{a}_{\mu}
                                \partial^{\mu} \pi^{b}
                )
        $
\\
\hline
        $ {\rm DE}_{\pi N, \pi N} \{ \Delta, \Delta \} $
                        &
\\
        $
           [ {\pi \Delta \Delta} ]
        $
             &
        $
                g_{\pi \Delta \Delta}
                \bar{\Delta}^{a}_{\mu}
                        \gamma_{\nu} \gamma_{5}
                                A_{a b c}
                                \bar{\Delta}^{b}_{\mu}
                                        \partial^{\nu} \pi^{c}
          \; ,
        $
\\
             &
        $
        { A_{a b c} }
              =
                \frac{4}{9}
                (
                 - 2 {\delta}_{b c} \tau_{a}
                 - 2 {\delta}_{a c} \tau_{b}
                 + 8 {\delta}_{a b} \tau_{c}
                 - 5 i {\epsilon}_{a b c}
                )
        $
\\
\hline
\hline
        $ {\rm DE}_{\pi N, \pi \pi} \{ N, \sigma \} $
                        &
\\
        $
           [ {\sigma N N} ]
        $
             &
        $
                g_{\sigma N N}
                \bar{N}
                                N
                                \sigma
        $
\\
\hline
        $ {\rm DE}_{\pi N, \pi \pi} \{ N, \rho \} $
                        &
\\
        $
           [ {\rho N N} ]
        $
             &
        $
                g_{\rho N N}^{\rm V}
                \bar{N}
                        \gamma_{\mu}
                            \tau^{a}
                                N
                                \rho^{\mu}_{a}
              +  g_{\rho N N}^{\rm T}
                \bar{N}
                        \sigma_{\mu \nu}
                            \tau^{a}
                                N
                           \partial^{\mu} \rho^{\nu}
        $
\\
\hline
        $ {\rm DE}_{\pi N, \pi \pi} \{ N_{*}, \sigma \} $
                        &
\\
        $
           [ {\sigma N N_{*}} ]
        $
             &
        $
                g_{\sigma N N_{*}}
                \bar{N}
                                N_{*}
                                \sigma
                + \{ {\rm H.C.} \}
        $
\\
\hline
        $ {\rm DE}_{\pi N, \pi \pi} \{ N_{*}, \rho \} $
                        &
\\
        $
           [ {\rho N N_{*}} ]
        $
             &
        $
                g_{\rho N N_{*}}^{\rm V}
                \bar{N}
                        \gamma_{\mu}
                            \tau^{a}
                                N_{*}
                                \rho^{\mu}_{a}
              +  g_{\rho N N_{*}}^{\rm T}
                \bar{N}
                        \sigma_{\mu \nu}
                            \tau^{a}
                                N_{*}
                           \partial^{\mu} \rho^{\nu}
                + \{ {\rm H.C.} \}
        $
\\
\hline
        $ {\rm DE}_{\pi N, \pi \pi} \{ \Delta, \rho \} $
                        &
\\
        $
           [ {\rho N \Delta} ]
        $
             &
        $
                g_{\rho N \Delta}
                \bar{\Delta}_{\mu}^{a}
                            P_{ab}
                                N
                                \rho^{\mu b}
                + \{ {\rm H.C.} \}
        $
\\
\hline
\hline
        $ {\rm DE}_{NN, \pi \pi} \{ \pi, \sigma \} $
                        &
\\
        $
           [ {\pi \pi \sigma} ]
        $
             &
        $
                {g_{\pi\pi \sigma}}
                \sigma
                  \partial_{\mu} \pi^{a}
                  \partial^{\mu} \pi^{a}
        $
\\
\hline
        $ {\rm DE}_{NN, \pi \pi} \{ \pi, \rho \} $
                        &
\\
        $
           [ {\pi \pi \rho} ]
        $
             &
        $
                {g_{\pi\pi \rho}}
                  \partial^{\nu} \rho^{\mu}_{a}
                  \partial_{\nu} \pi^{b}
                  \partial_{\mu} \pi^{c}
                   {\epsilon}_{a b c}
        $
\\
\hline
        $ {\rm DE}_{NN, \pi \pi} \{ \omega, \rho \} $
                        &
\\
        $
           [ {\pi \rho \omega} ]
        $
             &
        $
                g_{\pi \rho \omega}
                  \omega_{\mu}
                  \partial_{\nu} \pi^{a}
                  \partial_{\alpha} \rho_{\beta}^{a}
                  i {\epsilon}^{\mu \nu \alpha \beta}
        $
\\
        $
           [ {\omega N N } ]
        $
             &
        $
                g_{\omega N N}^{\rm V}
                \bar{N}
                        \gamma_{\mu}
                                N
                                \omega^{\mu}
              +  g_{\omega N N}^{\rm T}
                \bar{N}
                        \sigma_{\mu \nu}
                                N
                           \partial^{\mu} \omega^{\nu}
        $
\\
\hline
        $ {\rm DE}_{NN, \pi \pi} \{ A, \sigma \} $
                        &
\\
        $
           [ {\pi \sigma A} ]
        $
             &
        $
                g_{\pi \sigma A}
                  A^{\mu}_{a}
                  \partial_{\mu} \pi^{a}
                  \sigma
        $
\\
\hline
        $ {\rm DE}_{NN, \pi \pi} \{ A, \rho \} $
                        &
\\
        $
           [ {\pi \rho A} ]
        $
             &
        $
               [
               g_{\pi \rho A}^{1}
                  \partial^{\mu} \pi^{a}
                    \partial_{ [ \mu} \rho_{\nu ] }^{b}
                  A^{\nu c}
               +
               g_{\pi \rho A}^{2}
                  \partial^{\mu} \pi^{a}
                  \rho^{\nu b}
                    \partial_{ [ \mu} A_{\nu ] }^{c}
               ]
                  i {\epsilon}_{a b c}
        $
\\
        $
           [ { A N N } ]
        $
             &
        $
                g_{A N N}^{\rm V}
                \bar{N}
                        \gamma_{\mu}
                        \gamma_{5}
                            \tau^{a}
                                N
                                A^{\mu}_{a}
              +  g_{A N N}^{\rm T}
                \bar{N}
                        \sigma_{\mu \nu}
                        \gamma_{5}
                            \tau^{a}
                                N
                           \partial^{\mu} A^{\nu}
        $
\\
\hline
\hline
\end{tabular}
\end{center}
\caption[DEx graphs]{
                Lagrangians for 3--particle vertices
                of the
     $ {\rm DE} $
                graphs.
                         }
\label{DE3v}
\end{table}


\subsection{
               Background Contribution to
            $\pi N \to \pi \pi N$
               Amplitude
}
\vspace{0.5cm}

                An important
                issue
                of the approach
\cite{BolokhovVS87,BolokhovVS91}
                developed for the near--threshold
                energy region
                is given by the linear background terms
                presented in the form that respects
                the symmetries of strong interactions,
                namely,
        $ P $,
        $ C $,
        $ T $,
        $ SU_{F}(2) $
                and crossing.
                To modify this ingredient of the model
                we need first to discuss the role of
                these smooth terms in the amplitude.

                Initially,
                such terms were added to the OPE ones
                to stand for all other possible mechanisms
                of the
        $ \pi N \to \pi \pi N $
                reaction.
                When taking into account only a part
                of contributions of particles listed in
                Table
\ref{SEres}
                one leaves enough room for the background
                terms standing for the rest resonances.

                Another reason for the presence of
                a background is connected with ambiguities
                which are specific
                to the off--shell interactions
                of high--spin particles
                (see, for example, the old discussion
\cite{Hoehler83,NathEK70,OlssonO75}).
                These interactions result in
                polynomial terms
                in the considered amplitude
                (and
                polynomial terms
                in the 4--particle vertices).
                The overall contribution of this kind
                coming from all resonances
                is constrained by the asymptotic conditions
                for the entire amplitude.
                In principle,
                this makes necessary to use
                the consistent theory for particle propagators
                and all vertices
                (like
        $ \pi N  N $,
        $ \pi N \Delta $,
                etc.).
                Leaving the parameters
                of the polynomial background free
                we are safe to use the propagators and
                the vertices in the form
                determined by the simplicity reasons
                and/or
                by the chiral--symmetry arguments.
                (In particular,
                we are using the simplest form
\begin{equation}
\label{Deltaprop}
                S_{M}^{\mu' \mu} ( k )
        =
         \frac{1}{3 M^{2}}
         \left[
                 -3 M^{2} g^{\mu' \mu}
                 + M^{2} \gamma^{\mu'} \gamma^{\mu}
                 + 2 k^{\mu'} k^{\mu}
                 - M (
                       k^{\mu'} \gamma^{\mu} -
                       \gamma^{\mu'} k^{\mu}
                     )
         \right]
\end{equation}
                for the nominator of the propagator
                of a
        $  (3^{+}/2) $
                particle.)

                Since we are regularizing the exchange
                graphs which have poles in the physical
                region by the
        $ i M \Gamma $
                shift in the propagators
                the polynomial background must have both
                real and imaginary parts.

                The discussed nature of the background terms
                forces us to modify the model of the papers
\cite{BolokhovVS87,BolokhovVS91}
                in two directions.
                First,
                we add the second order terms in variables
(\ref{vars})
                into the scalar structures
        $ S_{A} $,
        $ S_{B} $,
        $ S_{C} $,
        $ S_{D} $:
\ba
\nonumber
        S^{(2)}_{A}
                &
                \equiv
                &
        S^{(2)}_{A}
                ( \tau, \nu, \bar{\nu}, \theta, \bar{\theta})
\\
\label{SAnext}
                &
                =
                &
                 A_{12} \tau^{2}
               + A_{13} \bar{\nu} \nu
               + A_{14} \bar{\theta} \theta
               + A_{15} ( \nu^{2} + \bar{\nu}^{2})
               + A_{16} ( \theta^{2} + \bar{\theta}^{2})
               + A_{17} \tau ( \theta + \bar{\theta} )
                \; ;
\\
\label{SBCnext}
        S^{(2)}_{B}
                &
                =
                &
                S^{(2)}_{A}
                ( \tau,
                        \bar{\epsilon}\nu,
                                {\epsilon}\bar{\nu},
                        \bar{\epsilon}\theta,
                                {\epsilon}\bar{\theta})
                \; ;
                \; \;
        S^{(2)}_{C}
                =
                S^{(2)}_{A}
                ( \tau,
                        {\epsilon}\nu,
                                \bar{\epsilon}\bar{\nu},
                        {\epsilon}\theta,
                                \bar{\epsilon}\bar{\theta})
                \; ;
\\
\label{SDnext}
        S^{(2)}_{D}
                &
                =
                &
                 A_{18} i ( \nu \bar{\theta}
                                - \theta \bar{\nu})
                \; .
\ea
                (This makes the dimensions of the tensor
                and the scalar structures
                of the decomposition
(\ref{IAmp})
                balanced.)

                Second,
                all 18 terms
                of the
                {\it real}
                background
                (11 of which are described in the paper
\cite{BolokhovVS91}
                and the rest are given by eqs.
(\ref{SAnext})--(\ref{SDnext}))
                had been copied to provide the
                {\it imaginary}
                background of the amplitude.

                Let us now clarify the concept of
                {\it contribution}
                which we are widely using throughout the paper.
                From the point of view of Chiral Dynamics
                the usage of notions like
                the background contribution
                (a part of which the
                so called contact terms are),
                the contribution of OPE, etc.
                is meaningless
                since the field redefinition does not allow
                a separate graph to be well--defined
 ---
                see the relevant discussion in the book
\cite{DonoghueGH92}.
                The absence of the common solution
                on the role of the higher--spin baryons
                in ChPT
                makes the above notions even more ambiguous.
                Nevertheless,
                such quantities
                as the residues of the poles
                and the on--shell parameters of
        $  V_{4\pi} $
                are well--defined
                --- these very quantities are
                in the focus of our
                project.
                As for the off--mass--shell contributions,
                in particular, the OPE one,
                we take them as they are,
                the Lagrangian source and Feynman rules
                providing the model--dependent answer.
                The field redefinition then modifies,
                first,
                the polynomial background terms and,
                second,
                the parameters of SE graphs.
                It has been already pointed out
                that we are leaving
                the background parameters free
                (and keeping all spin--isospin structures of
                4--particle vertices being represented)
                ---
                this helps to avoid the dependence
                of the results on a particular model.

                To summarize,
                the considered amplitude contains
                36 free parameters
                of the polynomial background,
                4 free parameters of the real part
                of the OPE contribution
                and 5 formal parameters
                of its imaginary part
                (their relations with
                parameters of the real part are described
                in the paper
\cite{BolokhovEtAl96})
                and parameters from the lists of
                Table
\ref{SE4v}
                and Table
\ref{DE3v}
                discussed in the previous subsection.
                In the current paper we shall discuss the
                fittings performed with 21 parameters
                from the above lists
                coming from the exchanges
     $ {\rm SE}_{\pi N} ( N ) $,
     $ {\rm SE}_{\pi N} ( \Delta ) $,
     $ {\rm SE}_{\pi N} ( N_{*} ) $,
     $ {\rm SE}_{\pi \pi} ( \rho ) $,
     $ {\rm DE}_{\pi N, \pi N} \{ N, N \} $,
     $ {\rm DE}_{\pi N, \pi N} \{ N, \Delta \} $,
     $ {\rm DE}_{\pi N, \pi N} \{ N, N_{*} \} $,
     $ {\rm DE}_{\pi N, \pi N} \{ N_{*}, N_{*} \} $,
     $ {\rm DE}_{\pi N, \pi N} \{ N_{*}, \Delta \} $,
     $ {\rm DE}_{\pi N, \pi N} \{ \Delta, \Delta \} $
                only.
                For the purpose of the forthcoming discussion
                we need to enumerate the fitting parameters
                --- this is done in Table
\ref{FitParTab}
                below.
                The background parameters
     $ A_{1} $--$ A_{18} $
                and their analogs
     $ i_{19} $--$ i_{36} $
                in the imaginary part of the amplitude
                have been already discussed
                in the previous subsection;
                these parameters had been being processed
                in fittings
                as they are
                --- to save space
                we do not list them once more.
                Some factors were absorbed
                into the fitting parameters
                to improve running characteristics
                of the code.
                The list of expressions
                of the actual fitting parameters
                in terms of the constants
                of the interaction Lagrangian
                is presented in Table
\ref{FitParTab}.

\begin{table}
\begin{center}
\begin{tabular}{|c|c|c|}
\hline
 Graph
&
\mbox{
\begin{tabular}{c}
                  Fitting\\
                 Parameter
\end{tabular}
     }
&
                                       Expression
\\
\hline
\hline
   $ {\rm SE}_{N N} \{ \pi \} $
&
                  $ o_{1} $
&
                                $ 2 g_{\pi N N} \cdot g_0 $
\\
&
                  $ o_{2} $
&
                                $ 2 g_{\pi N N} \cdot g_1 $
\\
&
                  $ o_{3} $
&
                                $ 2 g_{\pi N N} \cdot g_2 $
\\
&
                  $ o_{4} $
&
                                $ 2 g_{\pi N N} \cdot g_3 $
\\
\hline
\hline
   $ {\rm SE}_{\pi \pi} \{ \rho \} $
&
                  $ r_{1} $
&
                       $ g^{1}_{\pi \rho N N} \cdot g_{\pi \pi \rho} $
\\
&
                  $ r_{2} $
&
                       $ g^{2}_{\pi \rho N N} \cdot g_{\pi \pi \rho} $
\\
\hline
   $ {\rm DE}_{\pi N, \pi \pi} \{ N, \rho \} $
&
                  $ r_{3} $
&
                       $ g^{V}_{\rho N N} \cdot
                                g_{\pi N N}
                                          \cdot g_{\pi \pi \rho} $
\\
\hline
\hline
   $ {\rm SE}_{\pi N} \{ \Delta \} $
&
                  $ D_{1} $
&
                       $ g_{\pi N \Delta} \cdot
                                g^1_{\pi \pi N \Delta} $
\\
&
                  $ D_{2} $
&
                       $ g_{\pi N \Delta} \cdot
                                g^2_{\pi \pi N \Delta} $
\\
&
                  $ D_{3} $
&
                       $ g_{\pi N \Delta} \cdot
                                g^3_{\pi \pi N \Delta} $
\\
&
                  $ D_{4} $
&
                       $ g_{\pi N \Delta} \cdot
                                g^4_{\pi \pi N \Delta} $
\\
\hline
   $ {\rm DE}_{\pi N, \pi N} \{ \Delta, \Delta \} $
&
                  $ D_{5} $
&
                       $ 16 (g_{\pi N \Delta})^2 \cdot
                                g_{\pi \Delta \Delta} / 9 $
\\
\hline
   $ {\rm DE}_{\pi N, \pi N} \{ N, \Delta \} $
&
                  $ D_{6} $
&
                       $ (g_{\pi N \Delta})^2 \cdot
                                g_{\pi N N} / 3 $
\\
\hline
\hline
   $ {\rm SE}_{\pi N} \{ N_* \} $
&
                  $ R_{1} $
&
                       $ g_{\pi N N_*} \cdot
                                g^1_{\pi \pi N N_*} $
\\
&
                  $ R_{2} $
&
                       $ g_{\pi N N_*} \cdot
                                g^2_{\pi \pi N N_*} $
\\
&
                  $ R_{3} $
&
                       $ 2 g_{\pi N N_*} \cdot
                                g^3_{\pi \pi N N_*} $
\\
&
                  $ R_{4} $
&
                       $ g_{\pi N N_*} \cdot
                                g^4_{\pi \pi N N_*} $
\\
\hline
   $ {\rm DE}_{\pi N, \pi N} \{ N_*, N_* \} $
&
                  $ R_{5} $
&
                       $ g_{\pi N_* N_*} \cdot
                                (g_{\pi N N_*})^2 / 8 $
\\
\hline
   $ {\rm DE}_{\pi N, \pi N} \{ N_*, \Delta \} $
&
                  $ R_{6} $
&
                       $ g_{\pi N \Delta} \cdot
                          g_{\pi N_* \Delta} \cdot
                                g_{\pi N N_*} / 3 $
\\
\hline
   $ {\rm DE}_{\pi N, \pi N} \{ N, N_* \} $
&
                  $ R_{7} $
&
                       $ g_{\pi N N} \cdot
                                (g_{\pi N N_*})^2 / 8 $
\\
\hline
\hline
   $ {\rm SE}_{\pi N} \{ N \} $
&
                  $ N_{1} $
&
                       $ g_{\pi N N} \cdot
                                g^1_{\pi \pi N N} $
\\
&
                  $ N_{2} $
&
                       $ g_{\pi N N} \cdot
                                g^2_{\pi \pi N N} $
\\
&
                  $ N_{3} $
&
                       $ 2 g_{\pi N N} \cdot
                                g^3_{\pi \pi N N} $
\\
&
                  $ N_{4} $
&
                       $ g_{\pi N N} \cdot
                                g^4_{\pi \pi N N} $
\\
\hline
   $ {\rm DE}_{\pi N, \pi N} \{ N, N \} $
&
                  $ N_{5} $
&
                       $ (g_{\pi N N})^3 / 8 $
\\
\hline
\hline
\end{tabular}
\end{center}
\caption[DEx graphs]{
                Expression of
                fitting parameters of the phenomenological
                amplitude
                in terms of the Lagrangian parameters.
                         }
\label{FitParTab}
\end{table}


\newpage
\vskip1.cm
\section{ Experimental Base }

\vspace{0.5cm}

                The experimental base for fitting
                the amplitude parameters
                was built of the total cross sections
                in all five channels
                of the considered reaction
                in the energy region
     $ P_{\rm Lab} \le 500 $~MeV/c
                and the distributions
                measured in the hydrogen bubble--chamber
                experiments at these energies
\cite{Kirz_++n62, Kirz_-+n63, Bloh_-+n70, Saxon-+n70, Jones-+n-0p74}.
                The bubble--chamber data were preferred
                since they satisfy the condition of
                the coverage of the reaction phase space
                in the most complete way
                (the only restrictions are
     $ P_{\rm proton} \ge 120 $~MeV/c
                and
     $ P_{\pi} \ge 30 $~MeV/c
                which correspond to
                5 mm of the flight distance).
                The losses occupy only a small part of the
                phase space
                (namely,
     $ \le 2 $ \%
                in the case of
     $ \{\pi^{-} \pi^0 p \} $
                and
     $ \le 3 $ \%
                in the case of
     $ \{\pi^{-} \pi^{+} n \} $
                at
     $ P_{\rm beam} = 400 $~MeV/c)
                and might be easily taken into account
                during the calculations of theoretical
                distributions.
                Besides,
                the systematic errors of the
                bubble--chamber experiments
                are also minimal.

                In the course of our fittings
                we faced the following problem.
                There are three works in the
     $ \{\pi^{-} \pi^{+} n \} $
                channel
                at close beam momenta:

                1. Kirz
        ($ P_{\rm Lab} = 485 $~MeV/c)
\cite{Kirz_-+n63},

                2. Blokhintseva
        ($ P_{\rm Lab} = 457 $~MeV/c)
\cite{Bloh_-+n70},

                3. Saxon
        ($ P_{\rm Lab} = 456 $~MeV/c)
\cite{Saxon-+n70}.

                It was found that results of these works
                are incompatible.
                If one tries to describe only distributions
                from these works
                without appealing to any other data
                (total cross sections or other channels' data)
                then the averaged
     $ \chi^{2} $
                per bin
     ($ \equiv \overline{\chi^{2}}_{p.b.}  $)
                for distributions by Kirz and Blokhintseva
                separately and for their sum as well
                is less than 1.
                Adding the distributions by Saxon
                we obtained
     $ \overline{\chi^{2}}_{p.b.}  > 1.5 $
                for every combinations
                and even
     $ \overline{\chi^{2}}_{p.b.}  = 2.1 $
                for the Saxon's distributions themselves.

                We should note here,
                that we have the results of the experiment
                by Blokhintseva
                as the collection of events.
                Hence, we can build any distribution
                and we did so for
                all kinds of
                distributions published in the Saxon's paper.
                Along with other spectra
                such distributions also
                represent the Blokhintseva work
\cite{Bloh_-+n70}
                in our fittings.

                The experiment by Saxon contains
                more events than the ones by Kirz
                and Blokhintseva together,
                so,
                at first glance,
                one should prefer to choose his results.
                However,
                the processing of films had been performed
                by the Saxon's group in a quite specific
                (nontraditional) way.
                In such films the elastic
     $ \{\pi^{-} p \} $
                reaction events
                look very much like the events of the
                considered reaction,
                their total number being ten times greater.
                To economize manual measurements
                the group had estimated visually
                the density of the positive tracks
                and, basing on the estimate, had selected
                the events with
     $ \pi^{+} $.

                However,
                besides the velocity of the charged particle
                the density of a track in the bubble chamber
                depends upon much more another factors like
                the moment of particle flight
                relative to the moment of
                liquid expansion start,
                the moment of snapshot
                relative to the moment of particle flight,
                the liquid superheating degree, etc.
                This dependence is extremely strong.
                The above parameters undergo stabilization
                but the latter never becomes ideal.
                The rest fluctuations can not prevent the
                ionization measurements of tracks
                but for the reliability of the determination
                of the particle velocity in the every
                snapshot
                the control measurements of the bubble density
                should be performed for the sample track.
                In the visual estimating
                the probability of an error is big
                especially in the case of the so large number
                of the processed tracks.

                One can estimate the systematic error of
                the Saxon's work in the following way.
                The elastic events accidentally selected into
                the list of the considered reaction
                anyway will be
                rejected after measurements by the reaction
                fit results.
                The inelastic ones
                for which the positive track was erroneously
                identified as the proton one
                are lost forever.
                In such a way the statistics become more poor,
                the parts of the phase space for which the
     $ \pi^{+} $
                velocities
                are small
                suffering
                the most significant losses.
                The measured value of the total cross section
                is given in the Saxon's work as
                1 mbn;
                since the results of all isotopic analyses
                of all set of total cross sections data
                provide the value of
                1.4 mbn
                one can deduce
                the 30\% level of losses of events in the
                ``economizing'' routine of the reaction analysis.

                So large value of systematic error forced us
                to withdraw the distributions of the paper
\cite{Saxon-+n70}
                from our fittings.

                Thus,
                the distributions liable to treatment
                belong to three
                channels of the considered reaction:
     $ \{\pi^{-} \pi^{+} n \} $,
     $ \{\pi^{-} \pi^{0} p \} $,
     $ \{\pi^{+} \pi^{+} n \} $.
                The data on distributions of works
\cite{Kirz_++n62},
\cite{Kirz_-+n63},
\cite{Saxon-+n70},
\cite{Jones-+n-0p74}
                were taken from the journal publications.
                The major part of distributions of the
                work
\cite{Bloh_-+n70}
                has never been published.
                These are the distributions
                in the following variables:
                the squares of invariant masses of all pairs
                of final particles
     $ W_{\pi^{-} \pi^{+}} $,
     $ W_{\pi^{-} n} $,
     $ W_{\pi^{+} n} $,
                the invariant variables
     $ \tau $,
     $ \nu_{R} $,
     $ \nu_{I} $,
     $ \theta_{R}$,
     $ \theta_{I} $
                introduced in
\cite{BolokhovVS87,BolokhovVS91},
                the cosine of the angle between
                the planes determined by
                CMS momenta of
                a) the beam and the recoil neutron;
                b)
     $ \pi^{+} $
                and
     $ \pi^{-} $
     ($ \cos \theta_{\pi\pi} $),
                the cosines of CMS angles of final particles
                with the beam
     $ \cos \theta_{\pi^{-} } $,
     $ \cos \theta_{\pi^{+} } $,
     $ \cos \theta_{n } $,
                the angle
     $ D \phi = \phi^{-} - \phi^{+} $
                of the planes determined by
                CMS momenta of
                a) the beam and the
     $ \pi^{-} $
                meson;
                b) the beam and the
     $ \pi^{+} $
                meson,
                the azimuth angle
     $ \phi_{\rm beam} $
                and the cosine of the polar angle of the beam
     $ \cos \theta_{\rm beam} $
                in the Saxon reference frame
                in which the
     $ x $
                axis is the direction of the neutron momentum,
                the
     $ z $
                axis being along the vector product
                of the neutron momentum with the momentum of
     $ \pi^{+} $
                meson,
     $ [{\bf p}_{\rm n}, {\bf p}_{\pi^{+}} ] $.

                The Jones paper
                (the
     $ \{\pi^{-} \pi^{+} n \} $
                and
     $ \{\pi^{-} \pi^{0} p \} $
                channels at
     $ P_{\rm Lab} = 415 $~MeV/c)
\cite{Jones-+n-0p74}
                provides distributions in the
                squares of invariant masses of all pairs
                of final particles
                (namely,
     $ \{\pi^{-} \pi^{+} n \} $:
     $ W_{\pi^{-} \pi^{+}} $,
     $ W_{\pi^{-} n} $,
     $ W_{\pi^{+} n} $;
     $ \{\pi^{-} \pi^{0} p \} $:
     $ W_{\pi^{-} \pi^{0}} $,
     $ W_{\pi^{-} p} $,
     $ W_{\pi^{0} p} $).

                The Kirz paper
                (the
     $ \{\pi^{-} \pi^{+} n \} $
                channel at
     $ P_{\rm Lab} = 485 $~MeV/c)
\cite{Kirz_-+n63}
                gives distributions in the invariant masses
     $ W_{\pi^{-} \pi^{+}} $,
     $ W_{\pi^{-} n} $,
     $ W_{\pi^{+} n} $
                of all pairs of final particles
                and in the cosine of angles of all final
                particles with the beam in CMS
     $ \cos \theta_{\pi^{-} } $,
     $ \cos \theta_{\pi^{+} } $,
     $ \cos \theta_{n } $.

                Another paper by Kirz
                (the
     $ \{\pi^{+} \pi^{+} n \} $
                channel at
     $ P_{\rm Lab} = 477 $~MeV/c)
\cite{Kirz_++n62}
                contains distributions in
                the CMS angles of final particles
     $ \theta_{\pi^{+} } $,
     $ \theta_{n } $,
                the
     $ \pi^{+} $
                CMS energy
     $ T_{\pi^{+} {\rm CMS} } $,
                the momentum
     $ P_{\pi^{+} } $
                of
     $ \pi^{+} $
                in the
     $ (\pi^{+}, \pi^{+}) $
                system
                and in
                the nonrelativistic momentum transfer
     $ P_{\rm transf} $.

                One should note
                that some of these distributions
                strongly differ from the distribution
                provided by the empty
                phase space
                while there are some
                with the insignificant difference.
                This becomes especially clear in terms of
                (normalized) quasi--amplitudes
                --- see pictures in Figs. 3--8
                where almost all distributions are reproduced.

                We are using
                the total cross sections data
\cite{Bari_+-n64},
\cite{Batu_-0p61},
\cite{Batu_-+n65},
\cite{Batu_+0p++n73},
\cite{Batu_+0p++n75},
\cite{Belk_00n80},
\cite{Bjork+-n80},
\cite{Blair-+n70},
\cite{Bloh_-+n63},
\cite{Bloh_-+n70},
\cite{Buni_-+n72},
\cite{Buna_00n77},
\cite{Deahl+-n61},
\cite{Jones-+n-0p74},
\cite{Kernl+-n89},
\cite{Kernl-0p89},
\cite{Kernl++n90},
\cite{Kirz_++n62},
\cite{Kirz_-+n63},
\cite{Krav_00n74},
\cite{Krav_++n78},
\cite{Lowe_00n91},
\cite{Perki+-n60},
\cite{PocanicEtAl94},
\cite{Saxon-+n70},
\cite{SeviorEtAl91},
\cite{Sober+0n-0n75}.
                94 points were selected
                on the grounds of
                the compatibility analysis
                of the paper
\cite{VVVSGSEtAl95}
                from the total list of 105
                experimental points.


                In some variants of fittings
                (namely,
                when the theoretical amplitude was not
                completely real ---
                see the next section
                where all variants
                are described in more details)
                there were used
                two experimental data points more.
                These points were fixing the phases of two
                (nonvanishing at the threshold)
                isotopic amplitudes by the known values
                of the elastic
        $  P_{31} $ $ (\approx -4^{\circ}) $
                and
        $  P_{11} $ $ (\approx 2^{\circ}) $
                phases
                in accordance with
                the final--state interaction theorem.

\vskip1.cm
\section{ Data Analysis }

                The analysis of the data described
                in the previous section required
                some preliminary steps which are described
                below. After then we discuss the main
                results in the subsect. 4.2.

\subsection{ Principal Steps of Analysis }

                To perform the analysis
                the following steps had been done:

                {\bf 1.}
                The contributions to the amplitude
                of every parameter of our
                model (discussed in the sect. 2)
                had been calculated analytically
                according to the Feynman rules
                of the tree approximation.
                (The
        $  i M \Gamma $
                shifts regularizing the terms with poles
                located in the physical region
                were made also for all cross terms
                to ensure correct crossing properties of
                the entire amplitude.)
                The contributions generally come
                to all 16 scalar--isoscalar form
                factors defined in eqs.
(\ref{IDec}),
(\ref{IAmp});
                the corresponding expressions had been
                obtained and transformed to the FORTRAN code
                with the help of the REDUCE package
\cite{Hern85}
                for
                analytic calculations in high energy physics.

                {\bf 2.}
                The second step was aimed to simplify
                the data fitting routine
                and save a lot of time
                during thousands of fitting iterations.
                For a given experimental point
        $ (\alpha) $
                the theoretical quantity
        $  \sigma_{(\alpha)}^{\rm Th} $
                confronting the experimental value
        $  \sigma_{(\alpha)}^{\rm Exp} $
                is the integral over the reaction phase space
                (or over its part in the case
                of the distribution data)
                of the squared modulus of the amplitude
(\ref{ChAmpform}).

                Since the set of all parameters
        $ \{ A_{\nu} \} $
                including the formal ones enters our amplitude
                linearly
                and the conditions for the formal parameters
                do not contain kinematics
                we can present the theoretical quantity
(\ref{cs})
                in the form
\begin{equation}
\label{corr}
  \sigma_{(\alpha)}^{\rm Th} =
                \sum_{\mu,\nu}
                        A_{\mu} A_{\nu} C_{(\alpha)}^{\mu\nu}
        \; .
\end{equation}
                Then we can build
                for every experimental point
        $  (\alpha) $
                the
                {\it correlator}
        $  C_{(\alpha)}^{\mu\nu} $
                performing all phase space integrations
                ones and forever.
                These calculations had been done for every
                bin of all distributions with the use of
                the standard high energy physics package
                based on the Monte Carlo integration.
                Since all distributions in question
                belong to the beam momenta
        415, 460, 477, 485 ~MeV/c
                few Monte Carlo runs were necessary
                (and sufficient)
                to provide the calculations.
                In the case of the total cross sections
                the correlator matrix
        $  \hat{C}_{(\alpha)} $
                for an experimental point
        $  (\alpha) $
                was being recovered
                during the fittings' run--time
                by the fast interpolation from 13 fulcrum
                matrices in every channel
                precalculated at beam momenta
        280,
        285,
        290,
        300,
        325,
        350,
        375,
        400,
        450,
        500,
        550,
        600
                and
        650
                MeV/c.
                The calculations of the fulcrum matrices
                had been performed with the fast and
                efficient program of Gauss integration
                described in the paper
\cite{BolokhovAPTS96}.

                {\bf 3.}
                The content of the third step was
                the data fittings themselves.
                Here, we point only the general
                features and specifics of the approach;
                more details will be revealed
                along with the discussion of results.

                The fittings had been performed by
                minimizing the value of
        $  \chi^{2} $
                defined as
\begin{equation}
\label{chits}
        \chi^{2} =
                \sum_{(\alpha)}
        \frac{
                (\sigma_{(\alpha)}^{\rm Th} -
                \sigma_{(\alpha)}^{\rm Exp} )^{2}
             }
        { ( \Delta\sigma_{(\alpha)}^{\rm Exp} )^{2}
        }
\end{equation}
                in the case of the total cross section data.
                However,
                to avoid the artificial increase
                of the statistical weight
                of the total--cross--section data point
                for which the numerous distributions also
                were undergoing fittings
                we were using the distributions' data
        $  \sigma_{(\alpha)n}^{\rm Exp} $
                which were
                {\it normalized}
                to 1 instead of the total cross section.

                The use of the precalculated correlator
                matrices
        $  \hat{C}_{(\alpha)} $
                and the simplicity of the expression
(\ref{corr}),
                being calculated in the course
                of the iterative minimization of
        $  \chi^{2} $,
                made it possible to perform thousands
                of fitting runs
                each of which
                executing hundreds or even few thousands
                of iterations.
                The same specifics of the approach
                provided the excellent flexibility
                in respect to variation of both the set
                of fitting parameters
                and the set of experimental points.
                The numerous variants of fittings will
                be discussed in the next subsection.


\subsection{ Major Results of Analysis }

                Already test runs of fitting the distribution
                data had shown
                that the simple model of the paper
\cite{BolokhovVS91}
                is unlikely to be capable to provide
                a satisfactory description.
                Therefore,
                the first question we were trying to find
                the answer to
                was if there exists a relatively simple
                model consistent with the selected data base
                (411 experimental points).
                The answer was ``no''
                and let us now discuss why.

                We had been grouping together parameters
                related to one or another mechanism
                of the considered reaction
                since it ought to be a hard task to test
        $  \approx 2^{61} $
                variants of all combinations of parameters.
                There were 7 such groups in total,
                the largest ones consisting of the sets
                of the background parameters.
                Below we are using
                the following symbolic notations
                for these groups:

                ``b'' --- parameters of the real background;

                ``i'' --- parameters of the imaginary
                          background;

                ``o'' --- parameters of the OPE contribution;

                ``r'' --- 3 parameters related to the
        $ \rho  $
                          meson;

                ``N'' --- parameters of the nucleon
                          contributions
                          (namely,
        $ {\rm SE}_{\pi N} \{ N \} $
                          and
        $ {\rm DE}_{\pi N, \pi N} \{ N, N \} $
                          ones);

                ``D'' --- parameters of the
        $ \Delta $
                          contributions
                          (namely,
        $ {\rm SE}_{\pi N} \{ \Delta \} $,
        $ {\rm DE}_{\pi N, \pi N} \{ N, \Delta \} $
                          and
        $ {\rm DE}_{\pi N, \pi N} \{ \Delta, \Delta \} $
                          ones);

                ``R'' --- parameters of the
        $ N_{*} $
                          contributions
                          (namely,
        $ {\rm SE}_{\pi N} \{ N_{*} \} $,
        $ {\rm DE}_{\pi N, \pi N} \{ N, N_{*} \} $,
\newline
        $ {\rm DE}_{\pi N, \pi N} \{ N_{*}, N_{*} \} $
                          and
        $ {\rm DE}_{\pi N, \pi N} \{ N_{*}, \Delta \} $
                          ones);

                For all possible combinations of the listed
                groups there had been performed
                the data fittings with
                at least 50 random starts.
                In general, no new solutions
                had been being revealed after 20--30 starts.
                In the cases when the number of distinct
                solutions was greater than usual
                we had been continuing the search
                with the increased number of random starts
                to 100 and more.
                With one exception
                this had not provided new solutions
                (apart the solutions with incomparably large
        $  \chi^{2} $).
                Only
                in the ``DRNbior'' variant
                when all groups had been being
                involved
                acceptable distinct solutions
                had been found after 100 starts.
                The total list of this variant contains 12
                solutions, not all of them are shown in the
                discussed Table.
                (It should be noted that the latter variant is
                the most difficult in fitting:
                the convergence is slow because of huge
                correlations between parameters,
                the parameter errors in the gained solutions
                being too large.)
                Therefore,
                we consider the probability of existence
                of a missed solution to be negligible.

                The values of
        $  \chi^{2} $
                of the best solutions are listed in the Table
\ref{chilist1}.
                A part of variants is withdrawn there
                to make the table more compact.
                The baryon--exchange mechanisms are combined
                with the sets of parameters ``b'', ``i'', ``o''
                and ``r''.
                For example, the box
                in the ``DR'' row and ``bio''  column
                corresponds to ``DRbio'' variant, etc.;
                in the column (row) with the empty
                title
                we collect the results corresponding
                to the ``pure'' mechanism.
                Initially,
                the ``r'' set was considered as a
                {\it perturbation}
                to the basic amplitude since
                the relatively narrow
        $  \rho $
                resonance
                might get into the reaction phase space at
        $  P_{\rm Lab} \approx 800 $~MeV/c
                only.
                (The use of the higher polynomial background
                in such a role
                requires to make calculations with about
                a hundred of the next order terms
                with free parameters
                or to find the reasons for rejecting the most
                of them.)
                However,
                the presented values of
        $  \chi^{2} $
                can not provide the inference that
                the ``r'' mechanism is unimportant
                (unless all baryon--exchange mechanisms
                are committed to action).

                It is the examination of the Table
\ref{chilist1}
                which leads to the following conclusions:

                {\bf 1.}
                There is no particle for which
                a simple exchange mechanism
                is capable to describe the data.

                {\bf 2.}
                Among the double--particle exchanges
                the participation of OPE does not look
                advantageous.
                (Even the dummy ``r'' mechanism looks
                more preferable sometimes.)

                {\bf 3.}
                The leading order
                ChPT
                fails to describe the data.
                (Indeed,
                all contact terms of Chiral Dynamics
                and tree--level graphs are contained in the
                variant ``Nbo'';
                the result of the latter even
                for parameters being free looks depressing.)

                {\bf 4.}
                The most significant improvement of
        $  \chi^{2} $
                is achieved when the imaginary background
                ``i''
                is being added;
                the participation of
        $  \Delta $
                and
        $  N_{*} $
                exchanges
                with the strong imaginary part
                of contributions is leading almost to the
                same effect.

                To make the above general conclusions
                the knowledge of the absolute
        $  \chi^{2} $
                values was sufficient,
                especially,
                since the number of experimental points
                considerably exceeds
                the (varying) number of free parameters.
                For a more subtle deduction one needs
                an information on
        $  \chi^{2}_{\rm DF} $
                (i.e.
        $  \chi^{2} $
                per degree of freedom).
                However,
                before calculating
        $  \chi^{2}_{\rm DF} $
                one first ought to look if there are
                undetermined and inessential parameters
                in the solution.

\newpage
\begin{table}[t]
\begin{center}
\begin{tabular}{|c||c|c|c|c|c|c|c|c|}
\hline
  &       &  o    &  b    &  bo   &  bi   & bio   & bir   &bior    \\
\hline
\hline
  &       &8078.5 &6993.9 &6799.1 &773.28 &595.53 &598.99 &550.87  \\
  &       &9083.8 &6996.4 &6798.8 &854.11 &597.10 &601.17 &556.10  \\
  &       &12115. &6998.0 &6800.8 &       &       &620.91 &563.68  \\
  &       &       &7016.9 &6813.2 &       &       &629.66 &570.10  \\
  &       &       &       &6823.0 &       &       &       &576.84  \\
\hline
D &9936.3 & 987.7 &634.49 &622.43 &546.35 &517.90 &527.54 &491.08  \\
  &       &1105.3 &651.55 &634.08 &546.52 &520.18 &531.11 &493.17  \\
  &       &       &748.07 &644.39 &548.19 &       &534.24 &        \\
  &       &       &       &672.37 &       &       &544.76 &        \\
\hline
R &6138.6 &1160.2 &746.19 &618.26 &511.88 &499.41 &468.06 &453.23  \\
  &       &1378.3 &756.55 &636.47 &515.07 &505.34 &484.49 &502.90  \\
  &       &3988.6 &761.74 &638.38 &       &566.52 &517.41 &523.12  \\
  &       &       &       &672.74 &       &       &       &        \\
  &       &       &       &676.47 &       &       &       &        \\
\hline
N &13693. &7069.7 &6660.6 &6649.4 &549.83 &525.73 &513.17 &496.87  \\
  &13709. &7341.0 &6660.8 &6649.9 &551.27 &526.72 &515.28 &497.00  \\
  &       &7388.9 &6668.7 &6653.0 &       &527.84 &516.82 &498.42  \\
  &       &9928.6 &6739.4 &6674.7 &       &528.84 &518.16 &500.98  \\
  &       &       &8508.0 &6707.2 &       &       &518.89 &501.18  \\
  &       &       &       &6731.2 &       &       &       &        \\
\hline
D &2758.5 &760.88 &578.68 &564.35 &471.09 &469.33 &448.73 &443.09  \\
R &       &870.24 &586.21 &566.70 &475.71 &469.96 &466.93 &450.85  \\
  &       &978.58 &659.71 &568.58 &483.21 &470.86 &469.82 &        \\
  &       &3544.5 &       &573.63 &494.95 &479.92 &473.89 &        \\
  &       &       &       &       &       &482.53 &       &        \\
\hline
D &4438.7 &647.74 &557.32 &550.95 &495.19 &485.30 &473.29 &444.11  \\
N &5023.7 &793.23 &562.20 &555.25 &496.55 &487.58 &481.65 &463.69  \\
  &       &       &572.59 &561.61 &513.20 &       &496.34 &        \\
  &       &       &619.67 &574.84 &       &       &       &        \\
  &       &       &       &576.49 &       &       &       &        \\
  &       &       &       &609.92 &       &       &       &        \\
\hline
R &2546.7 &907.85 &553.01 &540.30 &467.26 &455.12 &442.69 &426.66  \\
N &       &1067.6 &553.04 &568.76 &499.38 &487.41 &486.08 &480.20  \\
  &       &2997.9 &553.21 &       &519.16 &493.38 &       &        \\
  &       &       &555.16 &       &       &       &       &        \\
  &       &       &584.36 &       &       &       &       &        \\
\hline
D &1430.1 &588.65 &510.42 &506.13 &435.72 &430.88 &426.54 &412.84  \\
R &1430.7 &710.41 &516.79 &507.26 &449.27 &442.20 &432.59 &418.83  \\
N &1451.5 &       &       &       &449.62 &443.02 &435.85 &423.30  \\
  &       &       &       &       &450.10 &445.69 &447.12 &589.75  \\
  &       &       &       &       &       &       &       &******  \\
\hline
\hline
\end{tabular}
\end{center}
\caption[]{
\thispagestyle{empty}
                List of
     $ \chi^{2} $
                in distinct solutions obtained from random start
                (50 tests) with
     $ N_{\rm Exp} = 411 $
                experimental points.
                (In the truncated variant ``DRNbior'' there are
                12 solutions in total obtained in
     $ \approx 500 $
                runs.)
                         }
\label{chilist1}
\end{table}
\clearpage

                Now we must note the following feature
                of the discussed results.
                The numerous solutions in some variants
                reflect the existence of large correlations
                of parameters in the variant in question.
                The estimating of parameter errors
                by the fitting routine
                is not then precise enough and our simple
                algorithm for comparing solutions marks as
                different the solutions which are,
                in principle, identical.

                To make the situation clear and,
                what is more important,
                to reduce the errors of parameters
                we had performed the series of additional
                runs eliminating one by one those parameters
                which relative errors
                had been being greater than 1
                and fitting the data by the rest parameters
                (while the value of
        $  \chi^{2}_{\rm DF} $
                had been improving).
                This routine had been applied
                to all solutions.
                Unfortunately,
                there is no enough room
                to display the results.
                The Table
\ref{chiDFlist}
                presents the information on
        $  \chi^{2}_{\rm DF} $
                for some selected variants,
                the number of parameters in effect
        $  N_{\rm P} $
                and
                the independent lowest
        $  \pi \pi $
                scattering lengths.
                The results of eliminating parameters
                had no influence on the above
                conclusions
                {\bf 1.}--{\bf 4.}
                whereas
                the number of parameters in effect
                had been considerably reduced.
                The new inferences derived from the Table
\ref{chiDFlist}
                read:

                {\bf 5.}
                The considered data base requires
                a complicated model for its description.
                Below the level of
        $  \chi^{2}_{\rm DF} < 1.50 $
                the absolute minimum of the parameter number
                is 22
                (variant ``Nbi'');
                for the most part of the acceptable solutions
                the number is greater than 30.

                {\bf 6.}
                Assuming the consent that the sequence of
                signs
        $ + $,
        $ - $,
        $ + $
                of the
        $  \pi \pi $
                scattering lengths
        $a^{I=0}_0$,
        $a^{I=2}_0$,
        $a^{I=1}_1$
                is
                {\it physical},
                one finds that
                a half of all acceptable solutions with the
                OPE contribution falls to the
                {\it unphysical}
                sector.
                (The interpretation of the latter phenomenon
                in terms of threshold values of amplitudes
                will be discussed in the next section.
                Here,
                we simply note that the physical sequence
                is the one consistent with the predictions
                of Chiral Dynamics.)

\begin{table}
\begin{center}
\begin{tabular}{|c|| c c || c c r r r || c c || c c r r r|}
\hline
  & bi &   &  bio&  &   &     &     & bir &   & bior&  &    &     &   \\
  &$ \chi^2_{\rm DF}$
       &$N_{\rm P}$
           &$ \chi^2_{\rm DF}$
                 &$N_{\rm P}$
                    &$a^{I=0}_0$
                        &$a^{I=2}_0$
                              &$a^{I=1}_1$
                                    &$ \chi^2_{\rm DF}$
                                          &$N_{\rm P}$
                                              &$ \chi^2_{\rm DF}$
                                                    &$N_{\rm P}$
                                                       &$a^{I=0}_0$
                                                            &$a^{I=2}_0$
                                                                  &$a^{I=1}_1$
                                                                  \\
\hline
\hline
  &1.94&26 & 1.56&30&.04&-.151&-.016& 1.58&31 & 1.46&32&.009& .15 &-.040 \\
  &1.95&26 & 1.57&29&.03&-.155&-.011& 1.58&32 & 1.46&32&.014&-.16 & .020 \\
  &    &   &     &  &   &     &     & 1.64&31 & 1.50&30&.136&-.12 & .048 \\
  &    &   &     &  &   &     &     & 1.66&31 & 1.51&33&.044&-.16 & .048 \\
  &    &   &     &  &   &     &     &     &   & 1.53&33&.045&-.20 & .039 \\
\hline
D &1.45&33 & 1.37&24&.16&-.084&.040 & 1.40&30 & 1.31&30&.14 &-.14 & .056 \\
  &1.45&25 & 1.40&30&.11&-.187&.064 & 1.42&38 &     &  &    &     &      \\
  &1.45&27 &     &  &   &     &     & 1.42&36 &     &  &    &     &      \\
  &    &   &     &  &   &     &     & 1.44&31 &     &  &    &     &      \\
\hline
R &1.35&32 & 1.33&32&.11&-.004&-.026& 1.25&37 & 1.22&36&.038& .015& .000 \\
  &1.36&31 & 1.33&33&.12& .002&-.025& 1.38&37 & 1.35&39&.073&.0002& .008 \\
  &    &   & 1.51&35&.18&-.138&-.012&     &   & 1.40&34&.170&-.034& .028 \\
\hline
N &1.42&23 & 1.39&27&.07& .004&-.014& 1.36&33 & 1.32&30&.23 & .052& .011 \\
  &1.42&22 & 1.39&28&.07& .004&-.014& 1.37&27 & 1.32&30&.23 &-.012& .018 \\
  &    &   & 1.40&33&.16& .026&-.020& 1.37&31 & 1.32&34&.14 &-.012& .019 \\
  &    &   & 1.41&28&.16& .023&-.020& 1.37&28 & 1.33&29&.23 & .029& .017 \\
  &    &   &     &  &   &     &     & 1.37&27 & 1.33&26&.10 &-.014& .015 \\
\hline
D &1.25&34 & 1.25&30&.00& .000& .000& 1.20&36 & 1.21&41&.13 &-.026& .046 \\
R &1.27&33 & 1.26&35&.19&-.058& .038& 1.26&40 & 1.24&46&.12 &-.105& .057 \\
  &1.29&35 & 1.26&35&.10&-.048& .033& 1.26&39 &     &  &    &     &      \\
  &1.33&38 & 1.31&43&.10&-.016& .026& 1.28&38 &     &  &    &     &      \\
  &    &   & 2.01&40&.80& .21 & .100&     &   &     &  &    &     &      \\
\hline
D &1.31&31 & 1.29&32&.04& .025& .005& 1.27&34 & 1.19&35&.11 & .000& .045 \\
N &1.31&32 & 1.30&32&.15&-.022& .023& 1.29&34 & 1.24&31&.17 &-.104& .048 \\
  &1.36&26 &     &  &   &     &     & 1.33&33 &     &  &    &     &      \\
\hline
R &1.24&29 & 1.23&36&.22& .037&-.035& 1.19&35 & 1.15&43&.14 & .042&-.008 \\
N &1.34&39 & 1.31&36&.09& .009&-.015& 1.32&45 & 1.29&40&.18 &-.048& .005 \\
  &1.39&36 & 1.32&36&.20&-.047&.0002&     &   &     &  &    &     &      \\
\hline
D &1.18&41 & 1.17&39&.05&-.027& .090& 1.16&44 & 1.14&47&.15 & .041&-.026 \\
R &1.20&36 & 1.19&35&.02& .017& .005& 1.18&41 & 1.15&46&.07 & .025& .028 \\
N &1.21&36 & 1.22&38&.27& .005& .028& 1.19&40 & 1.16&42&.07 &-.056& .045 \\ 
  &1.22&37 & 1.22&38&.05&-.029& .015& 1.21&36 & 1.20&45&.07 &-.076& .047 \\ 
  &    &   &     &  &   &     &     &     &   & 1.20&41&.17 &-.053& .054 \\ 
  &    &   &     &  &   &     &     &     &   & 1.21&39&.19 &-.059& .053 \\ 
%
%
%
\hline
\hline
\end{tabular}
\end{center}
\caption[]{
                List of
     $ \chi^{2}_{\rm DF} $
                in distinct solutions obtained after deleting
                undetermined parameters;
     $ N_{\rm Exp} = 411 $
                experimental points.
                (To save the space only the best
                {\it unphysical}
                solutions of the ``DRNbior'' variant
                are shown.)
                         }
\label{chiDFlist}
\end{table}

                {\bf 7.}
                The results of fittings can not improve
                the precision of determinations of
        $ \pi \pi $
                scattering lengths.
                For example,
                the region
        $ 0.06 \le a^{I=0}_0 \le 0.19 $
                might be derived from the Table
\ref{chiDFlist}
                as the preferable one;
                however,
                even
                the current experimental value
\cite{Dumbrais83}
        $  a^{I=0}_0 = 0.26 $
                can not be rejected on the ground of the
        $  \chi^{2}_{\rm DF} $
                criterion.


                {\bf 8.}
                The predictions of ChPT
                in the next--to--leading order
\cite{GasserL82}
        $a^{I=0}_0 = 0.20$,
        $a^{I=2}_0 = -0.041$,
        $a^{I=1}_1 = 0.036$,
        $a^{I=0}_2 = 0.002$
                are compatible with the data base;
                the separate fittings with the OPE parameters
                being kept fixed by the above values of
                scattering lengths,
                resulted in solutions with the values of
        $  \chi^{2}_{\rm DF} $
                very close to the best ones;
                for example,
                in the ``DRNbior'' variant
                we get
        $  \chi^{2}_{\rm DF} $
                in the range
        $  1.175  $ ---
        $  1.200  $.

                {\bf 9.}
                The formal parameters of the imaginary
                part of the OPE contribution are insignificant
                at the explored energy region
                (this is the result of
                the separate investigations;
                eliminating these parameters we get
                some shifts in the parameters of the imaginary
                background and in the parameters of
        $ N_{*} $
                and
        $  \Delta $
                isobars,
                the real parameters of OPE
                remaining the same).
\vspace{0.2cm}

                The discussed above routine
                of eliminating parameters provided
                further support of the point
                {\bf 5.}:
                no one of the considered mechanisms
                had been rejected as a whole in the course
                of improving the
        $  \chi^{2}_{\rm DF} $
                value.
                The listings of this routine contain also
                a large volume of information about
                the significance of a given parameter
                in terms of the ``statistical'' frequency
                of its participation.
                For example,
                in all variants the parameters
        $ A_{2} $,
        $ A_{3} $
                and
        $ A_{4} $
                of the linear background of ref.
\cite{BolokhovVS91}
                and the parameters
        $ A_{13} $
                and
        $ A_{15} $
                of eqs.
(\ref{SAnext}),
(\ref{SBCnext})
                were found to be necessary
                whereas the parameter
        $ A_{11} $
                entering the tensor structure of the amplitude
        $ \hat{D} $
                had been always ignored.
                In respect to the OPE parameters
                the content of the Tables
\ref{chiDFlist},
\ref{chilist1}
                already shows that the parameters
                are not of the top significance
                since there are many acceptable solutions
                without the OPE contribution at all.
                The parameter
        $ g_{1} $
                of the paper
\cite{BolokhovVS91}
                is found to be
                relatively important,
                the
        $ D $--waves
                parameters
        $ g_{2} $,
        $ g_{3} $
                being much less necessary.

                For illustrations we have chosen the solution
                from the ``DRNbior'' variant with
        $  \chi^{2}_{\rm DF} = 1.16 $
                (see Table
\ref{chiDFlist}).
                The data on total cross sections
                and the theoretical curves in terms of the
                quasi--amplitude
(\ref{qamp})
                are drawn in Fig. 2.
                The most intriguing feature
                of the discussed curves is expressed
                by the practical coincidence
                of the theoretical results for the
        $ \{ - \, 0 \, p \} $
                and
        $ \{ + \, 0 \, p \} $
                channels
                --- this is clearly seen
                from the separate picture in Fig
                2
                which we draw
                for the combined data.
                The picture makes it obvious that
                the discussed phenomenon is strongly implied
                by the experimental data.
                This means
                that the isotopic amplitude
        $ \hat{D} $,
                being the only origin of the difference
                of the theoretical cross sections of these
                channels
                (see eqs.
(\ref{ChAmp})),
                must vanish indeed.

                At
                Figs. 3--8
                one can find
                the theoretical curves of the same solution
                for the distribution data discussed
                in the sect. 3
                (in terms of the normalized
                quasi--amplitude
(\ref{qamp})).
                A part of the data
                did not enter the fittings ---
                in such cases the curves
                actually represent our predictions.
                The normalized quasi--amplitude measures
                the deviation from the empty--phase--space
                pattern.
                In its terms the behavior
                of the experimental data themselves
                appears to be quite different.

                For example,
                almost all angular spectra look flat.
                The explanation is simple:
                unlike the case of elastic reactions
        $ 2 \to 2 $
                the 1D distributions are given by
                the remaining (3--dimensional) phase space
                integrals in our case of the
        $ 2 \to 3 $
                process.
                The averaging over polarizations together with
                this averaging over the reaction phase space
                make the angular dependence so weak.
                However,
                this can not be true for sections of the phase
                space and/or
                for higher dimensional distributions.

                In contrast,
                the distributions in the invariant variables
(\ref{vars})
                are nontrivial;
                this proves the choice of variables
                of the paper
\cite{BolokhovVS91}
                to be the characteristic one for the dynamics
                of the considered reaction.


                The results of data analyses described
                in the current section
                make it possible to use the obtained amplitude
                for modeling two other approaches
                discussed in the Introduction,
                namely,
                the one by Olsson and Turner
                and the Chew--Low extrapolation.
                What we learn from these tests is discussed
                in the sect. 5. and 6.


\newpage
\vskip1.cm
\section{ Modeling the Olsson--Turner approach }


                The original idea
                of the Olsson--Turner approach
                was to relate the threshold values of the
        $ \pi 2 \pi $
                amplitude with the
        $ \pi \pi $
                scattering lengths.
                Therefore,
                there are two principal steps in the discussed
                approach:
                {\bf 1)}
                determination of the threshold limits
                of amplitudes
                of the independent isospin channels;
                {\bf 2)}
                calculation of
        $ \pi \pi $
                scattering lengths
                with the account of other contributions
                to the above threshold limits.

\subsection{ Threshold Amplitudes }

                Let us discuss the first step.

\subsubsection{ Database }

                We already know
                that the threshold amplitudes of 5
                experimentally observable channels
                are not independent.
                To provide the experimental test of the relations
        (\ref{R431})
                the threshold limits
        $ | M^0_X | $
                must be determined from fittings of the data on
                total cross sections.
                The results of the similar
                fittings were already reported
                starting from
                publication
\cite {BurkhardtL91}
                (see also
\cite {Pocanic94Dub}).
                The most important conclusion of the work
\cite {BurkhardtL91}
                is that the data in the region of
        $ P_{\rm Lab} \leq 400 $~MeV/c
                admit amplitudes
                which are linear in the CMS energy.
                Here,
                the data up to
        $ P_{\rm Lab} \leq 500 $~MeV/c
                will be exposed to linear fittings.
                Moreover,
                any preliminary selection will be excluded.
                Indeed,
                it is the unmotivated preference of one
                or another set of data which is the reason for
                contradictory results.

                The entire database
                contains along with the old data
                (and very old ones)
                the relatively new results.
                It should be noted,
                that before the OMICRON measurements
\cite{Kernl+-n89}--\cite{KernelEtAl91}
                only the
        $\{ - + n \}$
                channel allowed to obtain the definite
                results of the linear fit;
                the above mentioned OMICRON data
                provided the
                possibility to carry out the procedure
                also for the
        $\{ - \, 0 \, p \}$
                and
        $\{ + + n \}$
                channels.
                The authors of the paper
\cite{BurkhardtL91}
                took advantage of the precise data
\cite{Lowe_00n91}
                 on the
        $ \sigma_{\{00n\}}$
                (very close to threshold)
                and provided the simultaneous linear fit
                of all channels.

                            The recent experimental
                information
\cite {PocanicEtAl94}
                for the first time
                makes  possible
                to determine the threshold limit of cross
                section of the channel
        $\{+0p\}$
                and to test the prediction of the relation
\be
\label{R52}
               M^{0}_{\{+0p\}} \ = \
               M^{0}_{\{-0p\}} \ .
\ee
                (In fact,
                the approximate equality of cross sections
                of these two channels is observed
                along all energy interval considered.)

                The separate attention we pay to the
        $\{++n\}$
                channel.
                        There are two sets of the near
                threshold data in the channel
        $\{++n\}$,
                namely,
\cite{Kernl++n90}
                and
\cite{SeviorEtAl91}
                which are in certain disagreement.
                They lead to different threshold limits
       $ | M^{0}_{\{++n\}} | $.

\subsubsection{ Procedure }

                Therefore,
                to provide the definite conclusion
                four basic variants of fitting were used.
                Their symbolic notations,
                entering the summary Table
\ref{RChT},
                are:
                ALL --- variant with all the data
                being included;
                OMI --- in this variant the data
\cite{SeviorEtAl91}
                are omitted;
                TRI --- variant with the exclusion of the data
\cite{Kernl++n90};
                X --- the variant in which both sets
\cite{SeviorEtAl91}
                and
\cite{Kernl++n90}
                are excluded.
                The notation IND is used for the case
                when the data in each channel are fitted
                separately
                (the channel
        $\{ + + n \}$
                was treated in accordance with the selection
                adopted in ALL, OMI, TRI and X variants).
                This helps to understand the trend dictated
                by data of individual channel in the
                simultaneous fit.

                There were two types of fits
                in all four basic variants.
                First,
                the independent threshold values
        $ A^0 $
                and
        $ B^0 $
                are treated as being real
                (or, the same, as obeying the trivial
                relative phase).
                We use 7 parameters for simultaneous
                linear fittings of 5 channels:
                2 independent parameters for threshold values
                (namely,
        $ A^0 $
                and
        $ B^0 $)
                and 5 independent slopes in the
                invariant kinetic energy
\be
              T_{\rm K} \ = \
               \sqrt{s} - T_0 \ .
\ee
                (In the course of calculations the latter
                quantity was taken to be
$
              T_{\rm K} \ = \
               \sqrt{ ( p + k_1 )^2 }
                - ( m_f + \mu_2 + \mu_3 ) \
$
                since isospin splitting in the particle masses
                can not be processed as a correction
                because of the
                nonanalytic dependence of the near--threshold
                phase space on masses
                --- the demonstration can be found in
\cite{Beringer92}.)


                Second, we add one parameter to describe
                the relative phase of the complex quantities
        $ A^0 $
                and
        $ B^0 $.

                There is the strong motivation to consider
                the parameters
        $ A^0 $
                and
        $ B^0 $
                real.
                The estimate of the imaginary part of the
        $ \pi 2 \pi $
                amplitude
                by the dispersion analysis
                of the paper
\cite{AitchisonB79}
                allows to consider the amplitude to be
                approximately real up to the energies
        $ P_{\rm Lab} = 0.50 {\rm GeV/c} $.
                From the general point of view of unitarity
                relations the value of the imaginary part
                collects contributions which are due to the
                following:

                a.
                \underline
                {Three--particle intermediate states}.
                This includes contributions of
                quasi--two--particle states
                when the third particle is present as
                a spectator
                (namely,
        $ N $
                for the
        $ \pi \pi $
                state
                and
        $ \pi $
                for the
        $ \pi N $
                state;
                the former configuration is characterized by
                the large imaginary part of isospin--zero
        $ \pi \pi $
                amplitude).
                 The three--particle phase space makes
                this contribution
                vanish at threshold.
                (In the paper
\cite{BernardKM94}
                with the use of
        $ ( 1 / m ) $
                expansion it was already verified that
        $ \pi \pi $
                loops did not contribute to the threshold
        $ \pi 2 \pi $
                amplitude.)

                b.
                \underline
                {Two--particle intermediate states}.
                In this case
                only the
   $ \pi N $
                system is allowed.
                The angular momentum conservation forces
                these particles to be in the
        $ P $   wave
                when the final particles at threshold
                are in the
        $ S $   wave
                according to Quantum Mechanics ---
                otherwise the
        $P$--parity
                properties of the
   $ \pi N $
                system mismatch those of the
   $ \pi \pi N $
                final state.
                The known phases of the
        $ P_{11} $
                and
        $ P_{31} $
                waves
                of the
   $ \pi N $--elastic
                amplitudes approve the neglect
                of the imaginary part of the
   $ \pi 2 \pi $
                amplitude at the threshold.

                c.
                \underline
                {Single--particle intermediate states}.
                The only possible one
                at the threshold of the
   $ \pi 2 \pi $
                reaction
                is the
   $ \Delta $
                isobar.
                It appears in the
   $ P $
                wave
                of initial particles
                and its decay into the
                threshold
                configuration of the final
   $ \pi \pi N $
                state
                has negligible width
\cite{PDG96}.

                Therefore,
                there should be no expectations
                for finding
                a physically meaningful imaginary part
                (or a nontrivial relative phase of
        $ A^0 $
                and
        $ B^0 $)
                at the
   $ \pi 2 \pi $
                threshold.
                Hence,
                the threshold identity
(\ref{R4m1p3})
                discussed in the subsect. 2.1.
                must be practically exact.
                (The identities
(\ref{R425})
                are the exact consequences of the
                isotopic invariance
                irrespective of the value
                of the imaginary part.)

\subsubsection{ Results}

                The experimental data are shown in
                Fig. 9
                where the values of quasi--amplitudes
        $\langle M_X \rangle$
                are plotted versus the invariant
                kinetic energy
        $ T_{\rm K} $
                and the threshold amplitudes
        $|M^{0}_X|$
                are collected in the Table
\ref{RChT}.

\begin{table}
\begin{center}
\begin{tabular}{|c||c|c|c|c|c||c|}
\hline
 Fit &
     $ \{-+n\} $ &
          $\{\ 0 \ 0 \ n\}$ &
                    $\{- \ 0 \ p\}$ &
                              $\{+ \ 0 \ p\}$ &
                                   $\{++n\}$ &
                                             $\chi^2_{\rm DF}$ \\
     &
     $ | A^{0} + B^{0} | / \sqrt{2} $
                 &
                 $ | A^{0} | / 2 $
                            &
                            $ | B^{0} | / 2 $
                                    &
                                    $ | B^{0} | / 2 $
                                              &
                                              $ |  B^{0} | $
                                                        &
                                                              \\
\hline
\hline
 {\rm ALL} &
       {\bf 398 $\pm$ 18 }   &
            {\bf 413 $\pm$ 12 }   &
                      {\bf 132 $\pm$  6 }   &
                                {\bf 132 $\pm$  6 }   &
                                     {\bf 264 $\pm$ 11 }   &
                                              {\bf 0.99 }
                                             \\
           &
       { {\sl 524 $\pm$ 24} }   &
            {\sl 252 $\pm$ 12 }   &
                      {\sl 118 $\pm$  5 }   &
                                {\sl 118 $\pm$  5 }   &
                                     {\sl 236 $\pm$ 11 }   &
                                              {\sl 1.56 }
                                             \\
           &
       { 401 $\pm$ 150 }   &
            { 409 $\pm$ 101 }   &
                      { 132 $\pm$  6 }   &
                                { 132 $\pm$  6 }   &
                                     { 264 $\pm$ 11 }   &
                                               1.00    \\
           &
   {\small ( 74 $\pm$ 123 ) } &
       {\small ( 52 $\pm$ 87 ) } &
                                        &
                                                   &
                                                        &
                                                        \\
\hline
 {\rm TRI} &
      {\bf 404 $\pm$ 18 }   &
            {\bf 406 $\pm$ 12 }   &
                      {\bf 121 $\pm$  6 }   &
                                {\bf 121 $\pm$  6 }   &
                                     {\bf 241 $\pm$ 11 }   &
                                              {\bf 0.78 }
                                             \\
           &
       { {\sl 518 $\pm$ 24} }   &
            {\sl 260 $\pm$ 12 }   &
                      {\sl 106 $\pm$  6 }   &
                                {\sl 106 $\pm$  6 }   &
                                     {\sl 212 $\pm$ 11 }   &
                                              {\sl 1.33 }
                                             \\
           &
       { 406 $\pm$ 26 }   &
            { 406 $\pm$ 18 }   &
                      { 119 $\pm$  6 }   &
                                { 119 $\pm$  6 }   &
                                     { 238 $\pm$ 12 }   &
                                               0.77     \\
           &
  {\small ( $10^{-4}$ ) } &
       {\small ( $10^{-4}$ ) } &
                                        &
                                                   &
                                                        &
                                                        \\
\hline
 {\rm OMI} &
      {\bf 372 $\pm$ 23 }  &
            {\bf 447 $\pm$ 13 }   &
                      {\bf 184 $\pm$ 10 }   &
                                {\bf 184 $\pm$ 10 }   &
                                     {\bf 368 $\pm$ 20 }   &
                                              {\bf 0.83 }
                                             \\
           &
       { {\sl 533 $\pm$ 32} }   &
            {\sl 241 $\pm$ 13 }   &
                      {\sl 136 $\pm$ 10 }   &
                                {\sl 136 $\pm$ 10 }   &
                                     {\sl 272 $\pm$ 20 }   &
                                              {\sl 1.60 }
                                             \\
           &
       { 401 $\pm$ 60 }   &
            { 410 $\pm$ 54 }   &
                      { 197 $\pm$ 11 }   &
                                { 197 $\pm$ 11 }   &
                                     { 395 $\pm$ 22 }   &
                                               0.80      \\
           &
  {\small ( 181 $\pm$ 23 ) } &
     {\small ( 128 $\pm$ 16 ) } &
                                        &
                                                   &
                                                        &
                                                        \\
\hline
 {\rm X} &
      {\bf 387 $\pm$ 26 }           &
             {\bf 427 $\pm$ 14 }   &
                      {\bf 154 $\pm$ 12 }   &
                                {\bf 154 $\pm$ 12 }   &
                                     {\bf 307 $\pm$ 24 }   &
                                               {\bf 0.76}
                                             \\
            &
       { {\sl 506 $\pm$ 36} }    &
             {\sl 275 $\pm$ 14 }   &
                      {\sl  83 $\pm$ 12 }   &
                                {\sl  83 $\pm$ 12 }   &
                                     {\sl 166 $\pm$ 24 }   &
                                              {\sl 1.36 }
                                             \\
           &
       { 402 $\pm$ 98 }             &
             { 410 $\pm$ 76 }      &
                      { 164 $\pm$ 14 }      &
                                { 164 $\pm$ 14 }      &
                                     329 $\pm$ 29          &
                                               0.76      \\
           &
  {\small ( 154 $\pm$ 40 ) } &
        {\small ( 109 $\pm$ 28 ) } &
                                        &
                                                   &
                                                        &
                                                        \\
\hline
\hline
 {\rm IND}  &
       {\bf 401 $\pm$ 19 }   &
       {\bf 409 $\pm$ 18 }   &
       {\bf 155 $\pm$ 25 }   &
       {\bf  97 $\pm$ 51 }   &
       {\bf 263 $\pm$ 11 }   & {\rm ALL} \\
          &
               &
                    &
                         &
                              &
              {\bf 237 $\pm$ 12 }   & {\rm TRI} \\
          &
               &
                    &
                         &
                              &
              {\bf 429 $\pm$ 26 }   & {\rm OMI} \\
          &
               &
                    &
                         &
                              &
              {\bf 358 $\pm$ 37 }   & {\rm X} \\
\hline
\hline
  Eq.(\ref{R4m1p3})   &
   -- &
     -- &
                 {\bf 125 $\pm$ 22 }   &
                           {\bf 125 $\pm$ 22 }   &
                                {\bf 250 $\pm$ 44 }   &
                                                       \\
\hline
  Eq.(\ref{R425})   &
   -- &
     -- &
          -- &
                           {\bf 155 $\pm$ 25 }   &
                                {\bf 310 $\pm$ 50 }   &
                                                       \\
\hline
\end{tabular}
\end{center}
\caption[Threshold amplitude]{
                Threshold values provided by data fittings
                in the variants:
                               ALL --- all data;
                               TRI ---  without
                                        \cite{Kernl++n90};
                               OMI ---  without
                                        \cite{SeviorEtAl91};
                                X --- excluding
\cite{Kernl++n90,SeviorEtAl91}.
                Bold--face numbers correspond to physical solutions
                of real fits with
        $ 7 = 5 + 2 $
                parameters,
                slanted ones --- to unphysical solutions.
                The roman font is used for fits with complex
                amplitudes.
                The numbers in brackets in the
                columns of the
        $\{ - + n \}$
                and
        $\{ 0 0 n \}$
                channels display the imaginary part of the
                resulting amplitude (provided
        $ B^0 $
                is real).
                             }
\label{RChT}
\end{table}

                This
                Table
                is organized in the following way.
                The first column
                contains the symbolic name of
                the selection used for fitting.
                Five  subsequent columns refer to the channel
                described in the column header.
                The last column provides the characteristics of
                the fit ---
        $ \chi^2_{\rm DF} $.
                To make the comparison easy
                the additional row IND contains the results of
                individual fits of every channel
                (in the case of the
        $\{++n\}$
                channel three additional fits
                with the exclusion of the
                controversial data sets, namely
\cite{SeviorEtAl91}
                and
\cite{Kernl++n90},
                were performed).
                The last two rows present
                the predictions of
                eqs.
  (\ref{R4m1p3})
                and
  (\ref{R425})
                based on the data of individual fits of
                the channels
        $\{ - + n \}$,
        $\{ 0 \, 0 \, n \}$
                and
        $\{ - \, 0 \, p \}$.

                The bold--face numbers in the boxes
                ALL, TRI, OMI, X display the results of the
                solutions in the main fit with
        $ 7 = 2 + 5 $
                parameters.
                Every time there was also present another
                solution of the fit
                (the slanted numbers in the Table
\ref{RChT}
                are used for resulting values).
                It originates from the sign ambiguity
                which was resolved in part by means of
                the phenomenological analysis
                of the subsect. 2.1.3.
                These auxiliary solutions choose the relation
        (\ref{R41p3})
                to be true one
                and it was called {\it unphysical}
                since it provides the equal signs of
                isospin--0 and
                isospin--2
   $ \pi \pi $
                scattering lengths.

                The third line in every box
                of variants ALL, TRI, OMI, X
                represents results of the fit with the
                additional parameter describing the relative
                phase
                of amplitudes
        $ A^0 $
                and
        $ B^0 $
                when
        $ A^0 $
                and
        $ B^0 $
                are considered as complex numbers.
                The resulting values of the imaginary
                parts of the amplitudes
        $ M^0_{\{-+n\}} $
                and
        $ M^0_{\{00n\}} $
                are given in brackets in the fourth line
                (the overall phase ambiguity is resolved
                be assuming the amplitude
       $ B^{0} $
                to be real).

\subsubsection{ Discussion }

                First of all,
                let us compare the results of the linear fit
                discussed here with the results
                of the previous section ---
                it is sufficient to examine the pictures
                on Figs.
                2, 9.
                One can deduce
                that the threshold limits provided by
                the linear fit are good for all channels
                but the
       $ \{ - + n \} $
                one.
                Because of the perceptible curvature
                revealed by solutions in the latter case
                the linear fit generally underestimates
                the value in question.

                The Table
\ref{RChT}
                does not allow to make
                an unambiguous conclusion
                on the ground of the quality of the fit only.
                According to the
   $ \chi^2 $
                criterion
                all solutions are
                practically on equal footing.
                Even the unphysical ones can not
                be formally rejected.

                The relatively low values of
   $ \chi^2 $
                of the unphysical solution are,
                first of all,
                due to rather large overall uncertainties of
                experimental data.
                The more discouraging reason
                is the existence of
                arrays of data
                (almost in every channel)
                supporting the discussed solution.
                Only the data of the
        $\{ 0 \, 0 \, n \}$
                channel as a whole reject this solution.
                The examination of the definitions
        (\ref{M0})
                displays that at the value of
        $ B^0 $
                being fixed by the data of
        $\{ - \, 0 \, p \}$
                and
        $\{ + + n \}$
                channels at the scale no less than
        $ 200 [ {\rm GeV ]^{-1} } $
                the threshold limits of
        $\{ - + n \}$ and
        $\{ 0 \, 0 \, n \}$
                channels can not be in balance
                provided
        $ A^0 $
                and
        $ B^0 $
                amplitudes are of equal signs.
                While the closest to threshold points
                of the OMICRON experiment
\cite{Kernl++n90}
                well agree with the high value of
        $ | M^0_{\{ - + n \}} | $
                the data of the
        $\{ 0 \, 0 \, n \}$
                channel
\cite{Lowe_00n91}
                leave no room for describing the cross
                sections of this channel at so small level
                (see Table
\ref{RChT}).

                We end up our discussion of the unphysical
                solution by noting
                that there is a striking feature of the
                entire database:
                the distribution data,
                being added to the total cross section ones,
                did not help to make the distinction
                between
                these two kinds of solutions.
                Moreover,
                the absolutely best solutions had been found
                to be the unphysical ones --- see Table
\ref{chiDFlist}.

                    In all variants the presence of the
                    imaginary part improves the fit.
                    In the variants ALL and TRI
                    the imaginary part
       $ {\rm Im}~A^{0} $
                remains consistent with zero:
       $ 104  \pm  174 $ --- ALL and
       $ 10^{-3}  \pm  10^{2} $ --- TRI
                (the amplitude
       $ B^{0} $
                is considered to be real).
                    In the rest variants the imaginary part
                    is found to be unreasonably large:
       $ 255  \pm  32 $ --- OMI ,
       $ 218  \pm  57 $ ---  X.
%
                Nevertheless,
                it can not help to get rid of the
                    contradiction of data in the
       $\{- \ 0 \ p\}$ and
       $\{++n\}$
                    channels,
                especially in the OMI variant of the individual
                fit.


                The examination of the threshold identities
(\ref{R425}),
(\ref{R4m1p3})
                clearly displays that the contradiction
                is between the OMICRON data in
       $\{- \ 0 \ p\}$
                and
       $\{++n\}$
                    channels
                themselves.
                Therefore,
                the analysis of the Table
\ref{RChT}
                provides the main conclusions
                that the most consistent
                from the point of view
                of threshold identities,
                the stability against the addition of the
                complex phase
                and a reasonable distinction from the
                unphysical solutions
                are the threshold amplitudes given in the
                TRI and ALL variants of the above Table.

\subsection{ Account of NonOPE Contributions }

                The discussed above threshold amplitudes
                are the keystone of the Olsson--Turner
                approach.
                (Therefore,
                it was the inconsistent input of the OMICRON
                analysis
\cite{Kernl+-n89}--\cite{KernelEtAl91}
                that provided a controversy in the derived
                values of
        $ \pi \pi $
                scattering lengths.)
                The original formulae
\cite{OlssonT686972}
                relating the
        $ \pi 2 \pi $
                threshold amplitudes with the
        $ \pi \pi $
                scattering lengths were based
                --- in modern terms ---
                on the leading order Lagrangian of ChPT.

                Now the improved formulae take into account
                the next--to--leading order terms of the
                Chiral
        $ \pi N $
                Lagrangian
\cite{OlssonMKB95,BernardKM94}.
                The series of papers
\cite{BernardKM95,Meissner95,BernardKM97}
                deals with various schemes of accounting
                the above terms in the framework of ChPT
                and/or Heavy Baryon ChPT
                and gives the predictions
                for the nonrelativistic quantities
        $ D_{1} $
                and
        $ D_{2} $
                which up to a factor coincide with
                the discussed above threshold amplitudes
        $ A^{0} $
                and
        $ B^{0} $:
\begin{equation}
\label{D1D2}
        \left\{
\begin{array}{c}
                D_{1} \\
                D_{2}
\end{array}
        \right\}
                        =
                - \frac{\sqrt{\frac{ 2m }
                                   { m + E_{0} }
                             }
                       }
                       { 4m \sqrt{ -\tau_{0} } }
        \left\{
\begin{array}{c}
                B^{0} \\
                A^{0}
\end{array}
        \right\}
          \; .
\end{equation}
                The difference in the predicted values
                (which are quoted in the Table
\ref{D1D2tab}
                below)
                makes it paramount to test the approximation
                schemes of the discussed papers
                and the hypotheses about the importance of
                various contributions.
                The results of our analysis are suitable for
                this purpose
                since the phenomenological amplitude
                determined by fittings is consistent at least
                with the treated data.

                The following Table
\ref{D1D2tab}
                contains the listing of
                (nonzero)
                contributions to the threshold quantities
        $ D_{1} $
                and
        $ D_{2} $
                found in the solution with
        $\chi^2_{\rm DF} = 1.16 $
                (see the variant ``DRNbior'' of the Table
\ref{chiDFlist}
                for values of scattering lengths).
                We also quote here the predictions of papers
\cite{BernardKM95,Meissner95,BernardKM97}
                and the results ALL and TRI of the linear fit
                from the Table
\ref{RChT}
                for an easy comparison.
                The underestimate of the
        $ D_{2} $
                value
                by the linear fits ALL and TRI
                is due to the pointed above
                tangible departure of the
       $\{-+n\}$
                channel's
                solution
                from the linear pattern.

\begin{table}
\begin{center}
\begin{tabular}{|c|| r c r || r c r |}
\hline
         &           &$D_{1}$&           &           &$D_{2}$&          \\
\hline
\hline
  $ A_1$ & 310.80    & $\pm$ &    35.44  &   310.80  & $\pm$ &    35.44 \\
  $ A_2$ & 194.66    & $\pm$ &     6.66  &   194.66  & $\pm$ &     6.66 \\
  $ A_3$ & 531.15    & $\pm$ &    13.61  & -1062.31  & $\pm$ &    27.23 \\
  $ A_4$ & 597.88    & $\pm$ &    21.26  &  1195.77  & $\pm$ &    42.52 \\
  $ A_6$ & -35.15    & $\pm$ &    95.36  &   -35.15  & $\pm$ &    95.36 \\
 $A_{12}$&   -69.56  & $\pm$ &     8.05  &   -69.56  & $\pm$ &     8.05 \\
 $A_{13}$&  -312.40  & $\pm$ &     7.48  &   624.80  & $\pm$ &    14.96 \\
 $A_{15}$&  -159.25  & $\pm$ &     4.08  &   318.50  & $\pm$ &     8.17 \\
 $A_{16}$&  -186.93  & $\pm$ &    16.28  &  -186.93  & $\pm$ &    16.28 \\
 $A_{17}$&  -170.52  & $\pm$ &    21.80  &   341.04  & $\pm$ &    43.60 \\
\hline
  $ o_1$ &   102.14  & $\pm$ &   251.12  &   102.14  & $\pm$ &   251.12 \\
  $ o_2$ &   488.02  & $\pm$ &    31.38  &  -976.04  & $\pm$ &    62.76 \\
  $ o_3$ &  -101.07  & $\pm$ &    33.53  &  -101.07  & $\pm$ &    33.53 \\
  $ o_4$ &  -161.90  & $\pm$ &    32.09  &   323.80  & $\pm$ &    64.19 \\
\hline
  $ r_2$ &   300.89  & $\pm$ &     7.75  &  -601.78  & $\pm$ &    15.50 \\
\hline
  $ D_1$ &   -46.75  & $\pm$ &    15.48  &    93.50  & $\pm$ &    30.96 \\
  $ D_2$ &  -238.69  & $\pm$ &    51.93  &    42.32  & $\pm$ &    55.87 \\
  $ D_3$ &    93.27  & $\pm$ &    23.77  &  -186.54  & $\pm$ &    47.53 \\
  $ D_4$ &   -14.91  & $\pm$ &    66.48  &  -115.75  & $\pm$ &   133.00 \\
  $ D_5$ &     5.88  & $\pm$ &    11.12  &   -19.53  & $\pm$ &    33.28 \\
\hline
  $ R_1$ &    52.87  & $\pm$ &   108.57  & -1079.45  & $\pm$ &   192.22 \\
  $ R_3$ &    65.55  & $\pm$ &    26.58  &  -131.09  & $\pm$ &    53.15 \\
  $ R_5$ &     0.46  & $\pm$ &    12.95  &   -90.90  & $\pm$ &    19.91 \\
  $ R_6$ &     2.41  & $\pm$ &   211.45  &    28.60  & $\pm$ &   505.56 \\
  $ R_7$ &    -7.91  & $\pm$ &     4.73  &   -40.14  & $\pm$ &    48.72 \\
\hline
  $ N_2$ &   128.70  & $\pm$ &    32.77  &     2.81  & $\pm$ &    22.75 \\
  $ N_3$ &   152.34  & $\pm$ &    45.49  &  -304.69  & $\pm$ &    90.97 \\
  $ N_5$ &   -12.29  & $\pm$ &    22.62  &    15.16  & $\pm$ &    21.78 \\
\hline
\fbox{Sum}
         & \fbox{
             313.92} &
                 \fbox{$\pm$}& \fbox{
                               101.37}   & \fbox{
                                           -1407.03} &
                                                 \fbox{$\pm$}& \fbox{
                                                                 229.23}\\
\hline
\hline
\cite{BernardKM95}
         &   339.69  & $\pm$ &    28.63  & -1186.96  & $\pm$ &   123.64 \\
\cite{Meissner95}
         &   344.89  & $\pm$ &    31.24  & -1179.15  & $\pm$ &   136.66 \\
\cite{BernardKM97}
         &   334.48  & $\pm$ &    13.01  & -1231.21  & $\pm$ &     7.81 \\
\hline
   ALL   &   327.78  & $\pm$ &    14.90  & -1025.54  & $\pm$ &    29.80 \\
   TRI   &   300.46  & $\pm$ &    14.90  & -1008.16  & $\pm$ &    29.80 \\
\hline
\end{tabular}
\end{center}
\caption[]{
                Contributions
                to the values of threshold
        $D_{1}$
                and
        $D_{2}$
        (in [GeV]${}^{-3}$).
                         }
\label{D1D2tab}
\end{table}

                It should be noted once more that there is no
                absolute meaning of the separate contributions
                because of the field redefinition freedom.
                However,
                the ``on--mass--shell'' parameters
        $g_{0}$,
        $g_{1}$,
        $g_{2}$
                and
        $g_{3}$
                of the
        $4 \pi$
                vertex
                are stable and the
        $\pi \pi$
                scattering lengths are well defined
                in our approach
                though the ``off--shell'' contributions
                of the parameters
        $g_{0}$,
        $g_{1}$,
        $g_{2}$
                and
        $g_{3}$
                in the Table
\ref{D1D2tab}
                are model dependent.
                Hence,
                the most general inferences which
                might be derived from this Table
                are the following:

                1.
                The resulting values of the threshold
                amplitudes and the quantities
        $D_{1}$
                and
        $D_{2}$
                are rather small differences of large
                contributions from various mechanisms.

                2.
                In both quantities the OPE mechanism
                meets the strong competition from all the
                rest ones,
                the
        $\Delta$
                being of importance for
        $D_{1}$
                while the
        $N_{*}$
                --- for
        $D_{2}$.

                3.
                Within the overall OPE contribution
                the influence of the
        $ D $--wave
                parameters
        $ g_{2} $,
        $ g_{3} $
                reaches 30\%
                in both quantities
        $D_{1}$
                and
        $D_{2}$.

\subsection{ Conclusion }

                We have already seen in the previous
                subsection
                that
                in the experimental side
                the resolution of an ambiguity in
                two threshold amplitudes
        $ A^{0} $
                and
        $ B^{0} $
                upon which
                the Olsson--Turner approach is relying
                is the matter of the precision
                of experimental data on
        $ \pi 2 \pi $
                total cross sections.
                We also noticed the evidence
                of the inapplicability of the linear fit
                for  the
       $\{-+n\}$
                channel
                --- this enlarges the systematic errors of
        $ A^{0} $
                and
        $ B^{0} $
                values.
                Here,
                one can see
                that there are also unknown systematic
                errors in the theoretical field.
                To fill in this gap the large amount of
                experimental and phenomenological information
                on isobar physics is necessary.
                Therefore,
                the approach is losing the advantage of
                simplicity which
                had been
                making it so attractive.

\newpage
\vskip1.cm
\section{ Chew--Low Extrapolation }

\vspace{0.5cm}

                The existence of the pole at
        $ \tau = \mu^2 $
                in the OPE contribution is the
                keystone of the Chew--Low approach.
                However,
                there are several routines for the
                extrapolation
                to this point
                --- the relevant discussion
                on the applications
                of the approach might be found in
                the review paper
\cite{Leksin70}
                (see also
\cite{Naisse64}
                for a phenomenological introduction);
                the tests of some variants performed
                long time ago were reported in the paper
\cite{Baton67}.
                The amplitudes provided by our fits present
                the possibility to test the approach by
                modeling the experimental data and to arrive
                at conclusions which are independent of the
                finite precision of the input.

\subsection{ Extrapolation Function }

                From the very beginning it has become evident
                that the extrapolation function constructed
                in terms of the total cross section could not
                provide a base neither for the linear
                extrapolation nor for the quadratic one
                (with the only exception for the case of the
        $ \pi 2 \pi $
                amplitude build of a
                {\it constant} OPE term plus a constant in the
                same spinor structure
        $ S $).
                Therefore,
                to save the space
                we discuss only the extrapolation function
        $  F_{M} ( \tau ) $
                defined in terms of the quasi--amplitude:
\ba
\label{FChL}
  \left[{ F }_{M}(\tau)\right]^2
      & \equiv &
        \frac{(\tau -\mu^2)^2}{(-\tau)( 2g_{\pi NN} )^{2}}
        \times
        \frac{d \sigma (||M||^{2})}
             {d \sigma(1)}
        \; .
\ea
                The principal feature
                of this extrapolation function
                is the one, reflecting the vanishing
                of the cross section at
        $  \tau = 0 $.
                It will be commented later on
                during the discussion of results.

                Since --- according to the general idea
                of the Chew--Low extrapolation ---
                only the OPE term of the
        $ \pi 2 \pi $
                amplitude
                contributes to the quantity
        $  F_{M} ( \mu^{2} ) $
                it is convenient to define the auxiliary
                function
\ba
\label{FChLOPE}
  \left[{F}_{\rm OPE}(\tau)\right]^2
      & \equiv &
        \frac{(\tau -\mu^2)^2}{(-\tau)( 2g_{\pi NN} )^{2}}
        \times
        \frac{d \sigma (||M_{\rm OPE} )||^{2}) }
             {d \sigma(1)}
  =
\frac{\int\!\!\!\int\limits_{\Omega(\tau)}
                           d \theta_{I}
                           d \nu_{I}
                              \left|{ V}_{4\pi}\right|^2
     }
     {\int\!\!\!\int\limits_{\Omega(\tau)}
                           d \theta_{I}
                           d \nu_{I}
     }
        \; .
\ea
                In the above equations
                both the LHS and the integrals
                of the RHS depend on the reaction energy
        $ s $
                and the dipion invariant mass
        $ s_{\pi\pi} $
                which are assumed to be fixed
                in the course of the extrapolation;
                the amplitude
        $  M_{\rm OPE} $
                is obtained from
        $  M  $
                by setting to zero
                all contributions but the OPE one.

                The analytic calculation of the function
        $  F_{M} ( \tau ) $
                is possible only for the case
                of the simplest models of the amplitude.
                Therefore,
                we calculate the function
        $  F_{M} ( \tau ) $
                obtained in various solutions for our
                phenomenological amplitude numerically.
                However,
                because of the collapse of the integration
                domain
        $  \Omega (\tau)  $
                outside the phase space of the
        $ \pi 2 \pi $
                reaction
                the numerical calculation of the function
        $  F_{M} ( \tau ) $
                at
        $  \tau = \mu^2 $
                also becomes impossible.
                (The Monte--Carlo based utilities of
                high energy physics can not generate
                events outside the phase space
                for the numerical integration.)

                The integration in the RHS of eq.
(\ref{FChLOPE})
                results in the rational function of
        $ \tau $
                and
        $ s_{\pi\pi} $.
                Neglecting the imaginary part of the OPE
                amplitude
                this function might be cast
                as the quadratic form in the OPE parameters
        $ g_{0} $,
        $ g_{1} $,
        $ g_{2} $,
        $ g_{3} $
                of the paper
\cite{BolokhovVS91}:
\be
\label{FOPE}
 \left[{ F}_{\rm OPE}\right]^2 =
\left(
\begin{array}{c}
         g_0\\
         g_1\\
         g_2\\
         g_3
\end{array}
\right)^{\rm T}
   \left(
          \hat{\Phi}
    \right)
\left(
\begin{array}{c}
         g_0\\
         g_1\\
         g_2\\
         g_3
\end{array}
\right)
        \; ,
\ee
                where the upper triangle
                of the symmetric matrix
        $ \hat{\Phi} $
                is explicitly given by
\begin{eqnarray}
\nonumber
\Phi_{00} & = &  2
       \; ; \;\;
\Phi_{01} \; = \;  \theta_{R}
       \; ; \;\;
\Phi_{02} \; = \;  2 (9 \theta_{R}^2 + A_{1})/9
       \; ; \;\;
\Phi_{03} \; = \;  (9 \theta_{R}^2 - A_{1})/9
        \; ;
\\
\nonumber
\Phi_{11} & = &  (3 \theta_{R}^2 + A_{1})/6
       \; ; \;\;
\Phi_{12} \; = \;  (9 \theta_{R}^2 + A_{1}) \theta_{R}/9
       \; ; \;\;
\Phi_{13} \; = \;  (9 \theta_{R}^2 - 7 A_{1}) \theta_{R}/18
        \; ;
\\
\nonumber
\Phi_{22} & = &  2 (45 \theta_{R}^4 + 10 \theta_{R}^2 A_{1}
                    + A_{1}^2)/45
       \; ; \;\;
\Phi_{23} \; = \;  (45 \theta_{R}^4 - A_{1}^2)/45
        \; ;
\\
\label{Phi0033}
\Phi_{33} & = &  (45 \theta_{R}^4 + 50 \theta_{R}^2 A_{1}
                  + A_{1}^2)/90
        \; ;
\\
\nonumber
A_{1} & = & \frac{ s_{\pi\pi} - 4 \mu^2}
                 { s_{\pi\pi} }
        [ (\theta_{R} - \tau)^{2} - 9 s_{\pi\pi} \tau /4 ]
        \; ; \;\;
             \theta_{R} \; = \; ( \tau + 3 \mu^2 - 3s_{\pi\pi}
                                ) / 4
        \; .
\end{eqnarray}
                The discussed function
(\ref{FChLOPE})
                is well defined outside the reaction
                phase space and at
        $  \tau = \mu^2 $
                gives the value in question
        $  F_{\rm OPE} ( \mu^{2} ) = F_{M} ( \mu^{2} ) $.

                In fact,
                the actual calculations had been being performed
                with the function
        $  \tilde{F}_{\rm OPE}(\tau) $
\ba
\label{FChLOPEw}
  \left[\tilde{F}_{\rm OPE}(\tau)\right]^2
      & \equiv &
 \left.
  {\int\limits_{s_{\pi\pi} - \Delta s_{\pi\pi}}^{s_{\pi\pi}
                                                 + \Delta s_{\pi\pi}}
        ds_{\pi\pi}~ \sqrt{ \frac{s_{\pi\pi} - 4\mu^{2}}
                                 {s_{\pi\pi}}
                          }
                \left[{F}_{\rm OPE}(\tau)\right]^2
     }
 \, \right/  \,
  {\int\limits_{s_{\pi\pi} - \Delta s_{\pi\pi}}^{s_{\pi\pi}
                                                  + \Delta s_{\pi\pi}}
        ds_{\pi\pi}~ \sqrt{ \frac{s_{\pi\pi} - 4\mu^{2}}
                                 {s_{\pi\pi}}
                          }
     }
        \, ; \,
\ea
                since the experimental
                (and the simulated)
                data are being
                presented
                for a strip
        $ s_{\pi\pi}^{(\alpha)} - \Delta s_{\pi\pi}
                \leq s_{\pi\pi} \leq
                        s_{\pi\pi}^{(\alpha)} + \Delta s_{\pi\pi} $
                in the Chew--Low plot in the
        ($ \tau $,
        $ s_{\pi\pi} $)
                variables with
                the non--negligible
                width
        $ 2 \Delta s_{\pi\pi} $.
                The analytic calculations result in much more
                complicated expressions than that of eqs.
(\ref{Phi0033}),
                so we do not present the final answers here.
                Because of the rapid growth of the
        $  \pi \pi $
                amplitude with
        $ s_{\pi\pi} $
                at the threshold
                the difference of the simple function
(\ref{FChLOPE})
                with the above one
(\ref{FChLOPEw})
                was found to be reaching 20\%
                at
        $ \Delta s_{\pi \pi} = 0.15 \mu^2 $.

                The nonnegligent spread in
        $ s_{\pi\pi} $
                has another important issue
                for the extrapolation.
                The specific bin
        $ (\alpha) $
                is then characterized by
                the rectangle
        ($ \tau^{(\alpha)} \pm \Delta \tau $,
        $ s_{\pi\pi}^{(\alpha)} \pm \Delta s_{\pi\pi} $)
                in the Chew--Low plane
        ($ \tau,  s_{\pi\pi} $).
                The cross section for the bin is given
                by the integral
\begin{equation}
        \sigma_{(\alpha)} ( \| M \|^{2} ; \tau ) =
        \int_{s_{\pi\pi}^{(\alpha)}
                - \Delta s_{\pi\pi}}^{s_{\pi\pi}^{\rm MAX}}
                ~ds_{\pi\pi} ~R ( \| M \|^{2} ; s_{\pi\pi}, \tau )
        \; ,
\end{equation}
		where
        $  R ( \| M \|^{2} ; s_{\pi\pi}, \tau )   $
		stands for
                the matrix element integrated
		over the rest 2 variables
		(which include the
	$ \pi \pi $
		scattering angle).
		The upper limit
\begin{equation}
        {s_{\pi\pi}^{\rm MAX}}
		\equiv {\rm MAX} \{
                                {s_{\pi\pi}^{(\alpha)} + \Delta s_{\pi\pi}},
                                s_{\pi\pi}^{+} ( \tau )
                                 \}
\end{equation}
		is independent of
	$ \tau $
                only for bins
        $ (\alpha) $
                for which the strip
        $ s_{\pi\pi}^{(\alpha)} - \Delta s_{\pi\pi}
                \leq s_{\pi\pi} \leq
                        s_{\pi\pi}^{(\alpha)} + \Delta s_{\pi\pi} $
                is going strictly inside the physical domain
                of the Chew--Low plot.
                Therefore,
                in the case of bins located at the boundaries
                of the physical interval
        [$ \tau^{-} ( s ), \tau^{+} ( s ) $]
                the physical space in the
        $ s_{\pi\pi} $
                variable is bounded not by the value
        $ s_{\pi\pi} + \Delta s_{\pi\pi} $
                independent of
        $ \tau $
                but by the curve
\begin{eqnarray}
\label{sata}
                s_{\pi\pi}^{+} ( \tau )
        =
                \frac{1}{  2 m^{2} }
   \Bigl \{
             \tau
             (  s + m^{2}  - \mu^{2} )
                + 2  m^{2} \mu^{2}
         +
                \sqrt{
                         \tau
                        ( \tau - 4  m^{2} )
                        ( s - ( m + \mu )^{2} )
                        ( s - ( m - \mu )^{2} )
                      }
     \Bigr \}
        \, . \,
\end{eqnarray}

                As a result,
                the value of the phase space
        $ \sigma_{(\alpha)} ( 1; \tau ) $
                for such bins does depend on
        $ \tau $
                as well as integrals of powers
        $ (s_{\pi\pi})^{n} $
                do.
                This makes necessary to withdraw such bins
                from the extrapolation base regardless of
                the kind of the extrapolation function.
                For example,
                both the cross section
        $ \sigma_{(\alpha)} ( \| M \|^{2} ; \tau ) $
                and the quasi--amplitude have the breaking
                points at 2 values of
        $ \tau $
                --- at the ones for which
        $ s_{\pi\pi}^{+} ( \tau ) =
          s_{\pi\pi} + \Delta s_{\pi\pi} $.
                This phenomenon is clearly seen in the
Fig. 10.
                In the course of a practical data treatment
                the selection of bins is an easy problem
                which is solved by calculating the empty
                phase space for the considered array of bins
                and keeping
                on
                the ones with the constant value of the phase
                space.


		Would one know in advance the
        $  s_{\pi\pi}  $
		dependence
		of the partially integrated matrix element
        $  R ( s_{\pi\pi}, \tau )   $
		it would be possible
		to make corrections for the bins
		intersecting the boundary curve
        $ s_{\pi\pi}^{+} ( \tau ) $
		and to include more points
                into the Chew--Low extrapolation.
                Our curves in
                Fig. 10
                had been corrected by the empty phase space
		--- evidently this is insufficient.
                (Since our solutions
		fit well the
        $ s_{\pi\pi} $
		distributions
		which are very distinct from the phase space
		there are no much room for wondering.)
                Another possibility to enlarge the base
                of the extrapolation by cutting more
                narrow strips in
        $ s_{\pi\pi} $
                depends completely on
                the experimental statistics.

\subsection{ Simulations of Chew--Low Extrapolation }

                The statistics of our data
                (see experimental points in
                Fig. 10)
                can not provide a confidence for the
                results of
                a real--data extrapolation.
                Therefore,
                we were simulating the distributions
                with the help
                of the theoretical amplitudes
                and constructing the extrapolation function
        $ F_{M} (\tau) $
                for a reasonable number of bins.
                The input data errors then become
                negligible,
                hence,
                the problem might be investigated
                in its pure state.

                Here,
                we discuss the simulations of
        $ \tau $
                distributions
                (for the fixed strip in
        $ s_{\pi\pi} $)
                performed for several energies,
                namely, for
        $ P_{\rm Lab}  = 335, \, 420 $
                and
        $ 460 $~MeV/c.
                The extrapolation functions calculated
                for three types of theoretical amplitudes
                at
        $ P_{\rm Lab}  = 460 $~MeV/c
                are shown in
                Fig. 10.
                The shown data
                are simulated with the binning
                and the precision which are only
                computer--dependent;
                the same binning of the available experimental
                data suffers from a lack of statistics ---
                this is clearly demonstrated by the empty
                experimental bins in the discussed pictures.

                The simulated data were subject to extrapolation
                to the point
        $\tau = \mu^2 $.
                The true limiting value
                for each amplitude is being calculated with
                the use of eq.
(\ref{FChLOPEw}).
                The linear
                ({\it lin})
                and the quadratic
                ({\it squ})
                extrapolation patterns are selected
                for demonstrations.
                The results of extrapolations are collected
                in the Table
\ref{P460}
                where the fulcrum numbers of the true limiting
                values specific
                for the considered amplitudes
                are given in the bottom boxes.
                We display here the variation of the
                extrapolated values with the choice of the
                left bound
        $\tau_{1}$,
                the right bound
        $\tau_{2}$
                being kept fixed.


\begin{table}
\begin{center}
\fbox{
        $ P_{\rm Lab} = 335 $~MeV/c,
        $ s_{\pi\pi} = 4.15 \mu^2 $
        }
\begin{tabular}{|c|c|c||c|c||c|c||c|c|}
\hline
$n$&$\tau_{1}/\mu^{2}$&$\tau_{2}/\mu^{2}$
             &$o_{lin}$&$o_{squ}$
                       &$g_{lin}$&$g_{squ}$
                                 &$x_{lin}$&$x_{squ}$
\\
\hline
 17& -5.519 & -1.155  &  0.6111 & 0.7889  &1.201 & 0.866    &   0.1612 & 0.3578 \\
 16& -5.262 & -1.155  &  0.6204 & 0.7883  &1.176 & 0.909    &   0.1712 & 0.3582 \\
 14& -4.749 & -1.155  &  0.6382 & 0.7871  &1.138 & 0.987    &   0.1898 & 0.3669 \\
 12& -4.235 & -1.155  &  0.6546 & 0.7892  &1.112 & 1.067    &   0.2087 & 0.3723 \\
 10& -3.722 & -1.155  &  0.6709 & 0.7875  &1.100 & 1.140    &   0.2286 & 0.3697 \\
  8& -3.208 & -1.155  &  0.6858 & 0.7864  &1.099 & 1.223    &   0.2479 & 0.3589 \\
  6& -2.695 & -1.155  &  0.6998 & 0.7906  &1.111 & 1.341    &   0.2607 & 0.3953 \\
\hline
   &        &         &
                   \fbox{0.7870}&\fbox{0.7870}&
                                    \fbox{0.6036}&\fbox{0.6036}&
                                              \fbox{0.0000}&\fbox{0.0000}
\\
\hline
\end{tabular}
\vskip 0.5cm
\fbox{
        $ P_{\rm Lab} = 420 $~MeV/c,
        $ s_{\pi\pi} = 4.15 \mu^2 $
        }
\begin{tabular}{|c|c|c||c|c||c|c||c|c|}
\hline
$n$&$\tau_{1}/\mu^{2}$&$\tau_{2}/\mu^{2}$
             &$o_{lin}$&$o_{squ}$
                       &$g_{lin}$&$g_{squ}$
                                 &$x_{lin}$&$x_{squ}$
\\
\hline
 17&   -10.01   & -1.283 &   0.4497 & 0.7097  & 0.4011 & 0.3829  & 0.4971 & 0.5755 \\
 16&   -9.497   & -1.283 &   0.4672 & 0.7090  & 0.3848 & 0.4435  & 0.4971 & 0.5965 \\
 14&   -8.470   & -1.283 &   0.4999 & 0.7098  & 0.3715 & 0.5568  & 0.5051 & 0.6259 \\
 12&   -7.444   & -1.283 &   0.5315 & 0.7055  & 0.3838 & 0.6502  & 0.5208 & 0.6395 \\
 10&   -6.417   & -1.283 &   0.5598 & 0.7028  & 0.4154 & 0.7361  & 0.5384 & 0.6496 \\
  8&   -5.390   & -1.283 &   0.5845 & 0.7070  & 0.4632 & 0.8261  & 0.5561 & 0.6688 \\
  6&   -4.364   & -1.283 &   0.6089 & 0.6990  & 0.5321 & 0.8703  & 0.5782 & 0.6715 \\
\hline
   &            &        &
                       \fbox{0.7054}&\fbox{0.7054}&
                                          \fbox{0.4751}&\fbox{0.4751}&
                                                 \fbox{0.0000}&\fbox{0.0000}
\\
\hline
\end{tabular}
\vskip 0.5cm
\fbox{
        $ P_{\rm Lab} = 460 $~MeV/c,
        $ s_{\pi\pi} = 4.45 \mu^2 $
        }
\begin{tabular}{|c|c|c||c|c||c|c||c|c|}
\hline
$n$&$\tau_{1}/\mu^{2}$&$\tau_{2}/\mu^{2}$
             &$o_{lin}$&$o_{squ}$
                       &$g_{lin}$&$g_{squ}$
                                 &$x_{lin}$&$x_{squ}$
\\
\hline
 21 &  -12.58 & -1.797 &  0.1849 & 0.4911  &-0.0071 & 0.908  & 0.2755 & 1.485 \\
 20 &  -12.06 & -1.797 &  0.2015 & 0.4916  & 0.0174 & 1.006  & 0.3477 & 1.462 \\
 18 &  -11.04 & -1.797 &  0.2337 & 0.4917  & 0.8569 & 1.200  & 0.4837 & 1.405 \\
 16 &  -10.01 & -1.797 &  0.2643 & 0.4917  & 0.1830 & 1.378  & 0.6084 & 1.325 \\
 14 &  -8.984 & -1.797 &  0.2928 & 0.4933  & 0.3074 & 1.535  & 0.7152 & 1.230 \\
 12 &  -7.957 & -1.797 &  0.3198 & 0.4951  & 0.4568 & 1.658  & 0.7997 & 1.131 \\
 10 &  -6.930 & -1.797 &  0.3459 & 0.4917  & 0.6242 & 1.737  & 0.8584 & 1.042 \\
  8 &  -5.904 & -1.797 &  0.3695 & 0.4895  & 0.8043 & 1.726  & 0.8938 & 0.969 \\
  6 &  -4.877 & -1.797 &  0.3906 & 0.4939  & 0.9711 & 1.653  & 0.9096 & 0.922 \\
\hline
    &         &        &
                    \fbox{0.4886}&\fbox{0.4886}&
                                      \fbox{0.3591}&\fbox{0.3591}&
                                            \fbox{0.0000}&\fbox{0.0000}
\\
\hline
\end{tabular}
\end{center}
\caption[]{
                Results
                {\it lin (squ)}
                of the linear (quadratic)
                Chew--Low extrapolation
                to the point
        $\tau = \mu^2 $
                for the varying left bound
        $\tau_{1}$.
                The theoretical amplitudes
                used for data simulations correspond to:
        $ o $
                --- the solution with OPE contribution only;
        $ g $
                --- the solution with all mechanisms;
        $ x $
                --- the solution with all mechanisms
                    excluding OPE.
                The numbers given in the bottom boxes
                show the true values at
        $\tau = \mu^2 $.
        }
\label{P460}
\end{table}

                The
        $ o $
                columns of the Table
\ref{P460}
                are the undoubted grounds for the
                crucial inference
                that
                even in the simplified case of the pure OPE
                mechanism
                the linear extrapolation method
                generally underestimates the value
                in question
                and
                results in the inappropriate systematic error
                of 25--35\%.
                The impression of some improvement with the
                shift to the extreme right position of the
                extrapolation database in the
        $  \tau $
                interval
                is misleading since
                in the conditions of
                the reduced extrapolation base
                the effect of nonzero errors
                of the real experimental data
                must make the result even more ambiguous.
                We must also note
                that the OPE amplitude
                (i.e.,
        $ o $)
                does not fit at all the overall data
                (see Table
\ref{chilist1}
                for values of
        $ \chi^{2} $).
                Besides,
                it was several times pointed out
                that only the ``gauge--covariant'' set
                of contributions makes sense
                due to the field--redefinition freedom.

                In the more realistic case
        $ g $
                when all mechanisms are being present
                (the amplitude
        $ g $
                of the solution with
        $ \chi^{2}_{\rm DF} \approx 1.16 $
                is at least compatible
                with the overall database)
                there are no advantages in both the linear
                and the quadratic extrapolation methods;
                the coincidence of the results with the exact
                answers seems to be of a random nature.
                The 200--300\% deviation makes it unreliable
                to use both the linear and
                the quadratic extrapolations
                even for estimations.

                What is really disappointing it is the
                examination of the columns
        $ x $.
                The nontrivial answers in this case raise
                suspicions that the extrapolations follow
                the dictate of the experimental data
                rather than the theoretical amplitude.
                Indeed, the theoretical amplitude
        $ x $
                fits well the data
                but
                {\it it has no pole at}
        $\tau = \mu^2 $
                {\it at all!}

\subsection{ Discussion }

                In view of the negative general conclusion
                on the applicability of the Chew--Low
                extrapolation approach
                at the considered energies
                we need to shed more light on its origin.

                The separate clarification is necessary
                for the case of the pure OPE mechanism
                since the application of the
                Chew--Low approach
                is based on the hypothesis
                of the OPE dominance.
                Indeed,
                if the nature follows the simplest pattern
                of the OPE dominance in the
        $ \pi 2 \pi $
                reaction
                then
                6--8 data points are enough for successful
                extrapolation ---
                the small discrepancy displayed in the Table
\ref{P460}
                (which must be growing with the energy)
                is due to different accounting of isospin
                breaking in the main program
                and in the extrapolation function
(\ref{FChLOPEw}).

                It is not so difficult to realize
                that the small departure
                of the extrapolation function
(\ref{FChLOPE})
                from the linear pattern
                is solely due to the participation of the
        $D$--wave
                parameters
        $ g_{2} $
                and
        $ g_{3} $
                --- see the quantities
        $ \Phi_{22} $,
        $ \Phi_{23} $,
        $ \Phi_{33} $
                given by eqs.
(\ref{Phi0033}).
                In the absence of the latter
                the internal integrations
                of the leading order
                ChPT amplitude
                in the formula
(\ref{FChLOPE})
                result in the linear function of
        $ \tau $.
                At the considered energies the influence
                of the quoted parameters on the
        $ \pi \pi $
                amplitude
                itself is negligible.
                The deviation of the extrapolation function
                from the linear shape is small but its effect
                on the results of extrapolations
                is found to be drastic.

                The off--shell appearance of the
        $ \pi \pi $
                amplitude
                in the
        $ \pi 2 \pi $
                reaction
                acts the part of the magnification lens
                in respect to
        $D$--wave
                parameters
                --- we have already seen this in the
                previous section when discussing
                their contributions
                to the threshold amplitudes.
                However,
                the above phenomenon does not present
                an obstacle
                by itself
                since the quadratic extrapolation for a pure
                OPE amplitude is proved to be exact and stable.

                It is the complicated form of the
        $ \pi 2 \pi $
                amplitude
                revealed by our data fittings
                which rules out the possibility of a reliable
                application of the Chew--Low extrapolation
                in the simplest manner.
                This conclusion is derived in terms
                of a particular ansatz of the extrapolation
                function
(\ref{FChL}).
                Let us now discuss
                why the conclusion is of more general nature.

                There is the difference of our function
        $ { F }_{M}(\tau) $
                with the ones defined in terms
                of cross sections
                --- we are extracting the square root
                of the eq.
(\ref{FChL}).
                The results of the extrapolation
                of the square of the function
        $ { F }_{M}(\tau) $
                in terms of the ansatz
\ba
\label{FChLansatz}
  \left[{ F }_{M}(\tau)\right]^2
      & = &
  \left[
        F_{0}  +
        F_{1} ( \tau - \mu^2 ) +
        F_{2} ( \tau - \mu^2 )^{2}
              \right]^2
        \;
\ea
                are completely equivalent to the ones
                displayed in the Table
\ref{P460}.
                At the same time the extrapolation via
\ba
\label{FChLansatz2}
  \left[{ F }_{M}(\tau)\right]^2
      & = &
        \tilde{F}_{0}  +
        \tilde{F}_{1} ( \tau - \mu^2 ) +
        \tilde{F}_{2} ( \tau - \mu^2 )^{2}
        \;
\ea
                provides worse results.

                Hence,
                we see no reason in keeping on
                the cross--section form.
                It is not the point
                which is capable to disapprove our conclusions.
                So let us discuss another feature
                which is implemented
                into our extrapolation function
(\ref{FChL}).

                There is the property
                of the pure OPE cross section
        $ \sigma_{|\tau \to 0} = 0 $
                which was displayed
                long time ago
                as by the nonrelativistic calculations
                (see, for example
                the textbook by K\"{a}llen
\cite{Kallen64})
                as well as by the relativistic ones
                (like that of the paper
\cite{Naisse64}
                by Naisse and Reignier).
                The work
\cite{Baton67}
                by Baton, Laurens and Reignier
                provided the phenomenological test
                of this property by the high energy
        ($ P_{\rm Lab} = 2.77 $~GeV/c)
                data
                --- since then it was being built into
                the applications of the Chew--Low procedure
                as a standard feature.
                Nevertheless,
                the known failures of applications of the
                Chew--Low approach were associated with
                the relying on the very property we are
                discussing here --- see the review
\cite{Leksin70}
                by Leksin.
                The more close analysis shows
                that in some cases
                the foothold on the property
        $ \sigma_{|\tau \to 0} = 0 $
                in the definition
(\ref{FChL})
                is the reason of overshooting
                of the quadratic extrapolations
                which are accurately following
                the extrapolated data in the physical region.

                Let us now briefly remind
                what is the theoretical status
                of the hypothesis that
        $ \sigma_{|\tau \to 0} = 0 $.
                Definitely,
                it is the exact property of the pure OPE
                mechanism.
                What the value of the matrix element
        $ \langle \pi_2 \pi_3 N(q) | S | \pi_1 N(p) \rangle $
                at
        $ p = q $
                is in the general case
                is the kinematical problem in part.
                (It should be noted
                that the point for which
        $ p = q $
                is located outside the physical region
                both for the
        $ \pi N \to \pi \pi N $
                reaction
                and for the
        $ 4 \pi $
                vertex
                since the condition
        $ p = q $
                implies
        $ s_{\pi\pi} = \mu^2 $.)

                The
        $ S $
                structure
                of the amplitude
(\ref{ChAmpform})
                gets the same multiplier
        $ (-\tau) $
                in the unpolarized matrix element
(\ref{SqAmp})
                as the OPE contribution does.
                This structure gives rise
                to spin--flip amplitudes which are
                the only amplitudes of the considered reaction
                surviving at the threshold.
                However,
                besides this spinor structure
                there are three more;
                kinematically,
                their contributions to the quantity
(\ref{SqAmp})
                are determined by the matrix
(\ref{Gmat}).

                Thus,
                to make the matrix element
(\ref{SqAmp})
                vanish at
        $ p = q $
                all entries of the matrix
(\ref{Gmat})
                must become zero simultaneously.
                The analysis of the explicit expressions
                (for which we have no room here)
                shows that three conditions are necessary:
                1)
        $
          s = ( m + \sqrt{s_{\pi\pi}} )^{2}
        $;
                2)
                collinear final pions
        $ k_{2} = k_{3} $;
                3)
                Chiral limit
        $ \mu = 0 $.
                The last condition is also
                the only general reason
                to make the considered structures vanish
                dynamically.

                Therefore,
                in the real dynamics
                the quantity
        $ \| M \|^{2}_{\tau \to 0} $
        ($ \sigma ( \tau = 0 ) $)
                stands for expressing
                the Chiral symmetry breaking
                which is similar to the
        $ \pi N $--elastic
        $ \Sigma $
                term.
                It depends on the energy but seems to be
                rather small.
                Nevertheless,
                it prevents to make the safe
                {\it simplification}
                in the definition of the extrapolation
                function,
                namely, to divide the quasi--amplitude
                by
        $ \sqrt{ -\tau} $
                in our case.
                Thus,
                we arrive at the conclusion
                that,
                from one side,
                the complicated dependence of the
                physical amplitude on
        $ \tau $
                makes useless the linear and the quadratic
                extrapolation methods even in terms of
                the quasi--amplitude,
                from the other side ---
                the presence of the Chiral symmetry breaking
                in the true amplitude
                forbids to
                soften
                the dependence in the extrapolation ansatz.


\subsection{    Remark on
        $D$--Wave
                Parameters }

                The parameters
        $ g_{2} $,
        $ g_{3} $
                of the cross--symmetric ansatz of the
        $ \pi \pi $
                amplitude
                (see
\cite{BolokhovVS88pi},
\cite{BolokhovVS91},
\cite{BolokhovEtAl96})
                determine the values of the
        $D$--wave
                scattering lengths,
                therefore,
                we call them the
        $D$--wave
                parameters here.
                Because of the crossing symmetry the same
        $D$--wave
                parameters
                determine also the slopes of the
        $S$--wave
                amplitudes
                (see eqs. (59), (60) of the ref.
\cite{BolokhovVS91}
                and eqs. (73)--(104) of the ref.
\cite{BolokhovEtAl96}).

                Due to the considerable growth of the
        $ I = 0 $
                amplitude
                the contribution of the slopes
                to the integrated over
        $ s_{\pi\pi} $
                cross section
                (extrapolated by the Chew--Low method)
                appears to be large.

                This very phenomenon makes it possible
                to extract the discussed parameters
                from the near--threshold
        $ \pi 2 \pi $
                experiment,
                otherwise
                it is simple to verify
                that
                at the considered energies
                the contribution
                to the
        $  \pi \pi $
                and
        $ \pi 2 \pi $
                cross sections
                by the
        $D$
                waves
                themselves
                is negligible.
                This also presents the additional motivation
                for the careful handling
                of the crossing properties of the amplitudes of
        $  \pi \pi $
                and
        $ \pi N \to \pi \pi N $
                reactions.


\vspace{0.5cm}

\section{ Conclusions }

                Throughout the paper
                we were making the inferable statements
                along the discussions.
                Here,
                we remind the most important ones and develop
                the general conclusions.

\subsection{
             Data
           }

                Our present work is devoted to the analysis
                of the near--threshold data on the
        $ \pi 2 \pi $
                reaction.
                The data base described in sect. 3. consists
                of the experimental total cross sections and
                1--dimensional distributions.
                The full--kinematics data
                of the work
\cite{Bloh_-+n63}
                also had been presented in the same form.
                This needs some comments.

                The available 1023 full--kinematics events
                of the quoted work constitute the solid
                ground for the total cross section;
                10--14 bins of
                a 1--dimensional distribution
                have a good filling
                with the averaged number of
                60--100 events per bin;
                the filling of
        $ 8 \times 8 \ $
                bins of a 2--dimensional distribution is
                satisfactory (15 events per bin in the average)
                while the filling of
        $ 6\times 6\times 6 \ $
                of the 3--dimensional ones and
        $ 4\times 4\times 4\times 4 \ $
                of the 4--dimensional bins is poor.
                In this conditions
                of the difficult choice
                between the poor filling of bins
                and the loss of the kinematical information
                we formed multiple
                1--dimensional
                projections for the data
                to bring to light
                the behavior in the crucial variables.

                We created some additional 1D projections
                of the data
\cite{Bloh_-+n63}
                and tested them with the obtained solutions
                --- the description was found to be excellent.

                The use of the numerous lower dimensional
                distributions
                acts the part of a kind
                of the tomography method.
                The fittings showed that at a distance from
                resonance poles this works.
                The improved statistics of the contemporary
                experiment
\cite{SmithK94,SmithEtAl95,KermaniChD94,SeviorChD94}
                definitely must make
                the direct use of the full--kinematics data
                more preferable.

                The general properties
                of the treated data base are found to be:

                1. The precision of data
                is insufficient to improve
                the accuracy of the determination of
                characteristics of the
        $ \pi \pi $
                scattering.
                However,
                this is the problem not only of the data.

                2. The coverage of the behavior of the low
                energy amplitude is good;
                practically,
                there were no losses of convergence in the
                course of fittings.
                The numerous solutions reflect the complexity
                of the reaction amplitude
                --- the current experimental setup can not be
                charged for this
                (we shall continue this discussion below
                in the subsects. 7.2., 7.3.).

                3. The distribution data of the
        $\{ + + n \}$
                channel
                are found to be extremely important
                for the resolution of parameter correlations.
                In the absence of these data the convergence
                of fittings becomes tremendously slow,
                the number of iterations being increased by
                several orders.


                To resume
                we state that the considered data base
                is in principle sufficient for determination
                of the phenomenological near--threshold
        $ \pi 2 \pi  $
                amplitude.
                Certainly,
                the need in the better quality of data
                in respect to the precision of
                the obtained parameters
                and the multiplicity of solutions
                is obvious.
                First,
                the quality of the newest data will be much
                better and,
                second,
                it seems oversimplified to charge
                only the data with
                this problem.
                The more detailed discussion of its origin
                will be given in the next subsection.

\subsection{
                Amplitude and Major Results
           }

                Our amplitude is built
                on the rather conservative basis.
                However,
                the orientation towards the modern ChPT
                approach in constructing the model
                might be lacking of an intrinsic tool for
                making the right explanation,
                if failed,
                whether an inconsistency of data
                or the neglect of higher order terms
                are responsible for the inappropriate fit.
                If successful,
                the approach provides the only conclusion
                that at the considered order
                ChPT is compatible with data ---
                a substantial estimate of the systematic
                error of determination of the low energy
                constants is impossible.

                The approach of HBChPT deserves the separate
                remark.
                Unlike the case of the
        $  \pi N $--elastic
                scattering
                this approach is,
                probably,
                inapplicable to the case of the
        $  \pi N \to \pi \pi N $
                reaction
                considered here.
                One should remind the
                basic keystones of
                HBChPT:
                1. Nonrelativistic limit;
                2. Small--pion--momentum expansion;
                3. Heavy baryon approximation.
                Then, let us consider the identity
\begin{equation}
\label{g5id}
                \bar{u}(q)
                        (
                                \hat{k}_{1} -
                                \hat{k}_{2} -
                                \hat{k}_{3}
                        )
                        i \gamma_5
                {u}(p)
        =
                - 2 m
                \,
                \bar{u}(q)
                        i \gamma_5
                {u}(p)
          \; ,
\end{equation}
                which is specific to the relativistic form
                of the
        $ \pi 2 \pi $
                amplitude.
                The identity can not support the above points
                2., 3. simultaneously.

                In the present paper we consider
                the amplitude of the
        $ \pi 2 \pi $
                reaction
                built of numerous resonance contributions
                (including the separately treated
                OPE mechanism)
                and the smooth polynomial background
                (see sect. 2.).
                Its complicated appearance reflects
                the influence of the
                (generally, off--shell)
                processes like
        $  \pi \pi \to \pi \pi $,
        $  \pi N \to \pi N $,
        $  \pi N \to \pi N_{*} $,
        $  \pi N \to \pi \Delta $
                on the near--threshold region
                of the discussed reaction.

                The fittings strongly confirmed
                the importance of all quoted
                exchange mechanisms.
                This result can not be considered as
                the totally new one.
                For example,
                the importance of
                isobars
                already had been stressed
                in the paper
\cite{JohnsonFS93}
                in terms of
                the sophisticated analysis
                of the amplitude form.

                The near--threshold region
        $ 280 \le P_{\rm Lab} \le 500 $~MeV/c
                can not be considered as the selfcontained one
                because of
                large
        $ \Delta $,
        $ N_{*} $
                widths.
                The isobars
                extend their strong influence up
                to the very
        $ \pi 2 \pi $
                threshold
                (this is demonstrated by the Table
\ref{D1D2tab}).
                In the absence of isobar contributions
                the considerable improvement of the fit due to
                the imaginary background serves as an indirect
                evidence of the importance
                of the discussed isobar mechanisms
                (see Table
\ref{chilist1}).
                The interrelation of these mechanisms
                with the OPE one will be discussed below.

                In view of the discussion of the role of
                background parameters
                (see subsect. 2.4.)
                it is not so surprising
                that
                only few of them
                are found important
                in the fittings
                when all exchange mechanisms
                are being present.
                The fact
                that the parameters of the isospin amplitude
        $ D $
                are found to be consistent with zero
                reflects the negligible influence of higher
        $ \tau $--resonances
                (like
        $ {\rm SE}_{N N}(\omega) $)
                on the amplitude
                in the considered energy region.
                Even the nonzero contributions to this
                isospin amplitude
                from isobar exchanges
                almost cancel each other ---
                this might be derived from an approximate
                equality of the resulting theoretical
                cross sections of
        $ \{ - 0 p \} $
                and
        $ \{ + 0 p \} $
                channels.

                The OPE contribution is in the center of our
                investigations.
                In all variants the improvement of
        $ \chi^{2}_{\rm DF} $
                with the inclusion of OPE is found to be
                statistically important
                (see Tables
\ref{chilist1},
\ref{chiDFlist}).
                The
        $ 4 \pi $
                vertex
                of the OPE graph
                is taken in the direct amplitude form which
                contains 4 parameters
        $ g_{0} $,
        $ g_{1} $,
        $ g_{2} $
                and
        $ g_{3} $
                and respects the isospin, crossing
                and approximate--unitarity properties of
                the
        $ \pi \pi $
                amplitude off the mass shell.
                In terms of the
        $ \pi \pi $--scattering
                lengths
                the values of these parameters
                in the best physical solutions are:
\begin{center}
\begin{tabular}{|c||c|c|c|c|c|}
\hline
    $\chi^{2}_{\rm DF}$
      &$ a^{I=0}_0 $& $ a^{I=2}_0$   & $ a^{I=1}_1 $ & $ a^{I=0}_2 $   & $ a^{I=2}_2 $    \\
\hline
 1.161&0.07$\pm$0.12&-0.056$\pm$0.036&0.045$\pm$0.017&0.0052$\pm$0.0031&-0.0005$\pm$0.0014\\
 1.203&0.07$\pm$0.11&-0.076$\pm$0.073&0.047$\pm$0.025&0.0023$\pm$0.0027&-0.0013$\pm$0.0014\\
 1.205&0.17$\pm$0.12&-0.053$\pm$0.039&0.054$\pm$0.022&0.0054$\pm$0.0025&-0.0002$\pm$0.0012\\
 1.212&0.19$\pm$0.11&-0.059$\pm$0.045&0.053$\pm$0.022&0.0062$\pm$0.0027&~0.0006$\pm$0.0015\\
\hline
\cite{Dumbrais83}
      &
       0.26$\pm$0.05&-0.028$\pm$0.012&0.038$\pm$0.002&0.0017$\pm$0.0003&~0.0001$\pm$0.0003\\
\hline
\end{tabular}
\end{center}
%
                Here, we list also the currently adopted
                experimental values of the compilation
\cite{Dumbrais83}.

                Generally,
                the solutions display that the precision
                of determination of the
        $ D $--wave
                parameters
        $ g_{2} $
                and
        $ g_{3} $
                is not worse than
                that of
        $ g_{0} $,
        $ g_{1} $
                parameters.
                This was attributed to the characteristic
                energy dependence of the isospin--zero
        $ S $
                wave
                (see subsect. 6.4.).
                The parameters of the latter via crossing
                and kinematics
                are connected to
        $ D $--wave
                scattering lengths ---
                this is clearly demonstrated by the
        $ a^{I=0}_{2} $
                errors in the above list.

                The poor precision of determination of the
        $ \pi \pi $--interaction
                parameters
                has its origin in 3 principal reasons:
                1) the data accuracy;
                2) the competition of other
                        (nonOPE) mechanisms;
                3) the incomplete nature of the current
                        experimental setup.

                The first point is evident ---
                the data accuracy will be improved soon
                (look
\cite{Pocanic94}
                for the survey of experiments).

                The second one stems mainly from the isobar
                exchanges --- this is clearly revealed
                by fittings:
                whenever an isobar mechanism is absent
                it is easy to find a solution with
        $ a^{I=0}_{0} $
                in the range
        $ 0.20 $ --- $ 0.30 $;
                it drops to the
        $ 0.00 $ --- $ 0.20 $
                range if all isobar mechanisms are involved
                (it is interesting
                that the effect of
        $ \rho $
                exchanges is quite opposite).
                In other words,
                the larger values of
        $ \pi \pi $--scattering
                lengths
                are gained
                when OPE is forced to stay for an essential
                but missed isobar
                contribution.

                The important question is about the nature of
                parameters
                which correlations with the OPE set
        $ g_{0} $,
        $ g_{1} $,
        $ g_{2} $,
        $ g_{3} $
                are so devastating.
                Unfortunately,
                these are not the parameters of the DE--type
                graphs which might be estimated from the
                decay characteristics or might be known
                (like the
        $ g_{\pi NN} $
                constant)
                from the low--energy
        $ \pi N $
                phenomenology.
                Instead,
                the root correlations are due to parameters
        $ R_{1} $--$ R_{4} $
                and
        $ D_{1} $--$ D_{4} $
                of the
        $ {\rm SE}_{\pi N} (N_{*}) $,
        $ {\rm SE}_{\pi N} (\Delta) $
                graphs
                (see Tables
\ref{FitParTab},
\ref{SE4v},
\ref{DE3v}).
                The explanation is simple.
                Parameters of the graphs of the SE type
                play the same role in respect
                to the ones of the DE type
                as the background parameters
     $ A_{1} $--$ A_{18} $
                (and their analogs
     $ i_{19} $--$ i_{36} $)
                do in respect to all exchange parameters.
                Indeed,
                outside the resonance region
                the contraction of any pole in the DE graph
                leads to the single--pole contribution,
                i.e. to the one described by the SE graphs.

                Here,
                one observes the interrelation of
                the point 2.
                with the first one
                since the data restricted to the region below
                resonances can not help much to fix up
                parameters of processes like
        $  \pi N  \to \pi N_{*} $,
        $  N_{*}  \to \pi \pi N $,
        $  \pi  N \to \pi \Delta $,
                etc.

                We found the numerous set of solutions
                describing the data at the acceptable level of
        $  \chi^{2} $.
                The origin of this phenomenon
                must be explained in part
                by the importance of all 4 spin structures
                of the
        $ \pi 2 \pi $
                amplitude
                (see eqs.
(\ref{IAmp}),
(\ref{ChAmpform})).
                This is the reason number 3 for the poor
                accuracy of our final results.

                Indeed,
                the unpolarized data measure only
                one combination of spin structures
                (namely, the combination given
                by the matrix element
(\ref{SqAmp}))
                leaving them almost free
                to stand one for another.
                Therefore,
                the absence of polarized measurements in the
                energy region
        $ P_{\rm Lab} \approx 500 $~MeV/c
                and the abundance of mechanisms
                specific to the considered reaction
                is the reason of huge correlations of
                OPE parameters with the rest ones
                on the available data base.

                Meanwhile,
                the extreme importance of the nucleon spin
                in the considered reaction
                at higher energies
                had been recently reported
                by Svec in the paper
\cite{Svec97ii}.
                In the results of our analysis and
                modeling the Chew--Low extrapolation function
                we see the clear signal
                of nontrivial spinor structures.
                This claims for the polarization measurements
                of
        $  \pi N  \to \pi \pi N $
                reactions
                at the discussed energies.
                Up to now the known polarization measurements
                of the
        $ \pi 2 \pi $
                reactions had been performed
                at considerably higher energies,
                for example,
                at 5.98 GeV/c and 11.85 GeV/c
\cite{Lesquenetal85}
                and at 17.2 GeV/c
\cite{Grayeretal74}.
                Their analyses
\cite{Svec92},
\cite{Becker79}
                already proved such measurements
                to be detailed sources of information on the
        $\pi \pi$
                interaction (at high energies).

                Certainly,
                the complete polarization experiment requires
                the analysis of the polarization
                of the final nucleon ---
                in the near future
                this is hardly to be carried out
                for such rare processes as
                the considered one.
                Nevertheless,
                the examination of the spinor structure
                of the considered amplitude
(\ref{IAmp})
                displays
                that the almost exhaustive information
                might be obtained already from the experiment
                with the polarized target.
                Indeed,
                there are two independent asymetries
\begin{equation}
\label{Asym}
             A_{\pm} ({\bf s})
        =
             \frac{
                    \sigma ( {\bf s} ;   {\bf k}_{\pm}^{\bot} )
                   -
                    \sigma ( {\bf s} ; - {\bf k}_{\pm}^{\bot} )
                  }
                  {
                    \sigma ( {\bf s} ;   {\bf k}_{\pm}^{\bot} )
                   +
                    \sigma ( {\bf s} ; - {\bf k}_{\pm}^{\bot} )
                  }
          \; ,
\end{equation}
                where
        $ {\bf s } $
                is the vector of the nucleon polarization
                and
        $ {\bf k}_{\pm}^{\bot} $
                are the projections of vectors
        $ {\bf k}_{\pm} = {\bf k}_{2} \pm {\bf k}_{3} $
                to the plane which is
                orthogonal to
        $ {\bf s } $.
                Their measurements must provide an information
                on two additional combinations of four
                spinor structures of the decomposition
(\ref{IAmp})
                which are independent from the combination of
                the matrix element
(\ref{SqAmp}).

%

\subsection{
                General Conclusion
           }

\vspace{0.5cm}

                In view of the results of modeling
                the Chew--Low extrapolation and
                the Olsson--Turner approach
                we conclude that only the approach
                based on the extensive phenomenological
                model
                ({\it a la}
                Vicente--Vacas)
                can help in investigations of the considered
                reaction.
                The investigations require to develop
                the common analyses of related processes like
        $  \pi N  \to \pi N_{*} $,
        $  \pi  N \to \pi \Delta $
                --- i.e. the processes described by the
                Lagrangian terms listed in Tables
\ref{SE4v},
\ref{DE3v}
                --- and,
                hence,
                to extend the energy region
                up to
        $ P_{\rm Lab} \sim 1 $~GeV/c.
                In other words,
                the problem of determination
                of the low--energy
        $ \pi \pi $--scattering
                characteristics
                is a part of the more comprehensive problem of
                investigation of
        $ \pi 2 \pi $
                dynamics
                at low and intermediate energies.

                The results of contemporary experiments
\cite{KermaniChD94,SeviorChD94,SmithK94}
                must provide much more precise determination
                of
        $\pi \pi$
                parameters since their correlations
                with the unknown parameters
                of isobar mechanisms originating
                from the discussed above spin structures
                will be resolved in part due to the improved
                data accuracy.
                In this respect the role of the polarized
                data is difficult to over--estimate.

                The amplitude which structure
                will be fixed by the analysis
                of statistically significant data
                might gain the wide range of applications
                beyond the testing of
        ChPT
                predictions.
                For example, it might be used for
                the common analysis of
        $ \pi N \to \pi \pi N $,
        $ \gamma N \to \pi \pi N $
                and
        $\pi N$--elastic
                data,
                for investigations of the
        $\eta$
                production in the process
        $\pi N \to \eta N$,
                for correcting experimental distributions
                obtained at devices with
                the restricted geometry
                and for other investigations at intermediate
                energies.
                This is ensured by the fundamental role of the
        $ \pi N \to \pi \pi N $
                reaction
                in nuclear and particle physics.

\newpage

\section{ Acknowledgments }

                This research was supported in part
                by the RFBR grant N 95-02-05574a.
                We thank
                        T.A.~Bo\-lo\-khov,
                        P.A.~Bolokhov,
                        V.A.~Guzey,
                        A.N.~Manashov,
                        V.V.~Vereshagin
                        V.L.~Yudichev,
                        A.Yu. Zakharov
                for checking the formulae,
                testing the code and for other help at
                various stages of the project.
                We are grateful to
                        P.~Amaudruz,
                        A.~Bernstein,
                        F.~Bonutti,
                        J.~Brack,
                        P.~Camerini,
                        G.A.~Feofilov,
                        E.~Frle\v{z},
                        N.~Grion,
                        G.~Hofman,
                        R.R.~Johnson,
                        M.~Kermani,
                        M.G.~Olsson,
                        O.O.~Patarakin,
                        D.~Po\v{c}ani\'{c},
                        R.~Rui,
                        M.~Sevior,
                        G.R.~Smith
                for helpful discussions.
                We especially
                thank the CHAOS team for presenting
                computer powers of ALPHA stations
                at TRIUMF (Vancouver) and INFN (Trieste).

%



%
\clearpage
\newpage
\thispagestyle{empty}

\begin{figure}[ht]
   \centerline{\epsffile{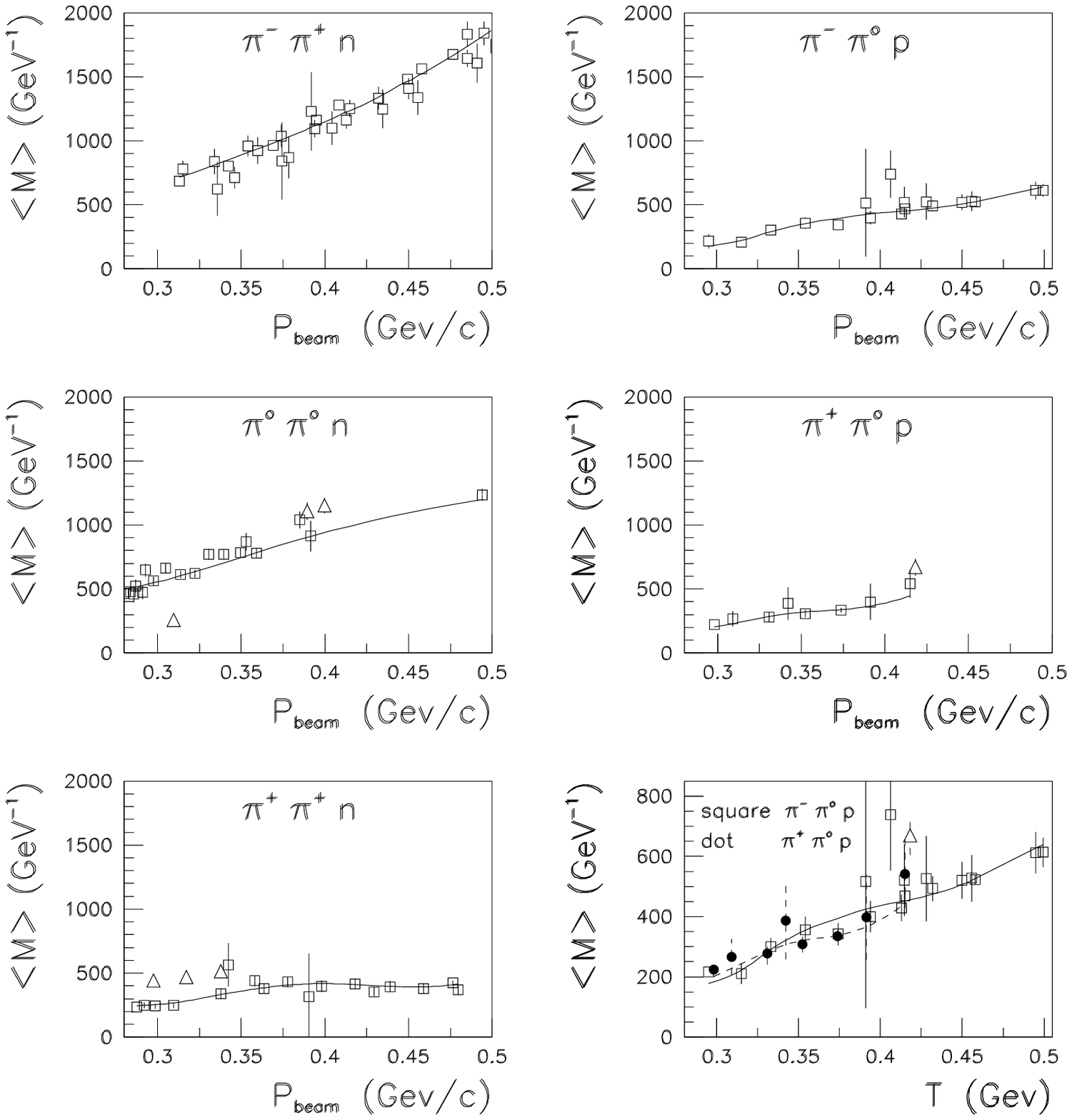}}
   \vspace{-2.5cm}
   \centerline{
               \parbox{13cm}{
                              \caption{
                                        \label{sigtot}
                Experimental points of total cross sections
                and the theoretical curve for
                the best physical solution
                (quasi--amplitude
        $ \langle M \rangle $
                in
        GeV${}^{-1}$).
                Triangle points were excluded from fittings.
                                      }
                             }
              }
\end{figure}

\clearpage
\thispagestyle{empty}

\vspace{1.5cm}

\begin{figure}[ht]
   \vspace{0.5cm}
   \centerline{\epsffile{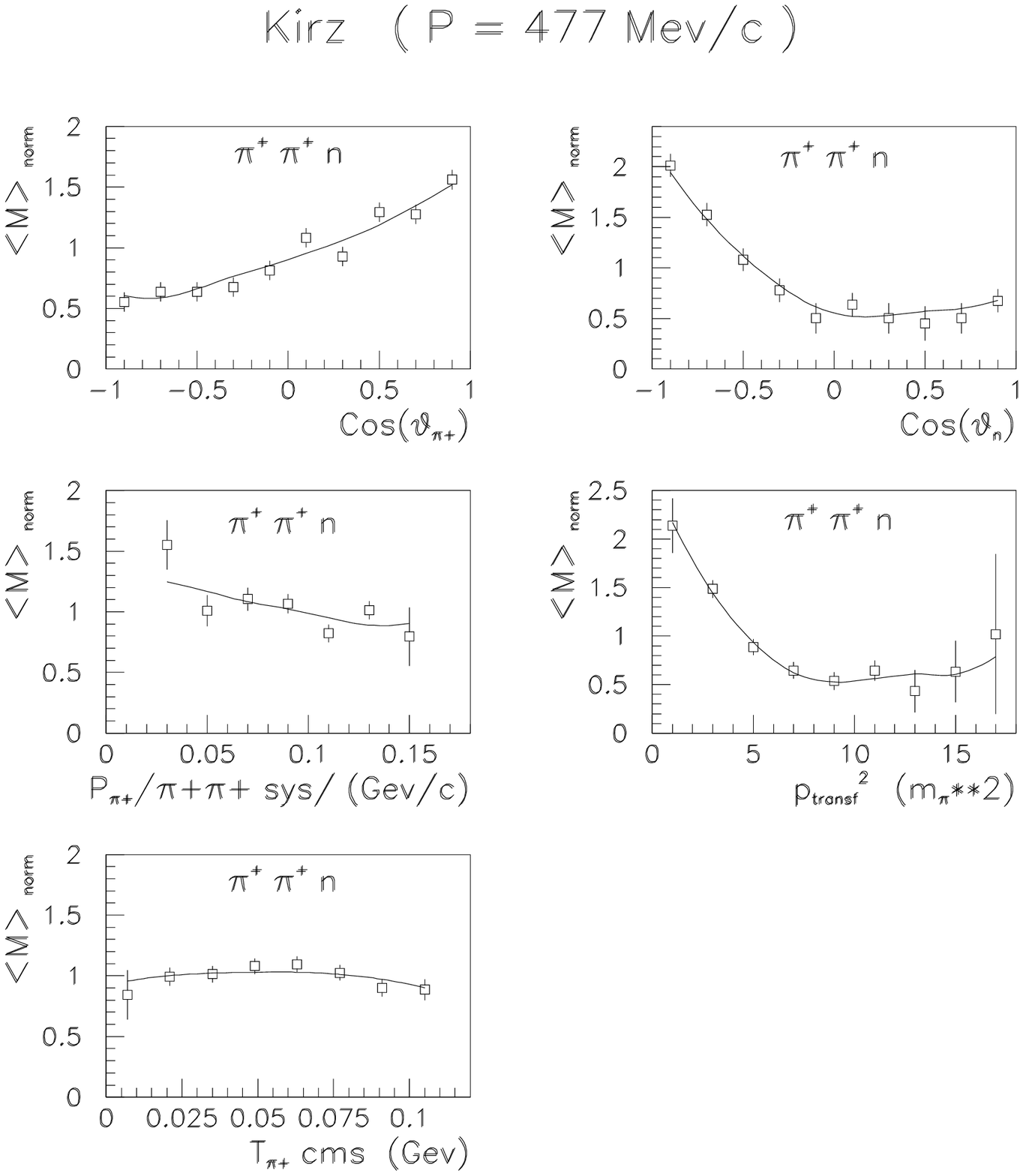}}
   \vspace{1cm}
   \centerline{
               \parbox{13cm}{
                              \caption{
                                        \label{krzppn}
                Experimental distributions for the
        $ \{ + + n \} $
                channel from the paper
\cite{Kirz_++n62}
                by Kirz and theoretical curves
                (normalized quasi--amplitude
        $ \langle M \rangle_{\rm norm} $).
                                      }
                             }
              }
\end{figure}

\clearpage
\thispagestyle{empty}

\vspace{1.5cm}

\begin{figure}[ht]
   \vspace{0.5cm}
   \epsfxsize=17cm
   \epsfysize=17cm
   \centerline{\epsffile{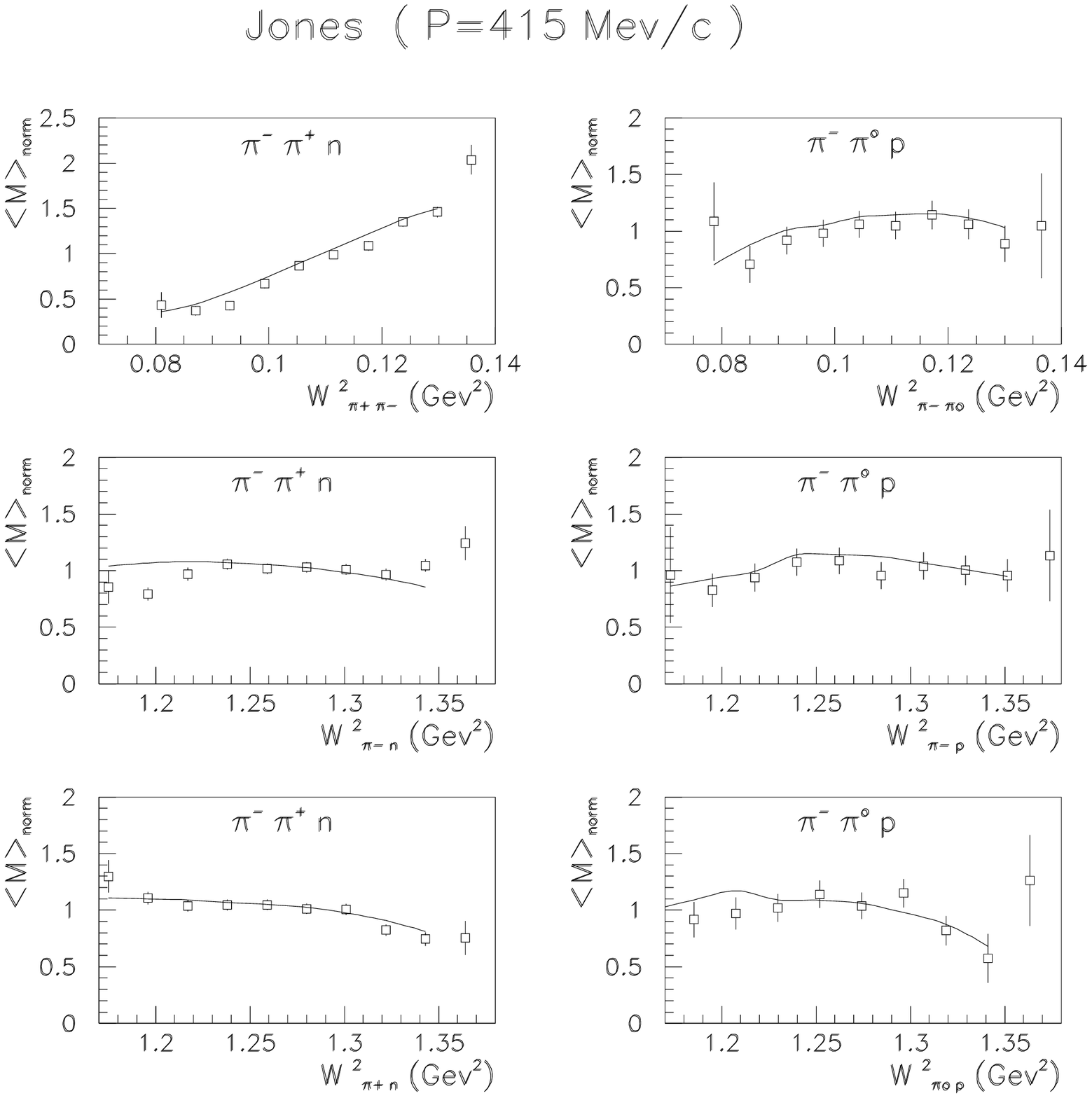}}
   \vspace{1cm}
   \centerline{
               \parbox{13cm}{
                              \caption{
                                        \label{jon415}
                Experimental distributions for the
        $ \{ - + n \} $
                and
        $ \{ - + n \} $
                channels from the paper
\cite{Jones-+n-0p74}
                by Jones and theoretical curves
                (normalized quasi--amplitude
        $ \langle M \rangle_{\rm norm} $).
                                      }
                             }
              }
\end{figure}

\clearpage
\thispagestyle{empty}

\vspace{1.5cm}

\begin{figure}[ht]
   \vspace{0.5cm}
   \epsfxsize=17cm
   \epsfysize=17cm
   \centerline{\epsffile{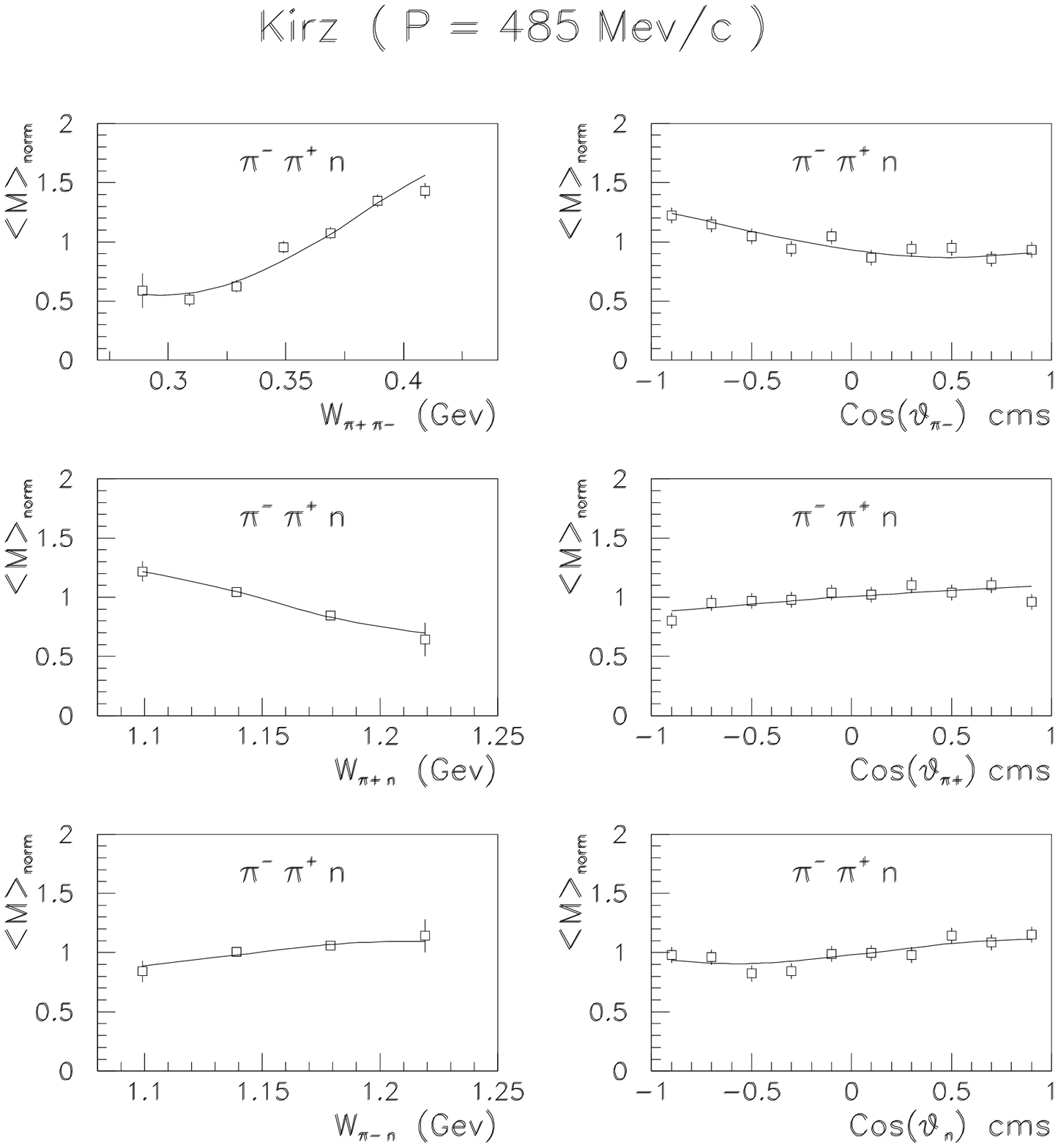}}
   \vspace{1cm}
   \centerline{
               \parbox{13cm}{
                              \caption{
                                        \label{krzmpn}
                Experimental distributions for the
        $ \{ - + n \} $
                channel from the paper
\cite{Kirz_-+n63}
                by Kirz and theoretical curves
                (normalized quasi--amplitude
        $ \langle M \rangle_{\rm norm} $).
                                      }
                             }
              }
\end{figure}

\clearpage
\thispagestyle{empty}

\vspace{1.5cm}

\begin{figure}[ht]
   \vspace{0.5cm}
   \epsfxsize=17cm
   \epsfysize=17cm
   \centerline{\epsffile{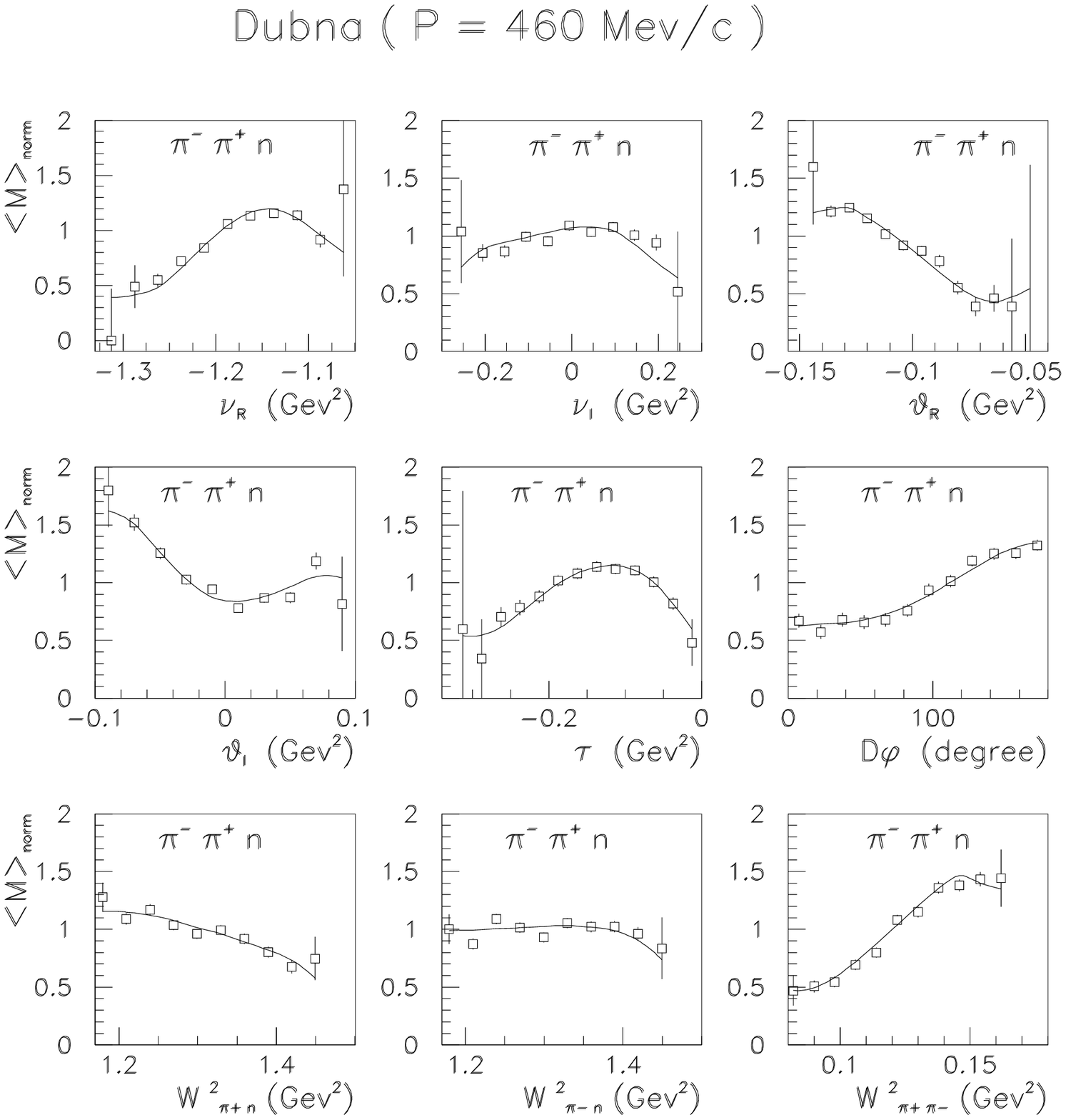}}
   \vspace{1cm}
   \centerline{
               \parbox{13cm}{
                              \caption{
                                        \label{dubinv}
                Some of experimental distributions for the
        $ \{ - + n \} $
                channel from the paper
\cite{Bloh_-+n70}
                by Blokhintseva and theoretical curves
                (normalized quasi--amplitude
        $ \langle M \rangle_{\rm norm} $).
                                      }
                             }
              }
\end{figure}

\clearpage
\thispagestyle{empty}

\vspace{1.5cm}

\begin{figure}[ht]
   \vspace{0.5cm}
   \centerline{\epsffile{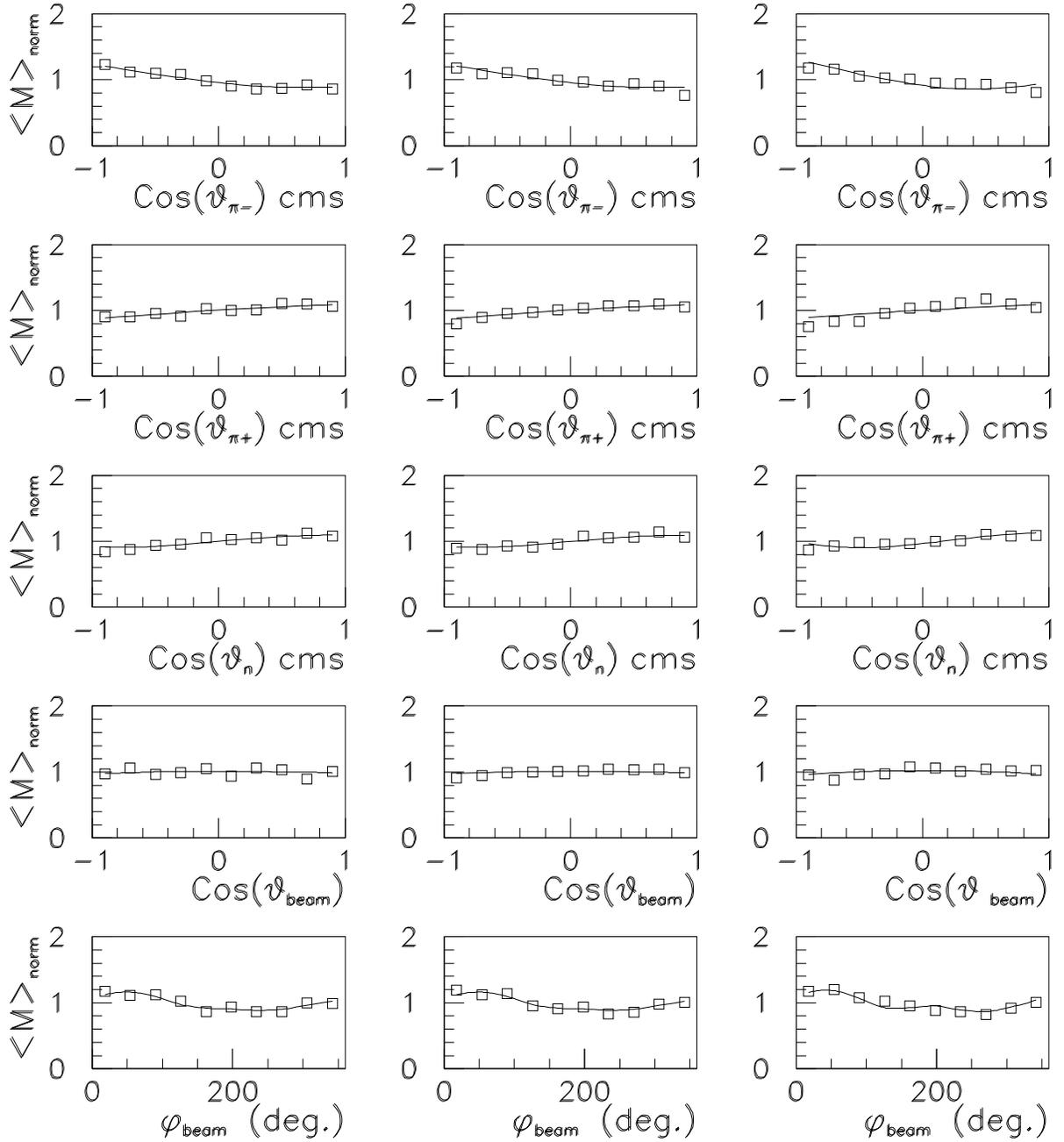}}
   \vspace{1cm}
   \centerline{
               \parbox{13cm}{
                              \caption{
                                        \label{dubang}
                Experimental distributions
                and theoretical curves
                for angular spectra
                of the
        $ \{ - + n \} $
                channel from the Saxon's paper
\cite{Saxon-+n70}
                and the same spectra build of
                the Blokhintseva data
\cite{Bloh_-+n70}
                (normalized quasi--amplitude
        $ \langle M \rangle_{\rm norm} $).
                The Saxon data had not been used
                in fittings.
                                      }
                             }
              }
\end{figure}

\clearpage
\thispagestyle{empty}

\vspace{1.5cm}

\begin{figure}[ht]
   \vspace{0.5cm}
   \centerline{\epsffile{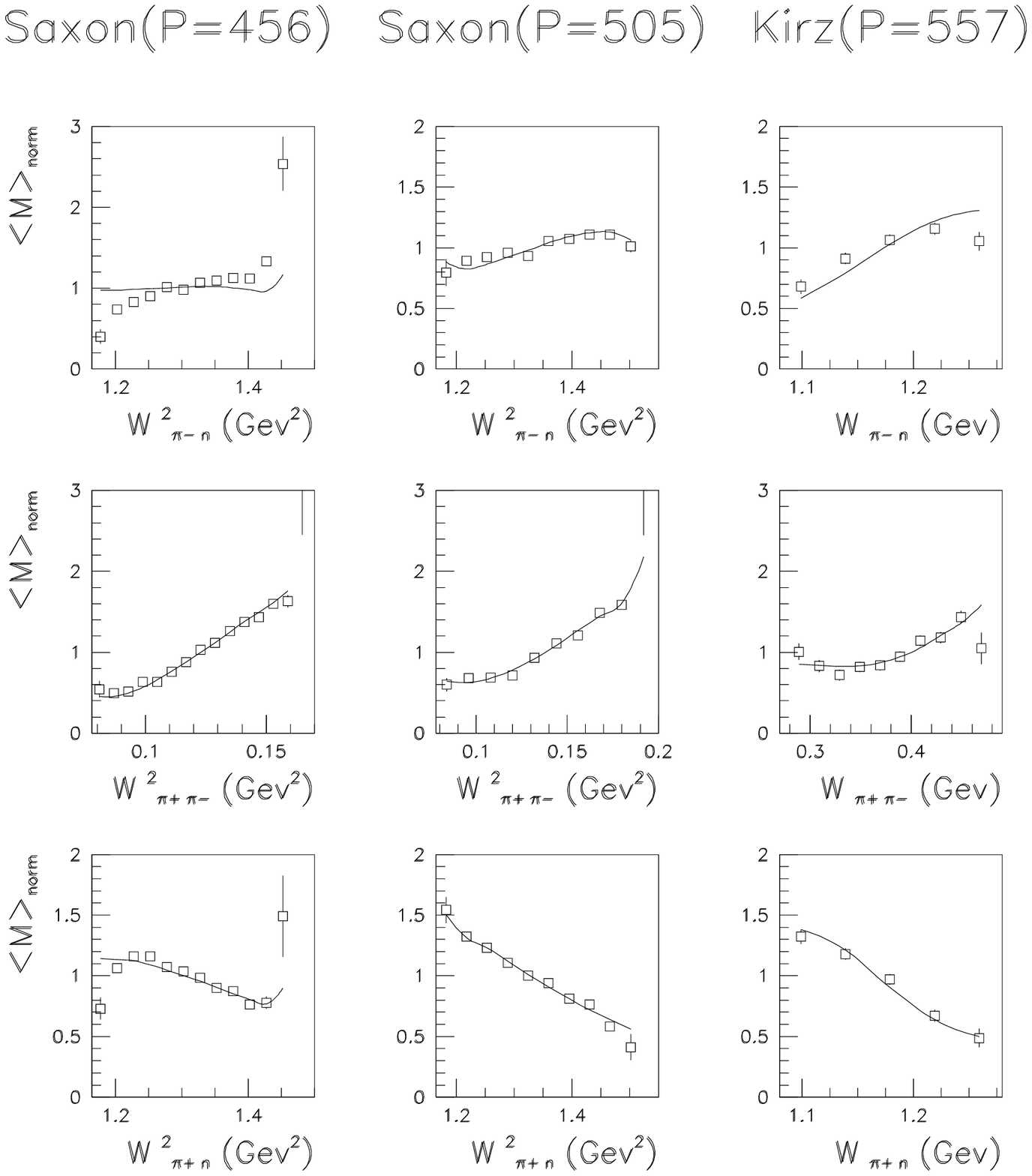}}
   \vspace{1cm}
   \centerline{
               \parbox{13cm}{
                              \caption{
                                        \label{masspred}
                Experimental distributions
                and theoretical predictions
                for spectra of the
        $ \{ - + n \} $
                channel from the paper
\cite{Saxon-+n70}
                by Saxon and from the paper
\cite{Kirz_-+n63}
                by Kirz
                (normalized quasi--amplitude
        $ \langle M \rangle_{\rm norm} $).
                                      }
                             }
              }
\end{figure}

\clearpage
\newpage
\thispagestyle{empty}

\vspace{1.5cm}
\begin{figure}[ht]
   \centerline{\epsffile{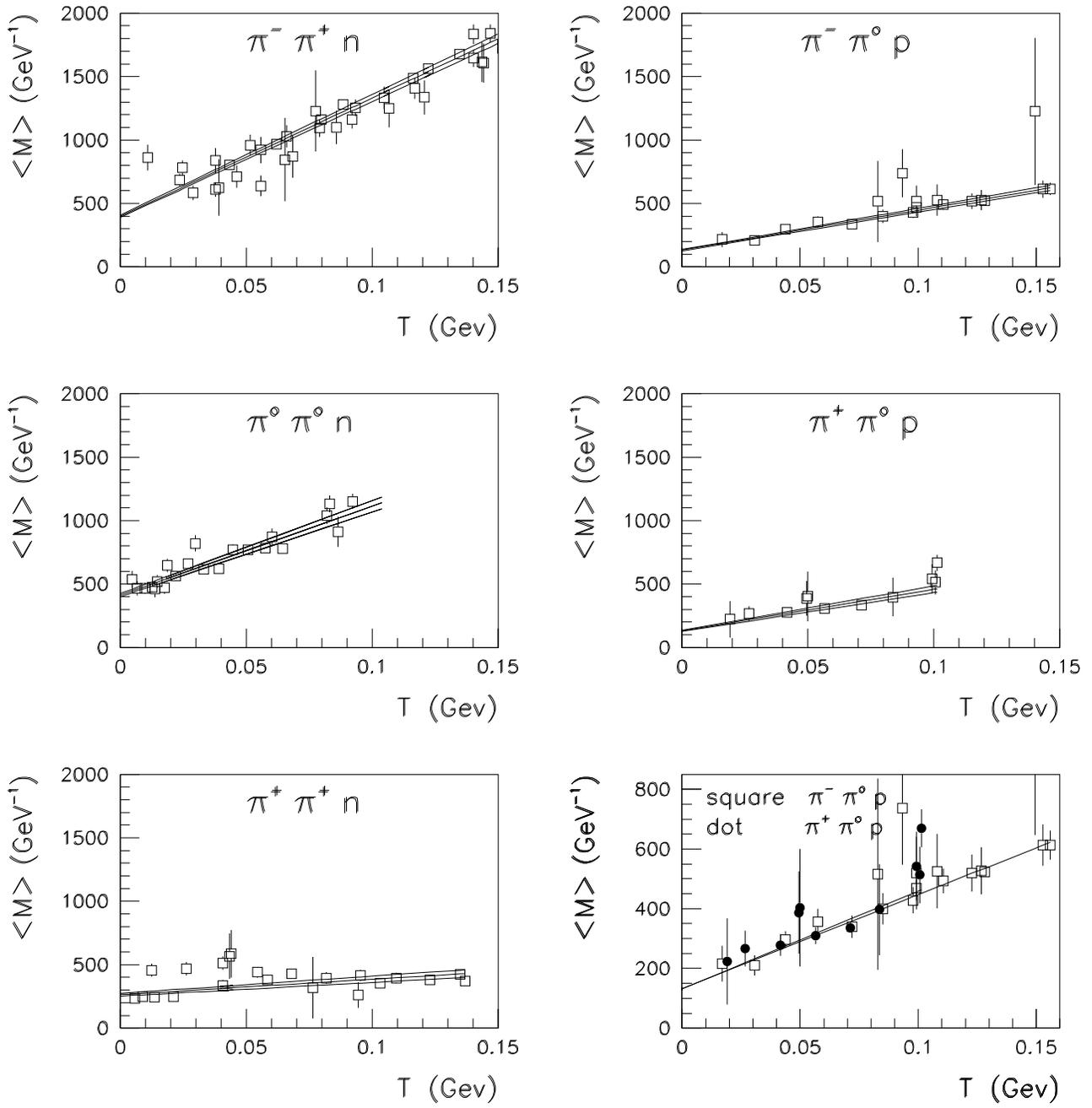}}
   \centerline{
               \parbox{13cm}{
                              \caption{
                                        \label{sigtotlin}
                Linear fit of total cross sections of
                all five channels with 7 parameters
                (quasi--amplitude
        $ \langle M \rangle $
                in
        GeV${}^{-1}$).
                                      }
                             }
              }
\end{figure}

\clearpage
\thispagestyle{empty}

\begin{figure}[ht]
   \epsfxsize=17cm
   \epsfysize=17cm
   \centerline{\epsffile{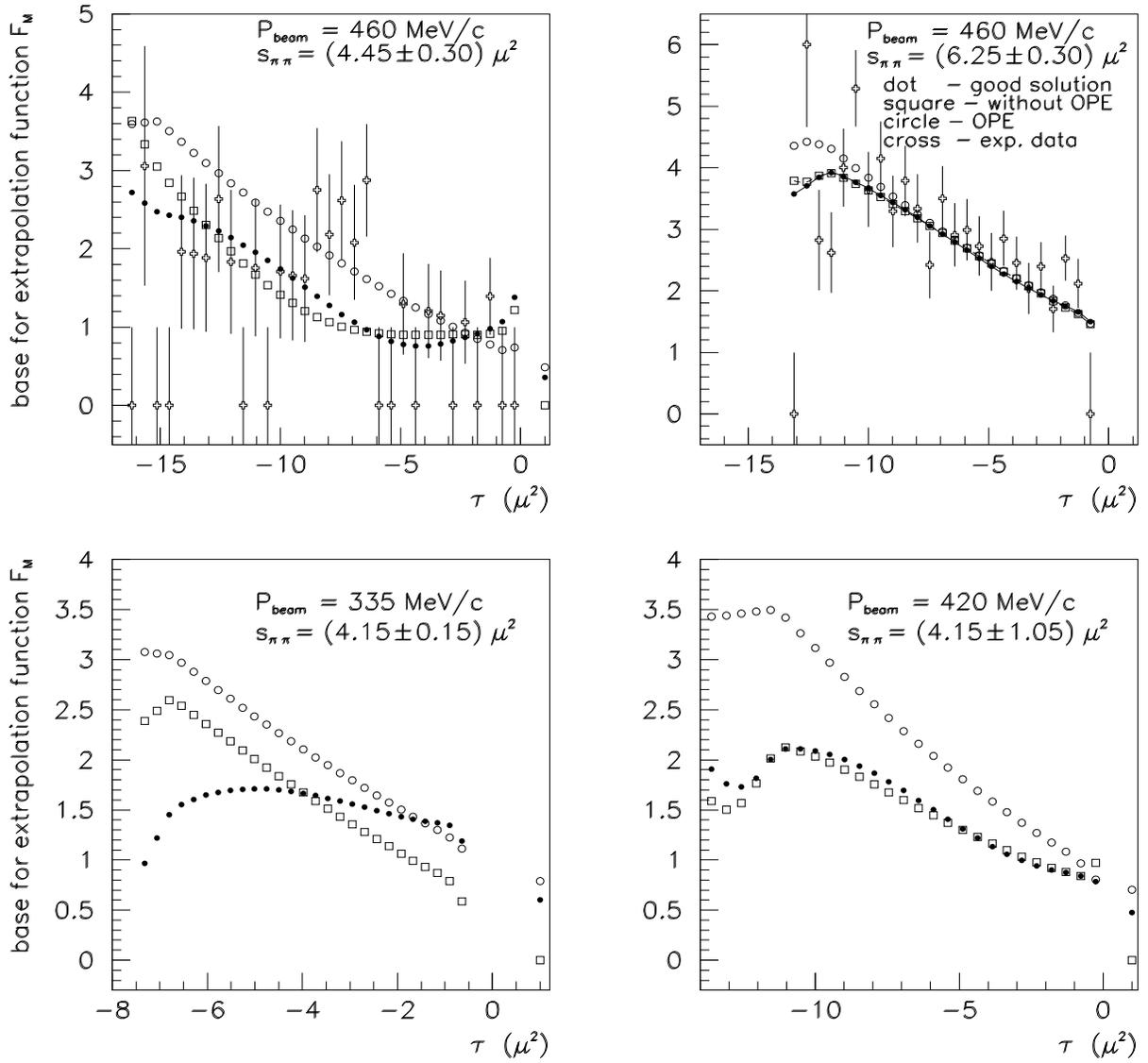}}
   \centerline{
               \parbox{13.cm}{
                              \caption{
                Simulations of extrapolation data
        ($ F_{M} $):
                dot    --- the amplitude of the best
                           physical solution;
                square --- the amplitude without
                           the OPE contribution;
                circle --- the pure OPE  amplitude;
                cross  --- the available experimental data of
                           the paper
\cite{Bloh_-+n70}
                           by Blokhintseva.
\label{ChL}
                                      }
                             }
              }
\end{figure}

\end{document}